\documentclass[%
 reprint,
 amsmath,amssymb,
 aps,
prstab,
floatfix,
]{revtex4-1}

\usepackage{graphicx}
\usepackage{dcolumn}
\usepackage{bm}
\usepackage{subfigure}
\usepackage{epstopdf}
\usepackage[english]{babel}

\begin{document}


\title{Beam Commissioning Results from the CBETA Fractional Arc Test}

\author{C. Gulliford}
\author{N. Banerjee}
\author{A. Bartnik}
\author{J. Crittenden}
\author{J. Dobbins}
\author{G. H. Hoffstaetter}
\author{W. Lou}
\author{P. Quigley}
\author{D. Sagan}
\author{K. Smolenski}
\author{V. Vesherevich}
\author{D. Widger}
\affiliation{Cornell Laboratory for Accelerator Based Sciences and Education,
\\ 
Cornell University,
\\
Ithaca, New York 14850, USA }

\author{J.{~}S. Berg}
\author{R. Hulsart}
\author{R. Michnoff}
\author{S. Peggs}
\author{D. Trbojevic}
\affiliation{Brookhaven National Laboratory,\\
Upton, NY 11973-5000, USA}

\author{B. Kuske}
\author{M. McAteer}
\author{J. V{\"o}elker}
\affiliation{Institute for Accelerator Physics,
\\
Helmholtz-Zentrum Berlin,
\\
Berlin, Germany}

\

\author{J. Jones}
\affiliation{STFC Daresbury Laboratory, 
\\
Warrington, Cheshire, WA4 4AD, UK
}

\author{D. J. Kelliher}
\affiliation{
STFC Rutherford Appleton Laboratory,
\\
Harwell Oxford, Didcot, Oxon, OX11 0QX, UK
}






\date{\today}

\begin{abstract}
This work describes first commissioning results from the Cornell Brookhaven ERL Test Accelerator fractional arc test.  These include the recommissioning of the Cornell photoinjector, the first full energy operation of the main linac with beam, as well as commissioning of the lowest energy matching beamline (splitter) and a partial section of the Fixed Field Alternating gradient (FFA) return loop featuring first production Halbach style permanent magnets. Achieving these tasks required characterization of the injection beam, calibration and phasing of the main linac cavities, demonstration of the required 36 MeV energy gain, and measurement of the splitter line horizontal dispersion and $R_{56}$ at the nominal 42 MeV.  In addition, a procedure for determining the BPM offsets, as well as the tune per cell in the FFA section via scanning the linac energy and inducing betatron oscillations around the periodic orbit in the fractional arc was developed and tested. A detailed comparison of these measurements to simulation is discussed.

\end{abstract}

\pacs{Valid PACS appear here}
\maketitle


\section{Introduction\label{sec:intro}}

The construction of a high energy, high luminosity, polarized Electron-Ion Collider (EIC) remains one of the highest priorities for the nuclear physics and accelerator communities and continues to drive research and development of many state-of-the-art accelerator technologies \cite{ref:LongRange1,ref:NAS}.  These include (but are not limited to): high-brightness, high current electron sources capable of delivering both polarized and unpolarized electrons, various electron-ion cooling schemes, as well as (multi-pass) energy recovery linacs (ERL).  In fact, both the electron - Relativistic Heavy Ion Collider (eRHIC) under design at Brookhaven National Lab (BNL), and the Jefferson lab Electron-Ion Collider (JLEIC) under design at Thomas Jefferson National Accelerator Facility (TJNAF) require high current electron beams for use in electron ion cooling, making the use of an ERL in either EIC design all but required \cite{ref:CBETAdr}.  Meeting the design parameters in either case will require significant development of the underlying accelerator technologies, as well as investigation of possible cost reducing mechanisms.  In the case of the eRHIC design(s), the use of Fixed Field Alternating-gradient (FFA) recirculating loop(s) \cite{ref:ffag1,ref:ffag2} may provide significant cost reduction by shortening the length of the linac, as well as minimizing the number of recirculating loops required.

\begin{figure*}[ht!] 
\includegraphics[width=0.95\textwidth]{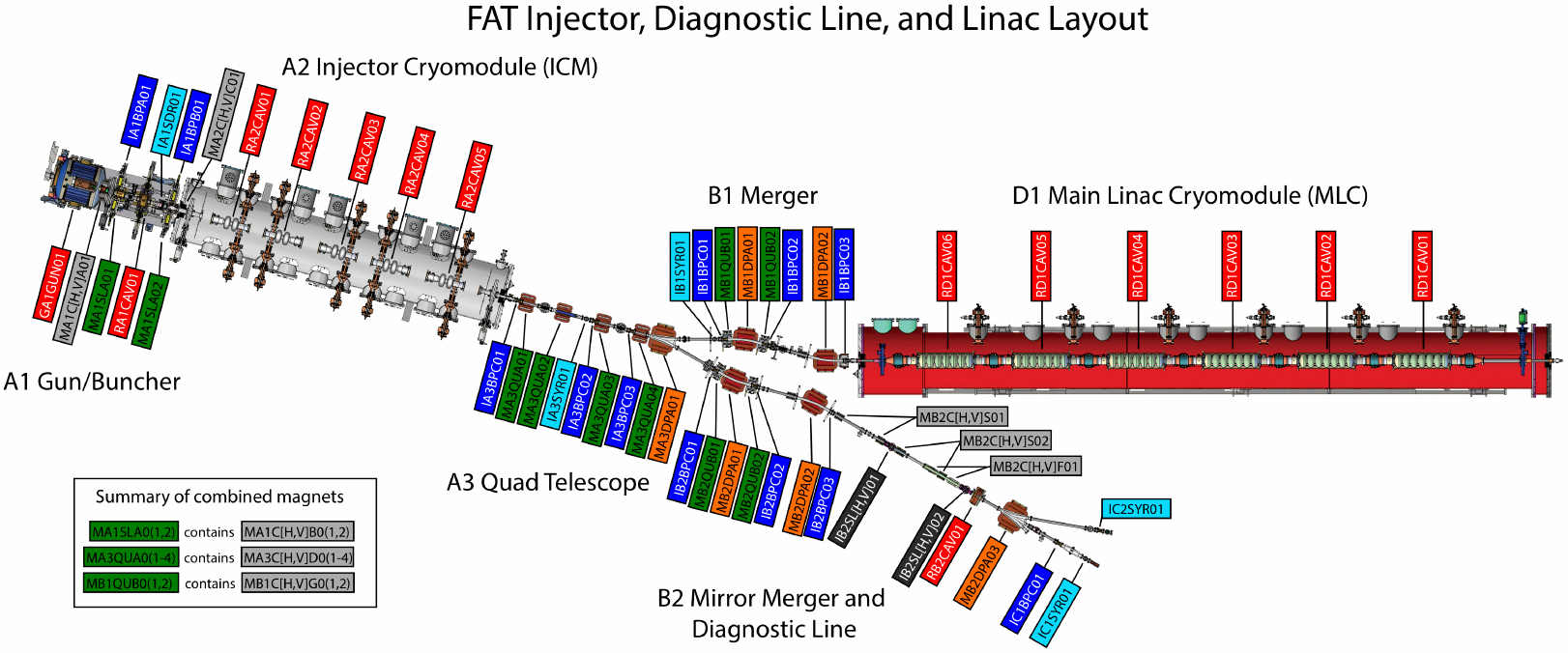}
\caption{\label{fig:fat_injector_layout}Layout of the injector, merger, linac, and diagnostic sections of the FAT experiment. The beam begins in the gun section (``A1''), accelerates to 6 MeV in the Injector Cryomodule (``A2''), and then merges into the Main Linac Cryomodule (``B1'', ``D1'').}
\end{figure*}

As part of this development effort, the Cornell-BNL ERL Test Accelerator (CBETA) \cite{ref:CBETAdr}, a 4-pass, 150 MeV ERL utilizing a Non-scaling Fixed Field Alternating-gradient (NS-FFA) permanent magnet return loop, is currently under design and construction at Cornell University through the joint collaboration of Brookhaven National Lab (BNL) and the Cornell Laboratory for Accelerator based Sciences and Education (CLASSE). Building on the significant advancements in high-brightness photoelectron sources and SRF technology developed at Cornell \cite{ref:hcrecord,ref:lowemitter,ref:lowemitter2,ref:lowemitter3,ref:mlcfab,ref:mlcperf}, as well as the FFA magnet and lattice design expertise from BNL \cite{ref:magdes,ref:1stprod,ref:halbach2,ref:halbach3,ref:halbach4,ref:ipac18,ref:ffacheck}, CBETA will establish operation of a multi-turn SRF based ERL utilizing a compact FFA return loop with large energy acceptance (a factor of roughly 3.6 in energy), and thus demonstrate one possible cost-reduction technology under consideration for the eRHIC design. Moreover, successful completion of the CBETA project requires the study and measurement of many critical phenomena relevant to both the EIC and ERL communities. Examples include the Beam-Breakup (BBU) instability, halo-development and collimation, as well as Coherent Synchrotron Radiation (CSR) microbunching and energy spread growth  \cite{ref:CBETAdr}.

In order to demonstrate the viability of the CBETA design, the Fractional Arc Test (FAT) was added to the CBETA commissioning sequence. Completed in the spring of 2018, this test brought together for the first time elements of all of the critical subsystems required for the CBETA project: the injector, the Main Linac Cryomodule (MLC), the low energy (S1) splitter line which includes several new electromagnets, a path length adjustment mechanism, and a new BPM system, as well as a first prototype production permanent magnet girder featuring 4 cells of the FFA return loop with its own corresponding vacuum system and BPM design.  Consequently, the main technical goals of the FAT included: recommissioning of the injector, full energy commissioning of the main linac with beam, commissioning of the S1 splitter line including beam-based calibration of the S1 magnets and diagnostics, testing of the splitter path length adjustment mechanism, calibration of the S1 and FA BPM designs, and beam based characterization of the permanent magnets in the fractional arc. 

\section{Experimental Setup\label{sec:expset}}

Construction of the CBETA machine at Cornell began in earnest with the disassembly and removal of the injector \cite{ref:hcrecord,ref:lowemitter,ref:lowemitter2,ref:lowemitter3,ref:FullIons} from its original experimental hall in early 2015. At this time, the Injector Crymodule (ICM) was temporarily removed for maintenance, and a short beamline constructed to study high current operation (up to 45 mA) from the original Cornell DC gun in the CBETA experimental hall. Between 2016-2017, after the maintenance on the ICM was completed, the injector was rebuilt and recommissioned, now with the Cornell Mark II DC gun, which features a segmented insulator design \cite{ref:jaredRSI,ref:redmte,ref:lasershaping2}. During that time the MLC was installed and tested without beam \cite{ref:mlcperf} in the CBETA experimental hall before being moved to its final location for CBETA. Work on the permanent magnets and girder design proceeded at BNL in parallel to this effort \cite{ref:magdes,ref:1stprod,ref:halbach2,ref:halbach3,ref:halbach4}, resulting in the first production permanent magnet girder being assembled in early 2018. Upon completion, this girder was sent to Cornell and installed along with the lowest energy splitter line in anticipation of the FAT experiment that spring.  

\subsection{Layout\label{ssec:layout} and Description}

\begin{figure*}[ht!] 
\includegraphics[width=0.8\textwidth]{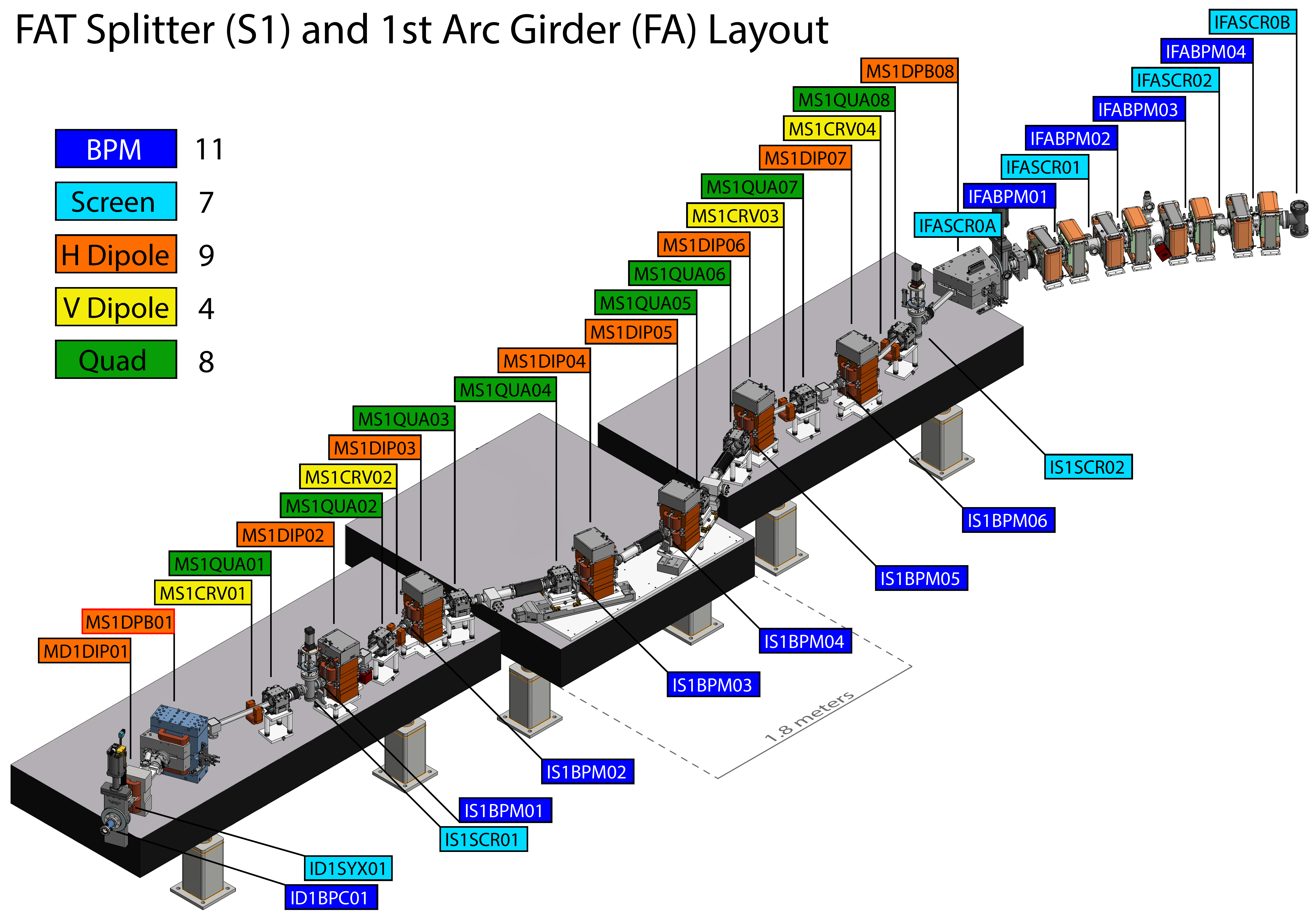}
\caption{\label{fig:fat_layout}Layout of the post linac section of the FAT experiment.  Beam exits the linac and enters the S1 splitter line at the bottom right and proceeds to the fractional FA arc (top right).}
\end{figure*}

Together, Fig.~\ref{fig:fat_injector_layout} and \ref{fig:fat_layout} show the beamline layout of the CBETA FAT.  In particular Fig.~\ref{fig:fat_injector_layout} shows the Cornell injector, merger section, diagnostic beamline, and main linac. For these measurements, the segmented Cornell DC gun was high voltage processed up to roughly 350 kV using the same techniques described in \cite{ref:jaredRSI}.  However, given the limited number of photocathodes currently available for the CBETA project, the gun voltage was set to 300 kV in order to eliminate any risk of degradation from possible vacuum activity at higher voltages. This work makes use of a single NaKSb photocathode similar to those used previously in the Cornell injector \cite{ref:hcrecord}.  At the time of these measurements, the cathode  quantum efficiency was roughly 4.5\%.  While the mean tranvsverse energy (MTE) of this cathode was not measured directly, similarly grown cathodes in the past typically had MTE's around 140 meV.  The drive laser used in conjunction with this cathode is the same 50 MHz, 520 nm laser system \cite{ref:GHzMHzLasers} used previously for low emittance/high bunch charge measurements \cite{ref:lowemitter,ref:lowemitter2,ref:lowemitter3}. This laser produces roughly 1 ps long pulses, which are shaped longitudinally using four rotatable birefringent crystals set to produce a roughly flat-top distribution with a 9.3~ps RMS duration.  The use of a pulse picking Pockels cell following shaping allows for reduction of the average beam power. Typical operation saw generation of pulse trains between 0.3-1.0 microseconds long, at rates from 0.3-2.0 kHz. 

Much of the CBETA injector layout following the gun remains the same as described before \cite{ref:lowemitter}: the beam is compressed transversely and longitudinally using a normal conducting 1.3 GHz bunching cavity and two emittance compensation solenoids located in the beamline immediately after the gun (labeled as ``A1''), before being accelerated to the CBETA injection energy of 6 MeV in the injector cryomodule (ICM).  Following the ICM the beam is passed through a four-quad telescope (``A3'') and a merger section comprised of a conventional three-dipole achromat (``B1''), and matched into the linac (``D1'').  For characterization of the injector beam and to verify this matching, the FAT layout features a diagnostic beamline line (``B2") comprised of a separate merger mirroring the merger into the linac, followed by a suite of diagnostics including the Cornell Emittance Measurement System (EMS). The placement of the EMS at roughly the corresponding position of the first MLC cavity allows for detailed characterization of the beam entering the MLC. The EMS installed here makes use of the same two-slit direct phase space measurement system described in detail in \cite{ref:bmsc,ref:heng}, and features a vertical deflecting cavity \cite{ref:defcav} for time resolution of the vertical phase space.  Following the EMS is a dipole spectrometer, which combined with the deflecting cavity allows for the direct measurement of the longitudinal phase space of the beam. 

Beams sent to the main linac nominally receive a total of 36 MeV from the combined voltage of the six 7-cell 1.3 GHz SRF cavities before being passed into the CBETA low energy splitter line (labeled ``S1''). Through the settings on its 8 dipoles and 8 quadrupole magnets manufactored by Elytt Magnets \cite{ref:Elytt}, the S1 splitter line is designed to match the beam orbit and Twiss parameter values required for proper transport through the FFA return loop. In addition, the S1 line features a pair of automated adjustment stages and bellows which provide the necessary path length control required to establish energy recovery in future CBETA experiments.

\begin{figure*}[ht!]
    \begin{center}
         \subfigure[\hspace{0.2cm}GPT (left of dashed line) and Bmad (right of dashed line) predictions for $\alpha$.]{%
            \label{fig:design_alpha}
            \includegraphics[width=0.3\textwidth]{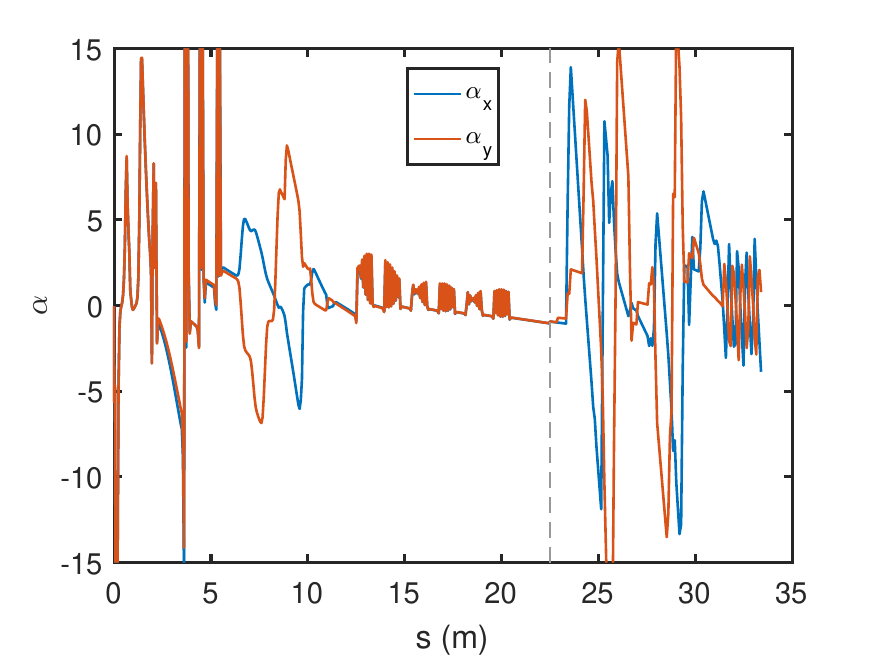}
        }%
        \subfigure[\hspace{0.2cm}GPT (left side) and Bmad (right side) predictions for $\beta$.]{%
           \label{fig:design_beta}
           \includegraphics[width=0.3\textwidth]{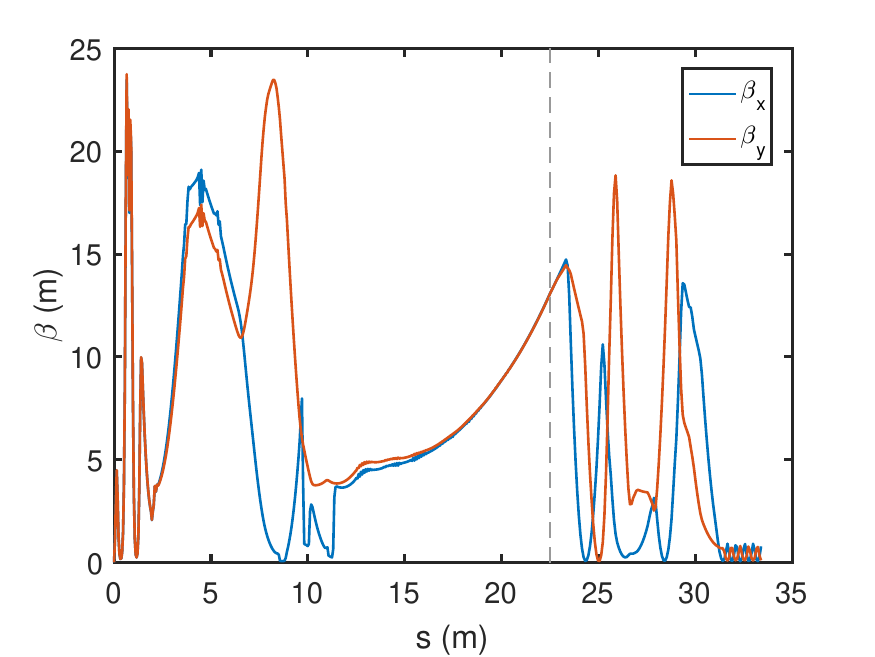}
        }
        \subfigure[\hspace{0.2cm}Bmad prediction for the dispersion $\eta$.]{
           \label{fig:design_eta}
           \includegraphics[width=0.3\textwidth]{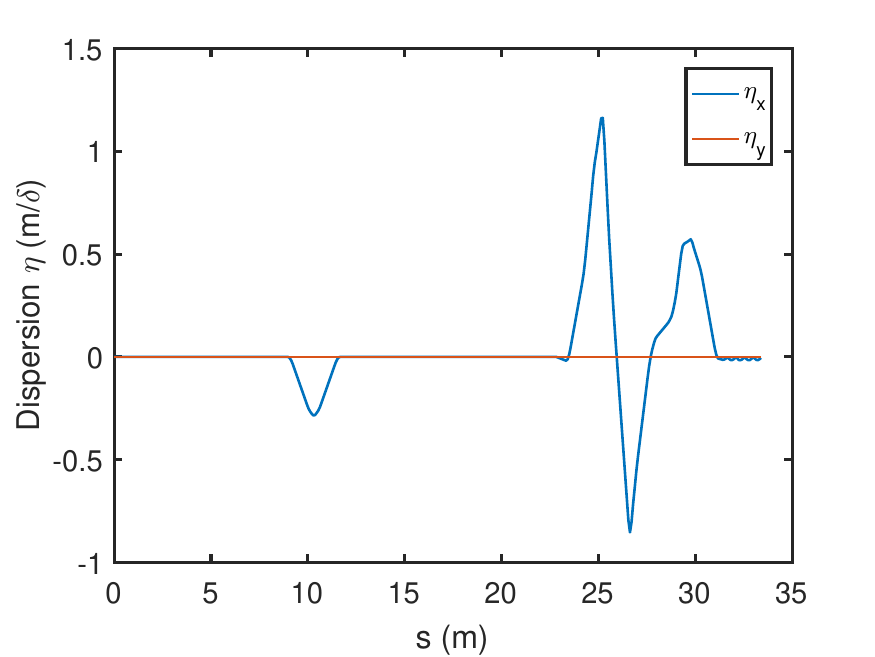}
        }
    \end{center}
    \caption{%
    \label{fig:design_twiss}
    Nominal simulated Twiss parameters: beam dynamics with space charge are simulated up to $s=22.5$ m (just after the linac) using GPT.  The resulting beam distribution is then tracked using Bmad to end of the fractional arc.  The dashed line in (a) and (b) shows the point where the two simulations are joined. In the plot of dispersion, only the Bmad simulation is shown for all $s$.}%
\end{figure*}

The fractional FFA arc (labeled ``FA") is a prototype example of the first arc girder for CBETA. The magnets are all Halbach style permanent magnets \cite{ref:halbach} of two types, a focusing quadarupole and a defocusing gradient dipole \cite{ref:magdes,ref:ipac18}. It begins with a half-length defocusing magnet \cite{ref:1stprod}, and then four cells of the periodic FFA arc lattice, each of which consists of a focusing and a defocusing magnet in a doublet configuration. There are four BPMs throughout the girder in periodic positions centered in the shorter drift of the cell, all of which were instrumented for the FAT, and four total viewscreens: two viewscreens within the arc (centered in the long drifts between cells 1 and 2, and 3 and 4), and one viewscreen at both the entrance and exit to the arc. The horizontal and vertical corrector dipoles that are wound around the permanent quadrupole magnets were installed but not powered during the FAT test.

\subsection{Modeling\label{ssec:model}}

The majority of the CBETA lattice design was done using Bmad, primarily within the Tao simulation environment \cite{ref:BMAD1,ref:TAO1}. In order to avoid complications from space charge, the design lattice begins after the MLC, where the beam energy significantly reduces the space charge forces. This approach allows the for fast evaluation of the single particle dynamics through the remainder of the CBETA lattice. Space charge modeling from the cathode up to that point was performed using General Particle Tracer (GPT) \cite{ref:gpt1,ref:gpt2} in conjunction with the user interface to injector's EPICS control system detailed in \cite{ref:lowemitter}.

In order to facilitate online simulation of the single particle dynamics using Tao, we developed a new application called the CBETA Virtual Machine (CBETA-V) \cite{ref:CBETAVM} which combines the functionality of the Tao environment with the ability to communicate with EPICS records\cite{ref:EPICS}.  When running, this software creates its own copy of the CBETA EPICS power supply and diagnostic records and links them to the corresponding beamline elements in Tao, allowing users to command virtual optical elements in the simulation via standard EPICS commands.  By changing any of these soft EPICS records, the CBETA virtual machine recomputes all relevant single particle tracking data (i.e. centroid orbit, dispersion, transfer matrix, etc), and publishes the results to its own EPICS diagnostic records, thus making the virtual machine data available to the user via EPICS in exactly the same manner as real machine data. This allows for simultaneous online display of both measured and simulated data for use by operators.  Additionally, the virtual machine provides the ability to develop automated measurement procedures to command and take data from both the real and virtual machine, in the later case prior to beam measurements, and to easily produce simulated predictions of measured data.

Many of the experimental procedures used in this work were developed and tested offline in this manner.  The software also provides a ``sync'' mode where the CBETA Virtual Machine continuously monitors the status of real CBETA EPICS records, and updates the simulated machine data upon detecting a change in the settings of the real machine, thus providing a useful online diagnostic as well.  For the low energy portion of the FAT layout (injector through MLC), the CBETA Virtual Machine uses the same fieldmap data used in the corresponding GPT injector model described above.  While necessary for capturing the low energy dynamics, tracking through field maps inevitably slows down the simulation, and leads to simulation times of on the order of a few seconds.

\section{Measurements\label{sec:meas}}

The measurements completed during the FAT break down naturally into the following categories: tune-up and characterization of the injector at 6 MeV, beam-based calibration of the MLC cavities, tests of the diagnostics in the S1 splitter line, tune-up and characterization of the beam dynamics in the splitter and FFA fractional arc at the nominal 42 MeV settings, and finally characterization of the beam dynamics through the S1 line and FFA fractional arc over a wide energy range (specifically 38 - 59 MeV).   

\subsection{Injector Tune-up and Characterization\label{ssec:meas:injector}}

Ultimately, the demonstration of the CBETA design parameters hinges on the production and transport of high quality beams through the injector.  For bunch charges within that a significant fraction of the 125 pC peak design charge, space charge forces dominate the beam dynamics in the injector (and to a lesser extent through the first pass of the linac), and thus require detailed modeling and optimization to ensure emittance preservation and proper matching into the main linac. For this work, we made use of the same Multi-Objective Genetic Algorithm optimization (MOGA) software used in \cite{ref:lowemitter,ref:lowemitter2,ref:lowemitter3,ref:coldgun,ref:RFgunUED}, and applied it to 3D space charge simulations of the beam passing through injector, merger, and MLC.  

\begin{table}[!htb] 
\caption{Simulated and Experimental Injector Settings}
\begin{tabular}{l | c | c | c}
\hline
\hline
Name & Simulation & Experiment & Units\\
\hline
Bunch charge & 6 & 6 & pC \\
Laser diameter (rms) & 0.28 & 0.28 & mm \\
Laser duration (rms) & 9.3 & 9.3 & ps \\
Gun voltage & 300 & 300 & kV \\
Solenoid 1 current & -3.07 & -3.15 & A \\
Buncher voltage & 30 & 30 & kV\\
Buncher phase & -90 & -90 & deg.\\
Solenoid 2 current & 2.27 & 2.28 & A \\
ICM 1 voltage & 1500 & 1500 & kV \\
ICM 1 phase & 0 & 0 & deg. \\
ICM 2 voltage & N/A & N/A & kV \\
ICM 2 phase & N/A & N/A & deg. \\
ICM 3 voltage & 1600 & 1600 & kV \\
ICM 3 phase & -15 & -15 & deg. \\
ICM 4 voltage & 1300 & 1300 & kV \\
ICM 4 phase & 20 & 20 & deg. \\
ICM 5 voltage & 1300 & 1300 & kV \\
ICM 5 phase & -20 & -20 & deg. \\
A3 Quad 1 current & 2.30 & 1.7 & A \\
A3 Quad 2 current & 1.21 & 1.0& A \\
A3 Quad 3 current & -2.54 & -1.8 & A \\
A3 Quad 4 current & -1.86 & -1.8 & A \\
B1 Quad 1 current & 6.0 & 6.0 & A \\
B1 Quad 2 current & 6.0 & 6.0 & A \\
\hline
\hline
\end{tabular}
\label{tab:machine_settings}
\end{table}

To ensure the highest beam quality and proper matching of the beam at the end of the MLC and entering the S1 splitter, the optimizer minimized the emittance at the match point ($s=22.5$ m) as a function of the maximum error in the match to the desired four Twiss parameters $\beta_{x,y} = 12.5$ m, and $\alpha_{x,y} = -1$ for several different bunch charges ranging up to 125 pC.  Table~\ref{tab:machine_settings} displays an example of the resulting optimized injector settings for a 6 pC bunch charge. We note that while the design Twiss values are specified at the end of the MLC, their only direct measurement is located in the EMS in the diagnostic line, which is equivalent to a measurement at the entrance to the first MLC cavity, not at the optimized match point. 

\begin{figure*}[!htbp] 
    \begin{center}
         \subfigure[\hspace{0.2cm}Measured horizontal phase space.]{%
            \label{fig:meas_x_phase}
            \includegraphics[width=0.3\textwidth]{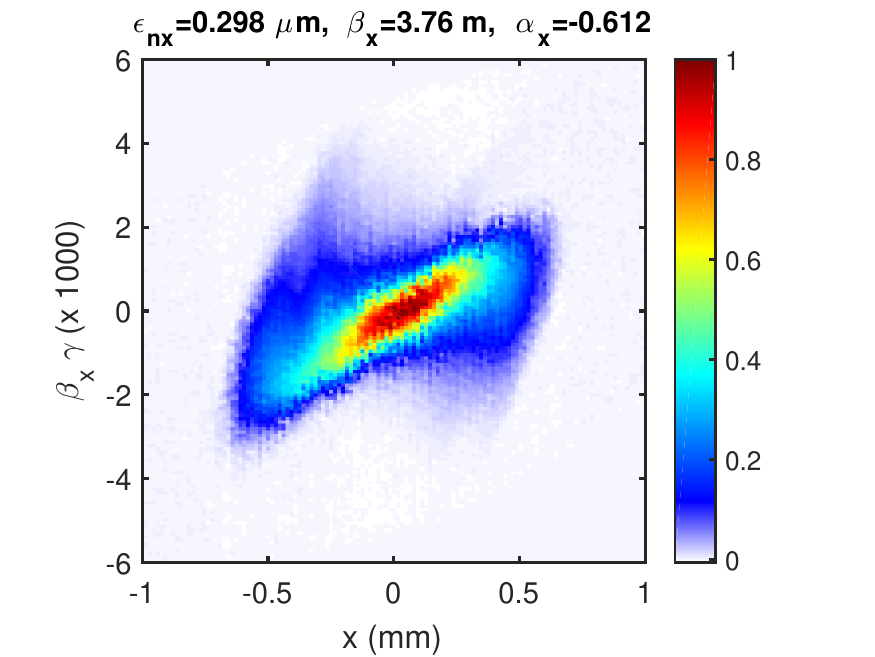}
        }%
        \subfigure[\hspace{0.2cm}Measured vertical phase space.]{%
           \label{fig:meas_y_phase}
           \includegraphics[width=0.3\textwidth]{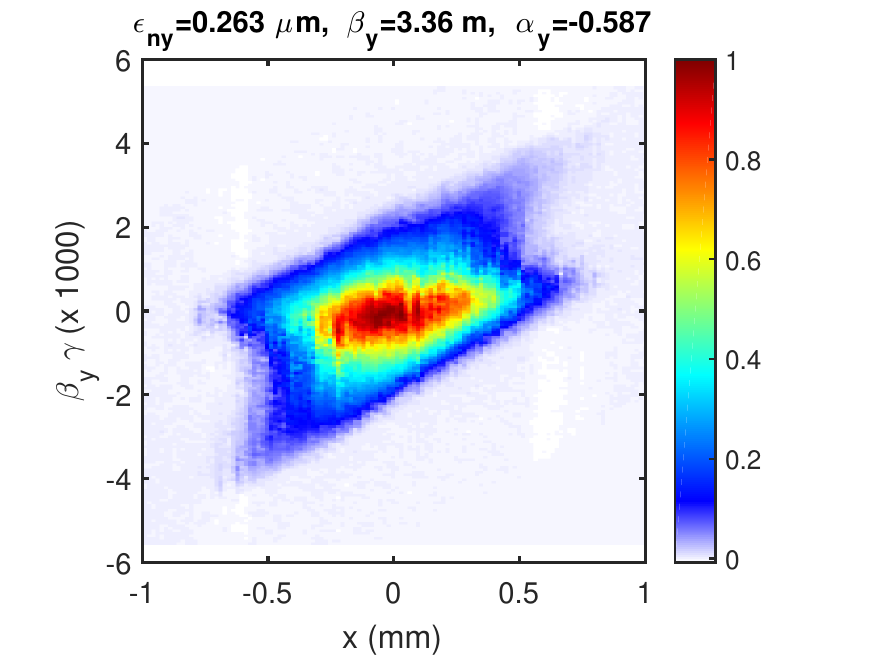}
        }
        \subfigure[\hspace{0.2cm}Measured transverse profile.]{%
           \label{fig:meas_profile}
           \includegraphics[width=0.3\textwidth]{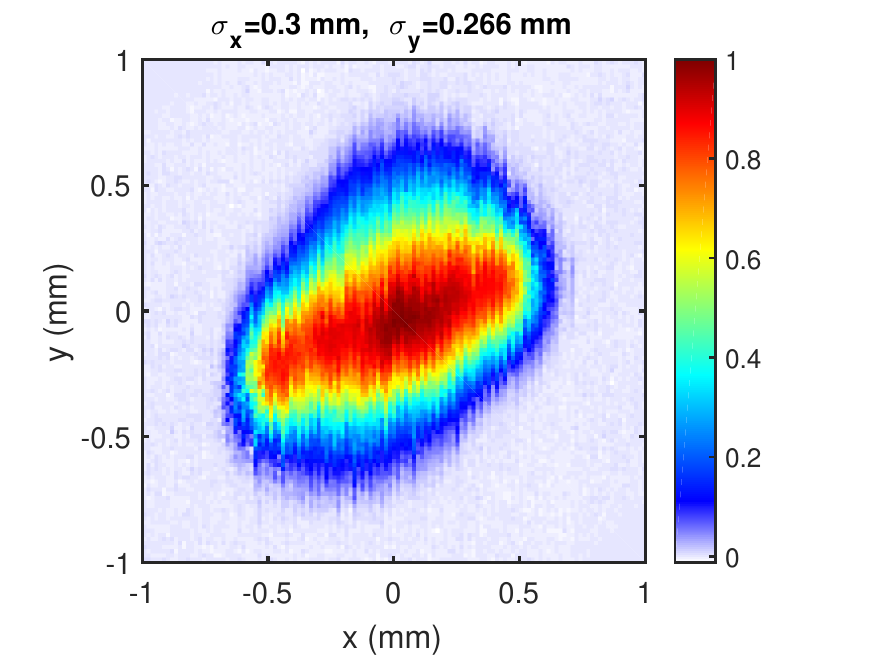}
        }\\ 
        \subfigure[\hspace{0.2cm}Simulated horizontal phase space.]{%
            \label{fig:sim_x_phase}
            \includegraphics[width=0.3\textwidth]{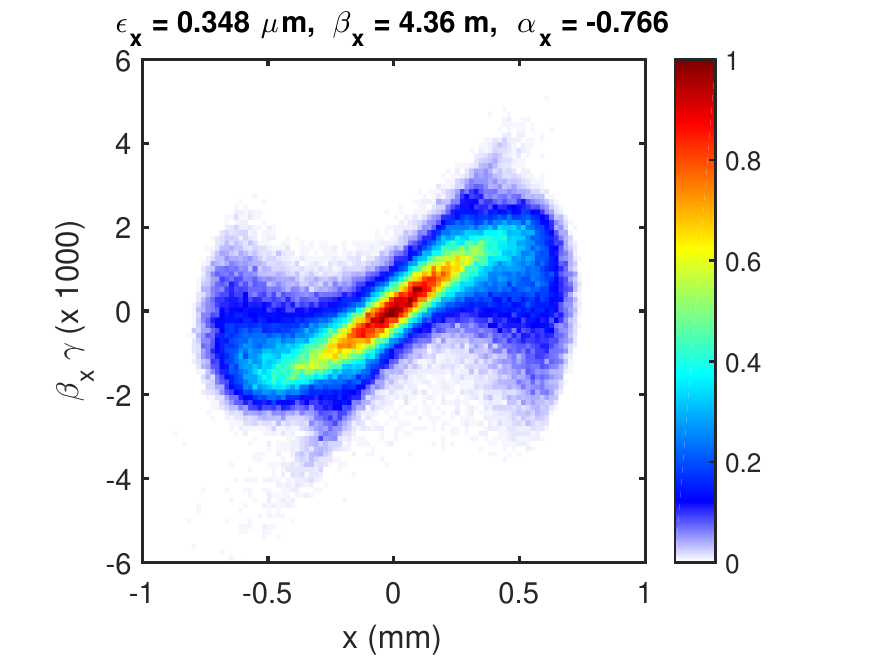}
        }%
        \subfigure[\hspace{0.2cm}Simulated vertical phase space.]{%
           \label{fig:sim_y_phase}
           \includegraphics[width=0.3\textwidth]{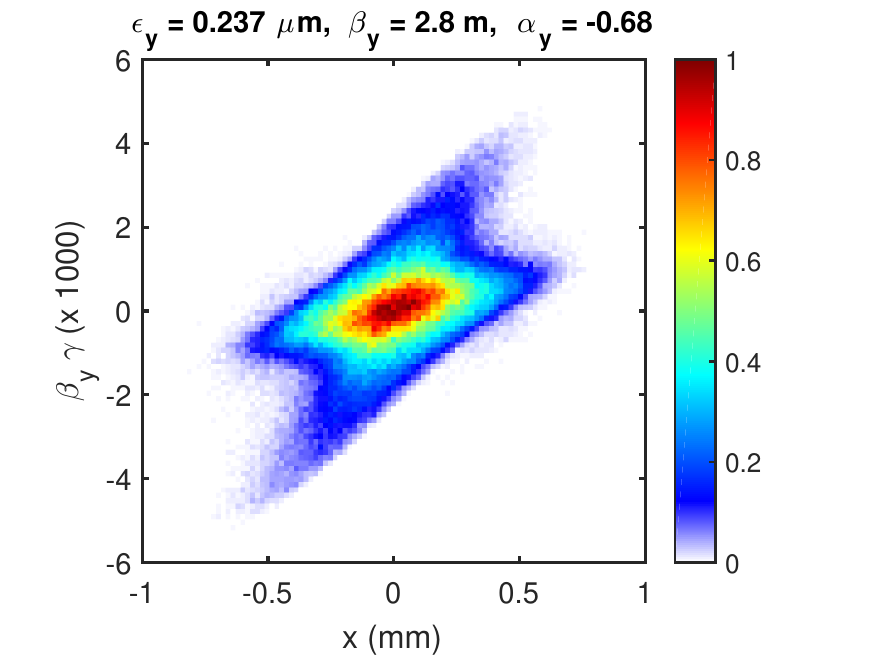}
        }
        \subfigure[\hspace{0.2cm}Simulated transverse profile.]{%
           \label{fig:sim_profile}
           \includegraphics[width=0.3\textwidth]{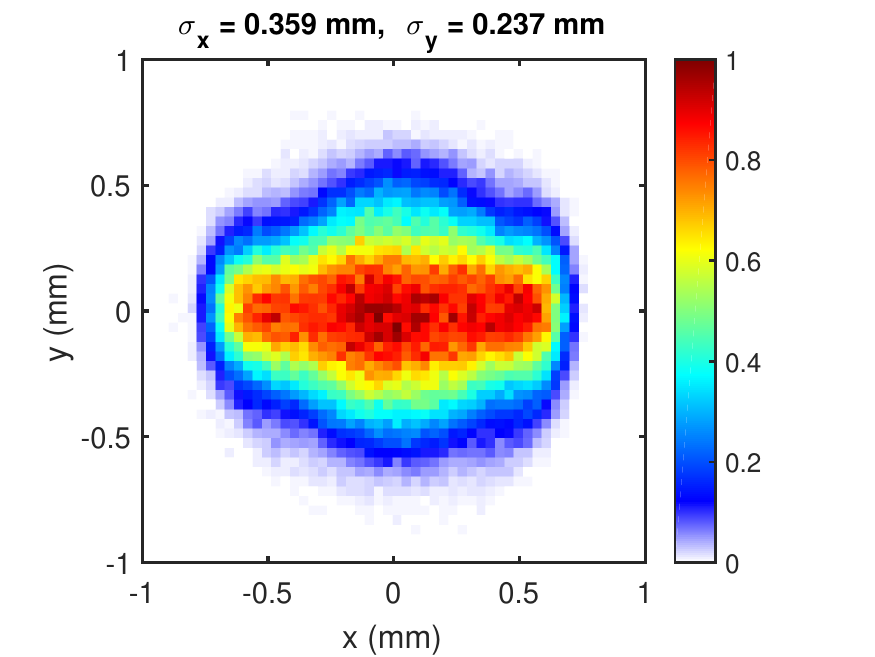}
        }
    \end{center}
    \caption{%
    \label{fig:injector_phase_space}
    Measured (top row) and simulated (bottom row) transverse phase spaces and beam profiles.}%
\end{figure*}

\begin{figure*}[ht!]
    \begin{center}
         \subfigure[\hspace{0.2cm}Simulated (lines) and measured (points) rms beam width.]{%
            \label{fig:inj_beam_size}
            \includegraphics[width=0.38\textwidth]{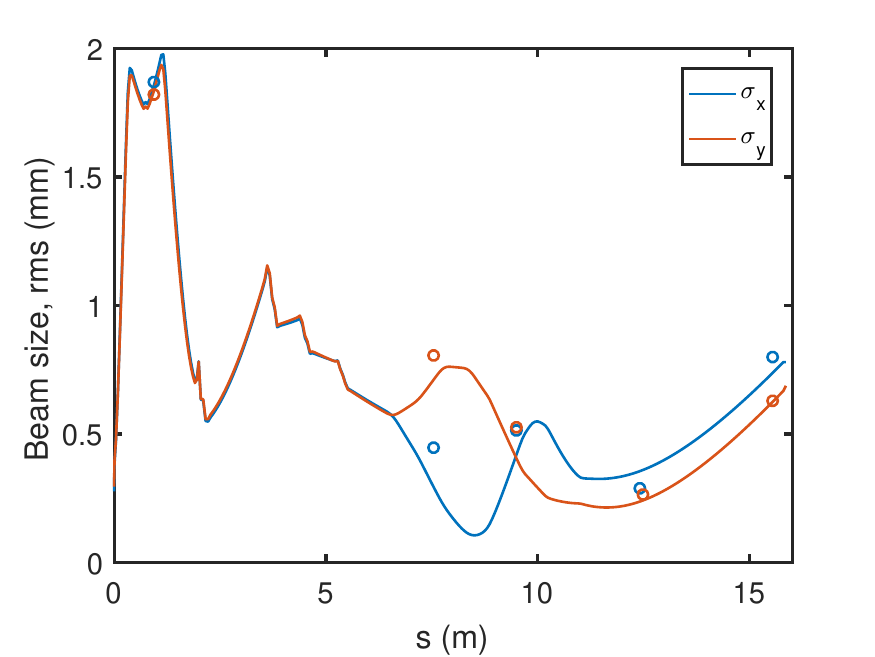}
        }%
        \subfigure[\hspace{0.2cm}Simulated (lines) and measured (points) rms beam emittance.]{%
           \label{fig:inj_emit}
           \includegraphics[width=0.38\textwidth]{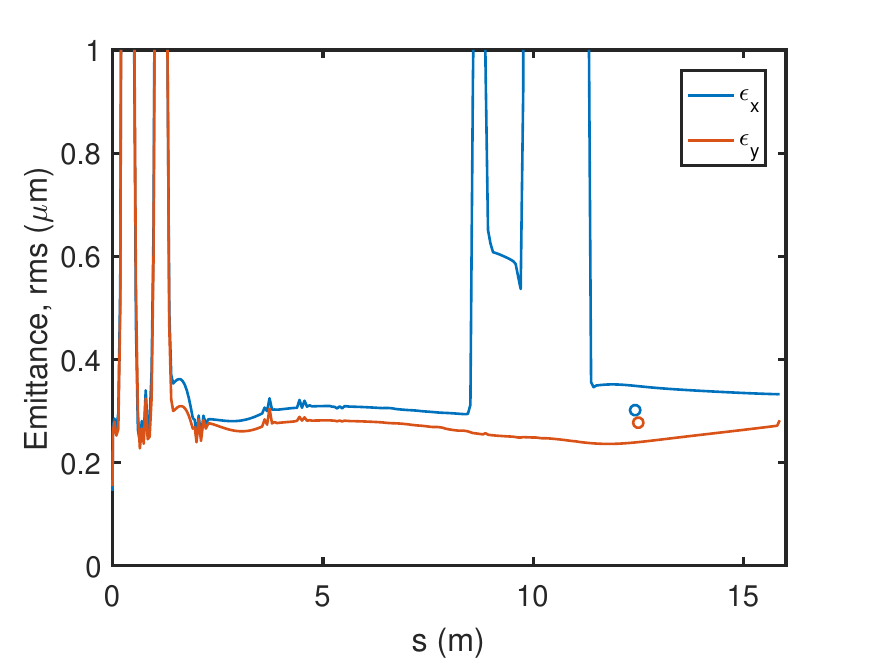}
        } 
        \subfigure[\hspace{0.2cm}Simulated (lines) and measured (points) alpha.]{%
            \label{fig:inj_alpha}
            \includegraphics[width=0.38\textwidth]{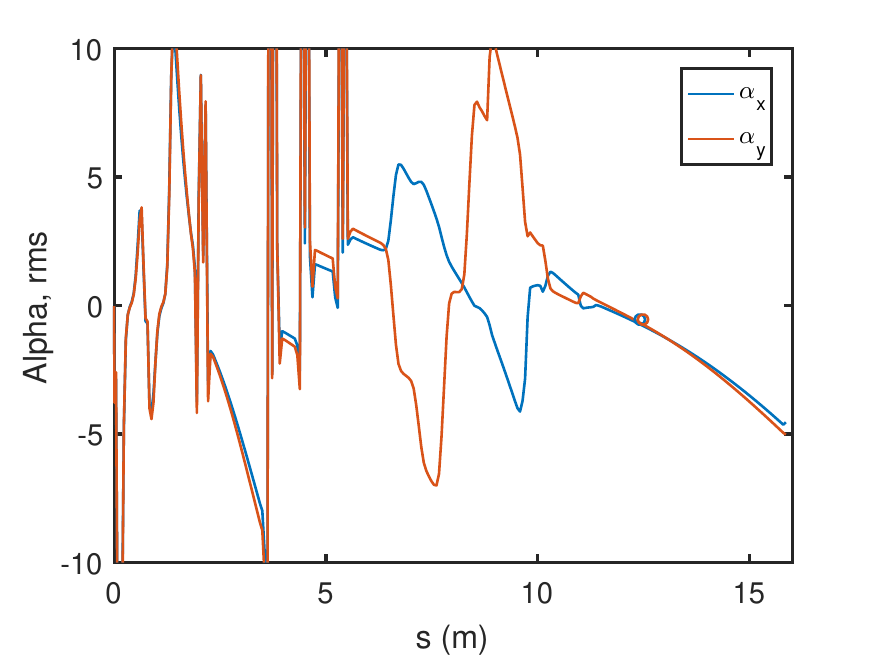}
        }%
        \subfigure[\hspace{0.2cm}Simulated (lines) and measured (points) beta.]{%
           \label{fig:inj_beta}
           \includegraphics[width=0.38\textwidth]{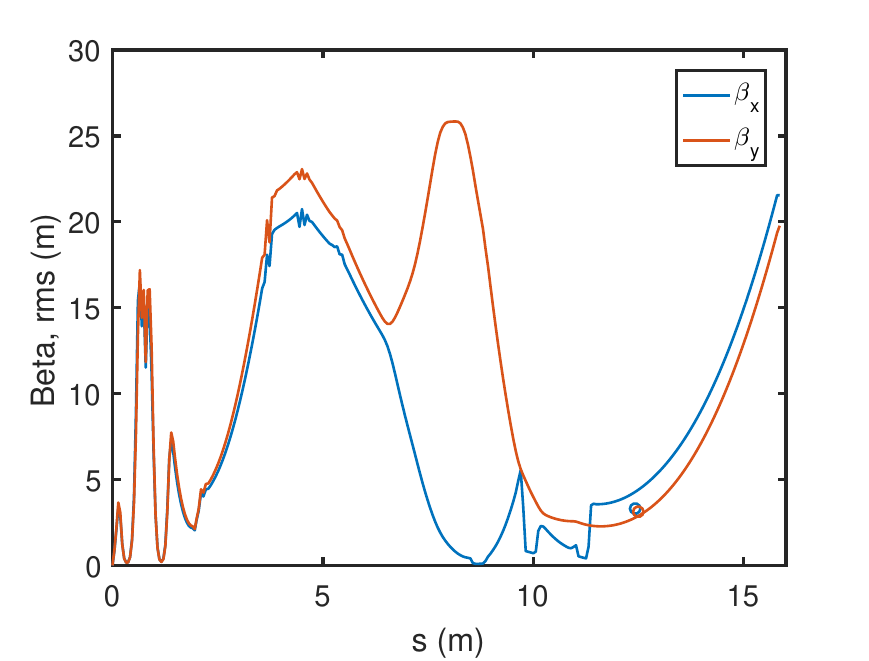}
        } 
    \end{center}
    \caption{%
    \label{fig:injector_twiss_measurements}
     Comparison between simulated (lines) and measured (points) properties of the beam through the injector and diagnostic line. The rms beam widths were measured on both viewscreens and with the EMS, while the other plots only show measured data from the EMS (at $s \approx 12.5$). Importantly, the GPT simulations shown here are taken through the diagnostic line, or equivalently, with the MLC turned off, which is why they disagree qualitatively with Fig.~\ref{fig:design_twiss} after $s\approx12.5$ m. }%
\end{figure*}

We began injector tuning by loading the machine settings from the 6 pC GPT simulation into the machine, and then tuning the settings manually from that starting point. The procedure involved first measuring both the horizontal and vertical phase spaces of the beam using the EMS, calculating the Twiss parameters from the measured distributions, and then manually adjusting machine parameters (primarily the quadrupoles in the A3 section and the solenoids in the A1 section) to try to optimize them. After we reached an acceptable machine setting, we re-ran GPT using the measured initial laser distribution and the used machine settings. Comparisons between the measured and simulated phase spaces are shown in Fig.~\ref{fig:injector_phase_space}, and their predicted Twiss functions in Fig.~\ref{fig:injector_twiss_measurements}. In addition, we also measured the beam size on all viewscreens throughout the injector at this machine setting, and summarized that data in Fig.~\ref{fig:injector_twiss_measurements}. In general, the phase space data shows strong qualitative agreement. Perhaps most interesting is the skew in the measured transverse profile of the beam at the EMS slits. The origin of this remains unresolved, but could be due to tilt errors in the alignment of the quadrupoles in the A3 section. Better agreement with simulation will require resolving this problem in future experiments.
It is important to emphasize here that the agreement shown in Figs.~\ref{fig:injector_phase_space} and \ref{fig:injector_twiss_measurements} represents the best attempt at injector tuning made thus far.  Unfortunately, subsequent attempts at reproducing these results did not achieve this level of agreement.  

\subsection{Main Linac Commissioning\label{ssec:meas:mlc}}


\subsubsection{MLC Cavity Energy Gain Calibration and Phasing\label{sssec:mlccal}}

Multiple methods were considered for determining the absolute energy gain calibrations of each MLC cavity. For example, a simple spectrometer consisting of a dipole immediately downstream of the MLC followed by a BPM allows for an accurate determination of the beam momentum, provided the spectrometer calibration is known (either by careful analysis of the dipole field or through an initial calibration measurement at a well known beam momentum). Unfortunately, the FAT layout features a sector magnet just after the MLC.  Use of a sector magnet as a spectrometer requires control of the beam orbit entering the dipole, as the integrated field through the magnet depends on the incoming beam position. 
This is difficult in the FAT layout as there is only a single BPM before this magnet. As an alternative, we opted to use the change in beam arrival phase on the first BPM after the MLC (ID1BPC01) as a measure of energy via the beam's average velocity.

In general, the phase on the BPM downstream of the cavity is given by
\begin{eqnarray}
\phi = \frac{\omega}{c}\int_{\text{cav}}^{\text{bpm}}\frac{ds}{\beta(V_c,\phi_b,E_0)} + \phi_{\text{offset}},
\label{eqn:genphase}
\end{eqnarray}
where $\beta=\beta(s)$ is the velocity of the beam along the trajectory from cavity to BPM determined by the cavity voltage $V_c$ (defined as the on-crest energy gain of a speed of light particle), phase relative to the beam $\phi_b$, and initial beam energy $E_0$.  The term $\phi_{\text{offset}}$ represents an unknown BPM specific offset. In this expression, the path length differential depends on the beam's instantaneous transverse angles with respect to the beamline axis: $ds=dz\sqrt{(1+{x^{\prime}}^2 + {y^{\prime}}^2)}$.  Note that the beam pipe aperture effectively limits the size of the angle terms. Assuming a $2^{\prime\prime}$ pipe diameter and minimum drift length from the last cavity to the BPM of roughly 2 meters effectively limits the angle terms to roughly  $x^{\prime}\lesssim 25$ mrad. This implies that the correction to the beam phase from the angle terms is on the order of $\delta\phi(x^{\prime})\lesssim 10^{-3}$. For a (maximum) beam phase change of $360^{\text{o}}$ this corresponds to a correction of about $0.36^{\text{o}}$, which is within noise level of the BPM phase reading ($0.3^{\text{o}}$ for nominal operation), and justifies the approximation $ds\approx dz$. Forming the phase change with respect to the on-crest beam arrival phase gives: 
\begin{eqnarray}
\Delta\phi &=& \frac{\omega}{c}\int_{\text{cav}}^{\text{bpm}}dz\left(\frac{1}{\beta(V_c,\phi_b)} -\frac{1}{\beta(V_c,\phi_b=0)}\right),
\label{eqn:dphi}
\end{eqnarray}
\begin{figure}[htb!]
    \begin{center}
         \subfigure[\hspace{0.2cm}First cavity (RD2CAV06) phase shift.]{%
            \label{fig:cavity_dE_6}
            \includegraphics[width=0.38\textwidth]{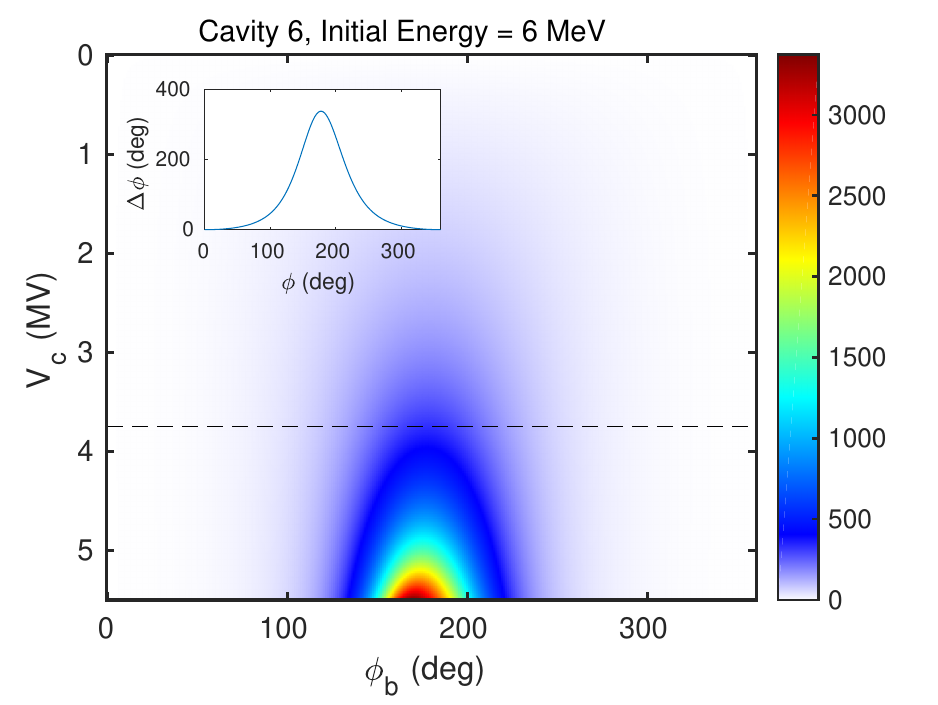}
        }\\%
        \subfigure[\hspace{0.2cm}Last cavity (RD2CAV01) phase shift.]{%
           \label{fig:cavity_dE_1}
           \includegraphics[width=0.38\textwidth]{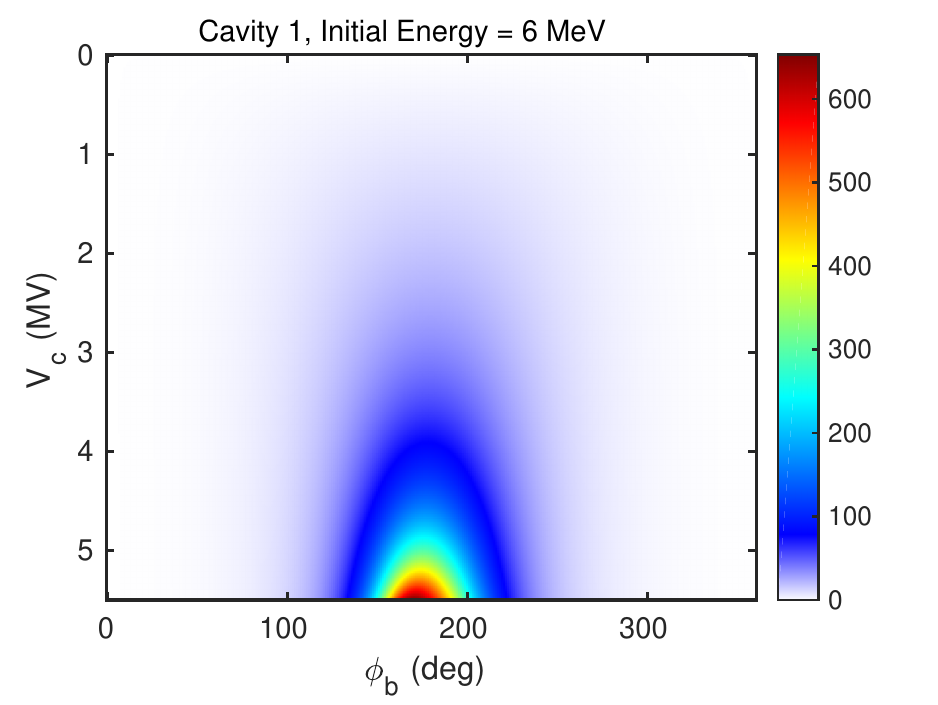}
        }\\
    \end{center}
    \caption{%
    \label{fig:cavity_dE}
    Relative BPM phase change simulated $\Delta\phi$ as a function of both cavity phase and cavity energy gain for the first cavity (a) and last cavity (b). The dashed line and inset shown in (a) display the voltage and \emph{predicted} phase change used in calibrating the first MLC cavity.}%
\end{figure}
In general, accurate evaluation of this expression requires particle tracking through the MLC cavity fieldmap (scaled for a desired cavity voltage and phase) and relies on knowledge of the correct drift length from cavity to BPM.

Fig.~\ref{fig:cavity_dE} shows the phase change computed numerically as a function of cavity voltage $V_c$ and beam phase $\phi_b$ for a 6 MeV beam entering the first (RD1CAV06) and last cavity (RD1CAV01), respectively.  While not shown, similar results for the other MLC cavities demonstrate that the phase change accurately scales with the distance from cavity to BPM.  For the examples shown, this implies that the data in Fig.~\ref{fig:cavity_dE_1} should be approximately given by scaling the data in Fig.~\ref{fig:cavity_dE_6} by roughly 9/2, which one can see holds true by examining the maximum in the each plot.

\begin{figure}[htb]
\includegraphics[width=0.38\textwidth]{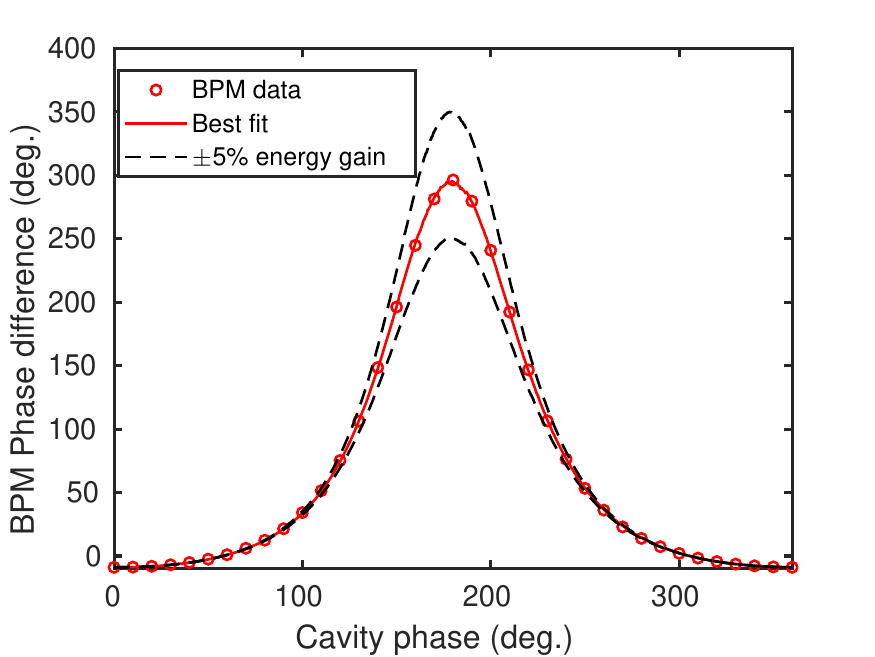}
\caption{\label{fig:cavity_calib}Change of the arrival time of the beam (shown as a phase change with respect to the RF clock) as a function of MLC cavity 6 phase set-point at a constant cavity gradient. Measured points are shown compared to the best fit model, and models that have $\pm5\%$ energy gain.}
\end{figure}

Inversion of Eq.~(\ref{eqn:dphi}) in the least squares sense provides a simple way of determining the cavity energy gain calibration $V_c$, as well as the initial beam energy entering the cavity $E_0$ (if not known) from the measured BPM phase change $\Delta\phi$ for each cavity. The two cavity parameters $V_c$ and $\phi_b$ in this expression suggest two methods for changing the energy gain through each cavity: ramping up the cavity voltage at constant phase, or changing the cavity phase at constant voltage. The voltage scan method suffers from the fact that it implies knowledge of the beam phase $\phi_b$ at each cavity setting, while the phase scan method requires no knowledge of $\phi_b$ (the on-crest phase is found by including it as another fit parameter) and still provides significant measured phase change, particularly when decelerating the beam.  After settling on this method, each MLC cavity was calibrated by first setting the voltage to a fixed value of roughly 2-4 MeV, and then slowly changing the cavity phase from 0-$360^{\text{o}}$. An example set of data for the first MLC cavity is shown in Fig.~\ref{fig:cavity_calib}. In the figure, the trend for the best fit energy calibration is shown, along with energy gains $5\%$ higher and lower, to give a sense of the measurement sensitivity.  From the random error in the BPM phases, we estimate an error of approximately $0.4\%$ for the final cavity calibrations. Assuming this represents the most significant source of error, this implies an overall error in the total MLC energy gain of roughly $\sqrt{6} \cdot 0.4\% \approx 1 \%$ for any given machine setting.

\begin{figure}
\includegraphics[width=0.38\textwidth]{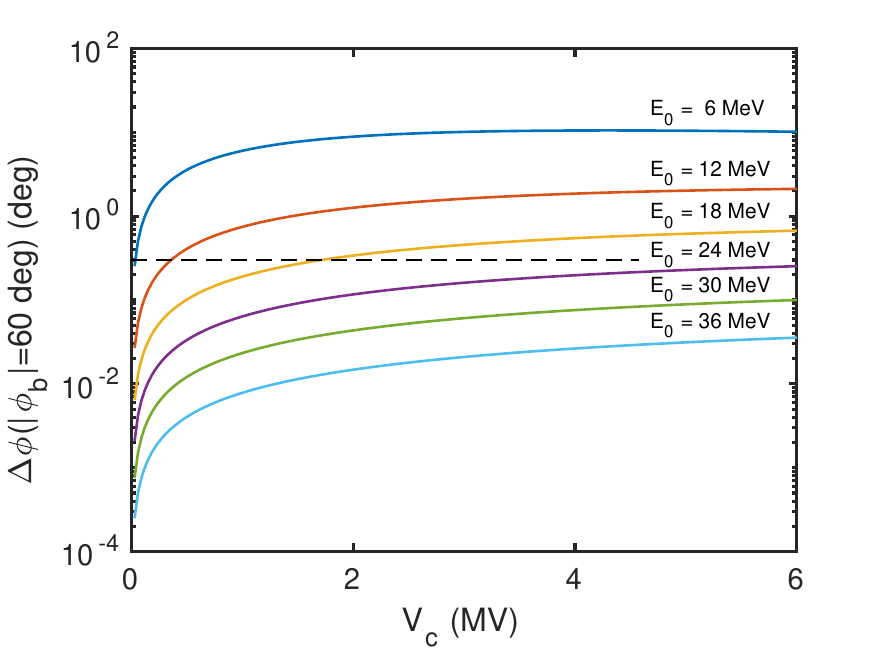}
\caption{\label{fig:phase_change}Change of the arrival time of the beam (shown as a phase change with respect to the RF clock) for each MLC cavity voltage turned on sequentially.  The input energy to the cavity being scanned assumes the cavities before it are set to the nominal 6 MV voltage.}
\end{figure}

In addition to the cavity energy gain calibrations, use of the beam arrival phase in principle allows for the determination of the on-crest phases. To test this, we re-purposed a single BPM arrival time phasing procedure typically used to phase the injector buncher cavity and last two ICM SRF cavities.  This method measures the arrival time at the downstream BPM while scanning the cavity phase by $\pm60^{\text{o}}$ in small steps from its starting set-point and finding the minimum phase change, as this occurs when the beam accelerates on-crest. Unfortunately, this method runs into practical limitations for initial energies well above 6 MeV.  Fig.~\ref{fig:phase_change} illustrates this point by plotting the phase change at $\phi_b =\pm 60^{\text{o}}$ as a function of voltage for each MLC cavity.  For each cavity in the MLC, all upstream cavities are set to their nominal 6 MV voltage, and any downstream cavities are off. The dashed line indicates the typical noise level of the BPM phase reading (roughly $0.3^{\text{o}}$).  These results predict the failure of this approach for all but the first two MLC cavities.

\begin{figure}[htb!]
    \begin{center}
         \subfigure[\hspace{0.2cm}First MLC cavity phasing with initial energy at 6 MeV.]{%
            \label{fig:cavity_phasing_6}
            \includegraphics[width=0.38\textwidth]{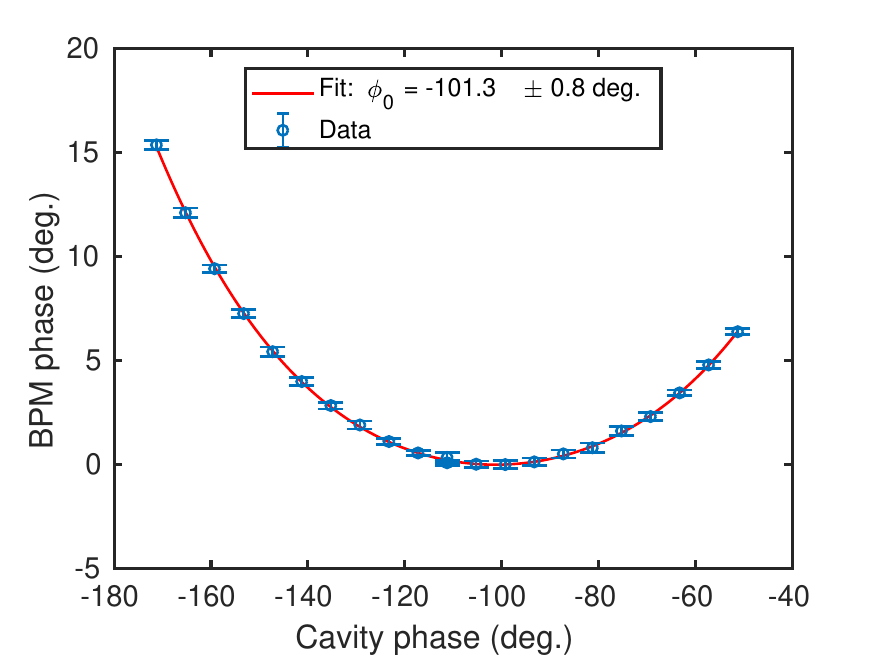}
        }\\%
        \subfigure[\hspace{0.2cm}Second MLC cavity phasing with initial energy at 12 MeV.]{%
           \label{fig:cavity_phasing_12}
           \includegraphics[width=0.38\textwidth]{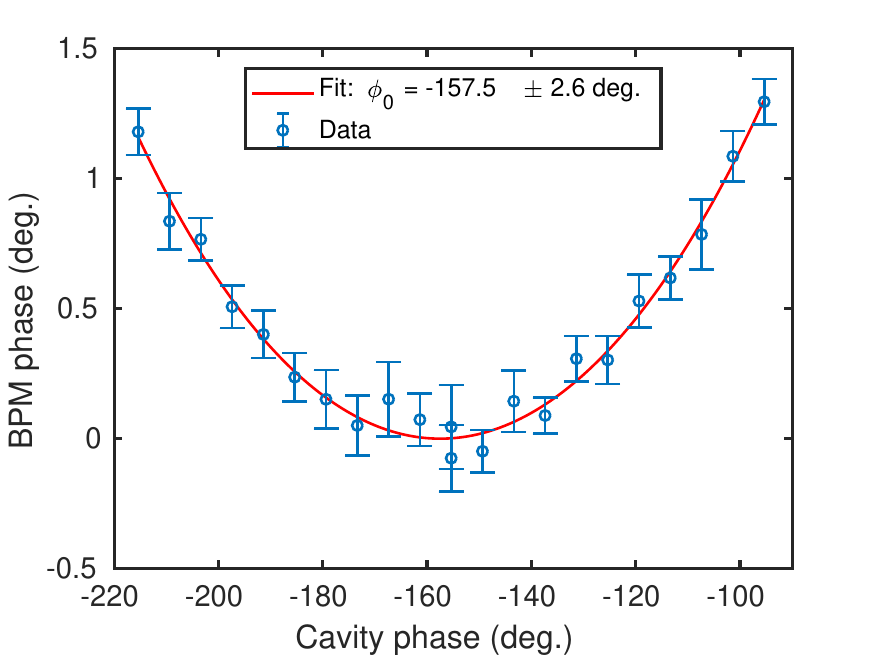}
        }
    \end{center}
    \caption{%
    \label{fig:cavity_phasing}
    Dependence of the arrival phase at BPM ID1BPC01 on the hardware phase of the first (a) and second (b) MLC cavity. The lower initial beam energy going into the first cavity results in larger BPM phase changes, and thus provides sufficient accuracy for the on-crest phase determination.}%
\end{figure}

Beam based measurements of the on-crest phase via this method confirm this prediction.  Fig.~\ref{fig:cavity_phasing_6} and \ref{fig:cavity_phasing_12} show the results of this procedure when phasing the first cavity with an incoming beam energy of 6 MeV, and the second cavity with an incoming beam energy of 12 MeV.  Note that use of the cavity hardware phase in these plots.  The location of the BPM phase minimum corresponds to the cavity hardware phase for on-crest acceleration ($\phi_b=0^{\text{o}}$). 
We estimate the resulting on-crest phase uncertainties to be roughly $\pm0.8^{\text{o}}$ and $\pm2.5^{\text{o}}$ for the two cases, respectively.  Data taken for the remaining cavities show worse signal to noise, and we conclude that this procedure only meets the desired phase requirements of $\phi_{\text{error}}<1^{\text{o}}$ when used to phase the first MLC cavity.

Due to the time constraints for the FAT, we decided on the following procedure to phase the remaining MLC cavities. Before turning on the cavity to phased, the beam is steered to the first splitter viewscreen, IS1SCR01, and the position on that screen and the BPM before it recorded. The cavity in question is then turned on to the desired voltage (nominally 6 MeV) and the phase adjusted until the beam returns to same location on the screen, thus determining one of the cavity zero-crossing phases. In the relativistic limit, the on-crest phase is roughly $\pm90^{\text{o}}$ from this value, with the sign chosen to ensure the resulting phase increases the energy.  During the FAT, operators performed this procedure by hand during tune up and the method thus currently lacks a systematic procedure for error estimates. Future continued use of this procedure requires both automation, as well as systematic characterization of the accuracy of the resulting on-crest phase. 

\subsubsection{MLC Vertical Offset\label{sssec:mlcoff}}

Initial attempts at steering the beam through the center of the main linac cavities indicated an offset of the MLC with respect the BPMs on either side of the linac (IB1BPC03 and ID1BPC01).  In particular, manual alignment of the beam in the first cavity suggested a vertical offset of roughly 5 mm.  Consequently, more detailed measurements were performed to better quantify these observations. These measurements proceeded as follows: each cavity was turned on individually (all other cavities turned off).  In each transverse direction a pair of corrector magnets was used to scan the beam position on the BPM just upstream of the MLC while keeping the beam's angle constant. For each incoming beam position in this scan, the phase of the cavity was then scanned from $0^{\text{o}}$ to $360^{\text{o}}$ in steps of $30^{\text{o}}$, and the transverse positions on the downstream BPM (ID1BPC10) measured. If the beam enters the cavity off axis, then the cavity focusing delivers a phase dependent kick, resulting in a periodic beam displacement on the downstream BPM.  The variance of the downstream positions on the BPM each direction gives was used to estimate the cavity offset, allowing for a rough centering of each position scan.  Horizontal scans were performed first, in order to minimize any horizontal beam offset going into the cavity, after which vertical scan data was taken for each cavity.  In order the partially compensate the change in drift lengths when taking data for different cavities, the voltage set-point of was varied linearly with the cavity index from 500 kV (RD1CAV06) for the first cavity to 1000 kV for the last cavity (RD1CAV01).  

\begin{figure}[htb]
\includegraphics[width=0.38\textwidth]{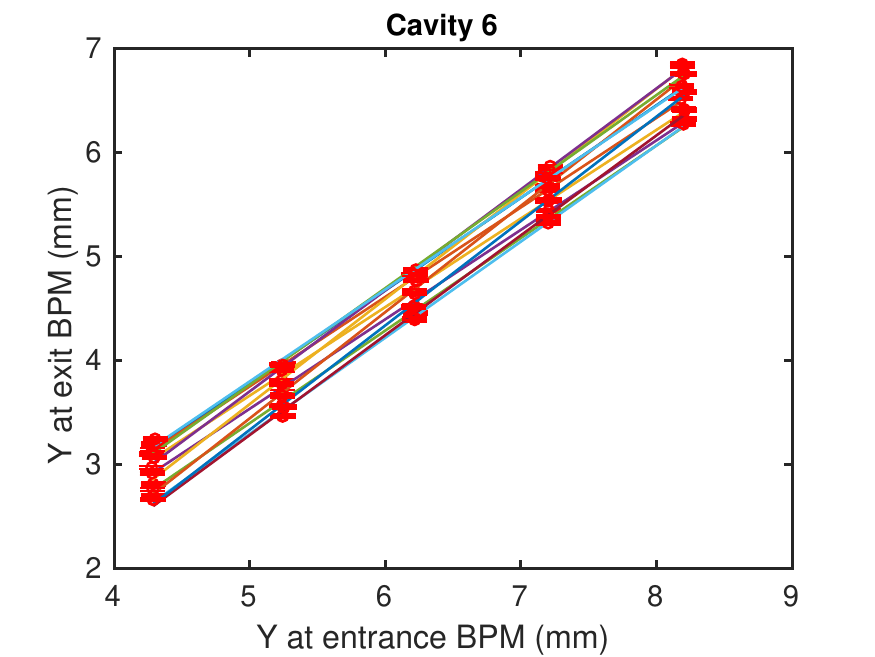}
\caption{\label{fig:cavslopes}Example BPM data taken for the first MLC cavity with best fit lines.  Each line corresponds to a different cavity phase.}
\end{figure}

Fig.~\ref{fig:cavslopes} shows the vertical position on the downstream BPM as a function of the vertical upstream BPM position for the various values of the first cavity (RD1CAV06) phase and a cavity voltage of 500 kV.  The data clearly imply a linear relationship between BPM readings.  

In this case, the linear transport of the beam centroid trajectory $\textbf{u} = (y, y^{\prime},1)^{\text{T}}$ through each cavity can be written as
\begin{eqnarray}
\textbf{u}_f=D(L_2)R_{\text{out}}T(\phi)D(L_1)R_{\text{in}}\textbf{u}_{i}\equiv M\textbf{u}_i,
\label{eqn:cavmat}
\end{eqnarray}
where $D(L)$ is the standard form for a drift transfer matrix and $L1$ and $L2$ are the drift lengths between the upstream BPM and cavity and cavity and downstream BPM respectively, $T(\phi)$ is the cavity transform matrix, and $R_{\text{out}}$ and $R_{\text{in}}$ transform the centroid position and angle into and out of the offset and tilted cavity coordinate system. In terms of the cavity offset $y_c$ and tilt $y_c^{\prime}$, these matrices take the form:
\begin{eqnarray}
R_{\text{in}} &=&
\begin{pmatrix}
1 & 0 & -y_c + \frac{L_c}{2}y_c^{\prime}\\
0 & 1 & -y_c^{\prime} \\
0 & 0 & 1
\end{pmatrix},
\label{eqn:rotate1}
\\
R_{\text{out}}&=&
\begin{pmatrix}
1 & 0 & y_c + \frac{L_c}{2}y_c^{\prime}\\
0 & 1 & y_c^{\prime} \\
0 & 0 & 1
\end{pmatrix},\label{eqn:rotate2}
\end{eqnarray}
 assuming the cavity is tilted about its center point.  Note the use of the third row in the above matrices and phase space vector, which allows for instantaneous shifts in coordinates at the cavity entrance/exit. The tilt $y_c^{\prime}$ is included the analysis appears as a titled cavity provides a transverse kick from the $E_z$ component of the cavity field, and thus contributes to phase dependent motion on the downstream BPM. 
 
 It turns out that the expression in Eq.~($\ref{eqn:cavmat}$) simplifies by writing the problem in terms of the ``effective thin lens" cavity matrix $\tilde T = D(-L_c/2) \cdot T \cdot D(-L_c/2)$, which parametrizes the problem in terms of one single drift length $\tilde{L} = L_2 + L_c/2$. Ignoring the initial angle of the beam (estimated here to be $\lesssim$ 0.1 mrad), the downstream position of the beam becomes:
 \begin{eqnarray}
 y_f = y_c + M_{11}(y_i-y_c) 
+\left[\tilde L\left(1-\tilde T_{22}\right)-\tilde T_{12}\right]y_{c}^{\prime},
\label{eqn:linrel}
\end{eqnarray}
where $M_{11} = \tilde{T}_{11} + \tilde{L} \tilde{T}_{21}$.  Note that the above expression takes the form of a line: $y_f= m\cdot y_i + b$ where the phase dependent slope and intercept are identified as $m(\phi) = M_{11}(\phi) = \tilde T_{11} + \tilde T_{21}\tilde L$ and $b(\phi) = (1-M_{11})y_c +\left[\tilde L\left(1-\tilde T_{22}\right)-\tilde T_{12}\right]y_{c}^{\prime}$.  In order to extract the offsets, these terms are Fourier expanded:  
\begin{eqnarray}
m(\phi) &=& \sum_{n=0}^{\infty}m_n^{(\text{c})}\cos(n\phi)+m_n^{(\text{s})}\sin(n\phi)
\\
b(\phi) &=& \sum_{n=0}^{\infty}b_n^{(\text{c})}\cos(n\phi)+b_n^{(\text{s})}\sin(n\phi) \\
\tilde T_{ij}(\phi) &=& \sum_{n=0}^{\infty}\tilde T_{ij,n}^{(\text{c})}\cos(n\phi)+\tilde T_{ij,n}^{(\text{s})}\sin(n\phi),
\label{eqn:fourier}
\end{eqnarray}
where $\phi = 0$ corresponds to the on-crest acceleration of the cavity. Substituting these expressions into Eq.~\ref{eqn:linrel} and collecting like Fourier coefficients gives the following expression for the cavity offsets (as a function of $n$):
\begin{eqnarray}
\begin{pmatrix}
y_c\\
y_c^{\prime}
\end{pmatrix}
=-\begin{pmatrix}
m_n^{(\text{c})} & \tilde L\tilde T_{22,n}^{(\text{c})}+\tilde T_{12,n}^{(\text{c})}
\\
m_n^{(\text{s})} & \tilde L\tilde T_{22,n}^{(\text{s})}+\tilde T_{12,n}^{(\text{s})}
\end{pmatrix}^{-1}
\begin{pmatrix}
b_n^{(\text{c})}
\\
b_n^{(\text{s})}
\end{pmatrix}.
\label{eqn:offsets}
\end{eqnarray}

\begin{figure}[htb]
\includegraphics[width=0.38\textwidth]{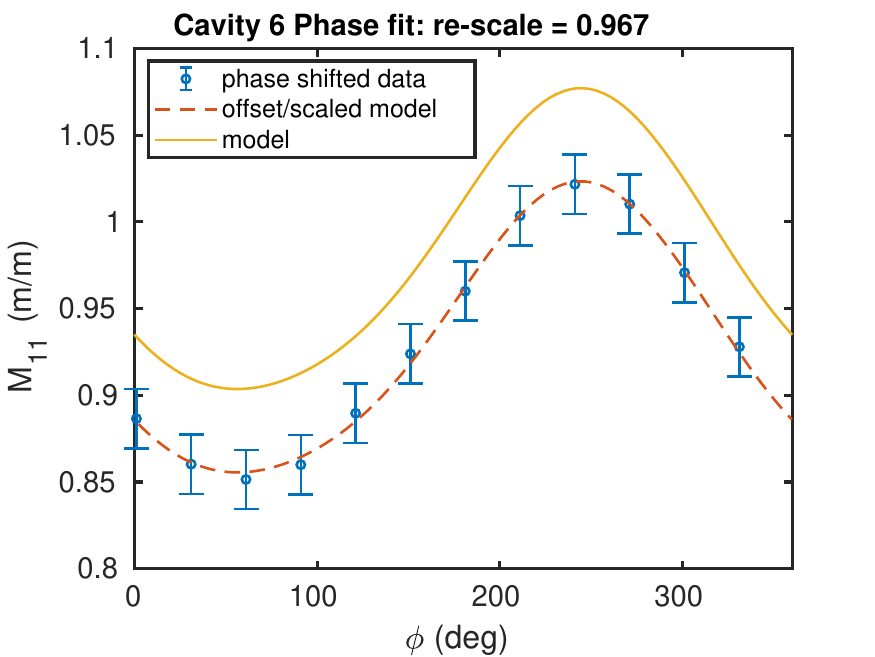}
\caption{\label{fig:phifit}Example of finding the on-crest phase by scaling, offsetting, and phase shifting the model Fourier components to match the measured data.  The results here show the measured data with the on-crest phase offset included, that is $\phi$=0 corresponds to on-crest in the data and model shown here.}
\end{figure}

Equation~\ref{eqn:offsets} finds the cavity offset and tilt using a combination of the data ($m(\phi)$ and $b(\phi)$) and matrix elements of the cavity ($\tilde T_{ij}(\phi)$), which we compute by integrating through a field map for the cavity. There may be a phase offset between our data and the model of the cavity through which we integrated, so our next step will be to find that phase offset. We do this by taking advantage of the fact that $m(\phi)=M_{11}(\phi)$. We first integrate through the cavity field map to determine $M_{11}(\phi)$, and fit the results to a Fourier expansion through the third harmonic. We then make a least squares fit of $m(\phi)$ to $AM_{11}(\phi+\phi_0)+B$ and find the parameters $\phi_0$, $A$, $B$. Fig.~\ref{fig:phifit} shows the result of this procedure for the slope data from the cavity~6.  For that cavity, the scaling factor $A$ was roughly 0.967 (the other cavities had $A$ ranging from 0.942 to 1.03). The purpose of this step is only to compute $\phi_0$; $A$ and $B$ are not used in subsequent calculations..

The uncertainties shown in Fig.~\ref{fig:phifit}, as well as those in all subsequent calculations arise from two sources: uncertainty in the underlying BPM readings, and systematic errors due to the model being an imperfect representation of our data. The systematic errors can be seen in the linear fits that determine $m$ and $b$ at each $\phi$. We estimate the systematic error by computing the $\chi^2$ per degree of freedom for the line fits, assigning the systematic error to be the square root of $\chi^2$ times the random errors, and adding that to the random error in quadrature to obtain our uncertainty estimate. The systematic errors are always larger than the random errors, and for cavity~1 by a large factor. 

\begin{figure}[hb!]
\includegraphics[width=0.38\textwidth]{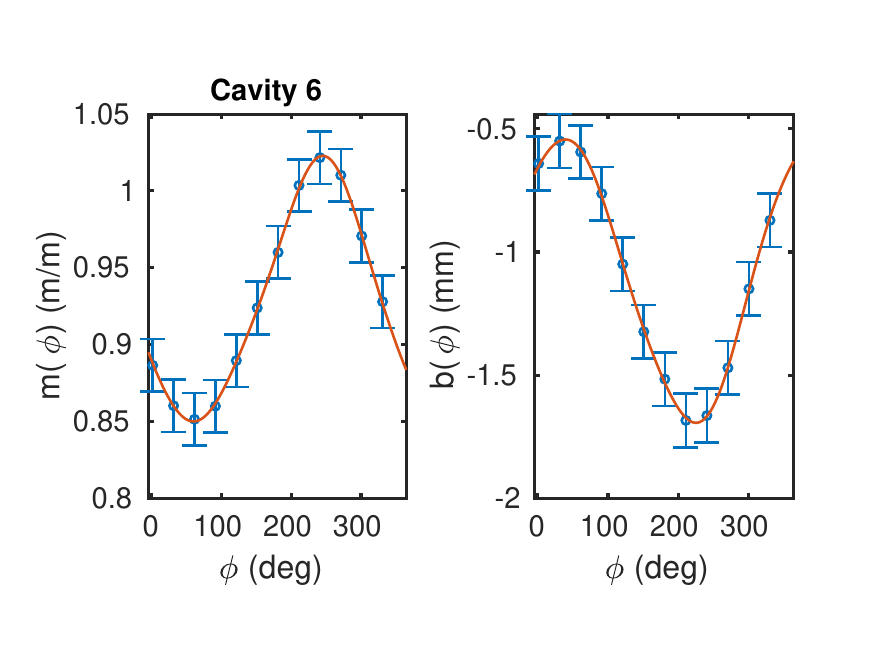}
\caption{\label{fig:phifitmb}Example of the final fit to the slope and offset terms in Eq.~(\ref{eqn:linrel}). $\phi$=0 corresponds to on-crest acceleration.}
\end{figure}

Now that we have the phase offset $\phi_0$, we perform a discrete Fourier transform on $m(\phi)$ and $b(\phi)$, phase shifted by $\phi_0$, to obtain $m_1^{(c,s)}$ and $b_1^{(c,s)}$ (Fig.~\ref{fig:phifitmb} shows the $m$ and $b$ data along with the Fourier series approximation to the third harmonic). We also fit $\tilde{T}_{12}(\phi)$ and $\tilde{T}_{22}(\phi)$ to a Fourier expansion to the third harmonic. We then apply Eq.~(\ref{eqn:offsets}), with $n=1$ (which is the dominant Fourier mode) to obtain $y_c$ and $y_c'$. 
\begin{figure}[htb!]
    \begin{center}
         \subfigure[\hspace{0.2cm}Vertical cavity offsets.]{%
            \label{fig:cavoffs}
            \includegraphics[width=0.38\textwidth]{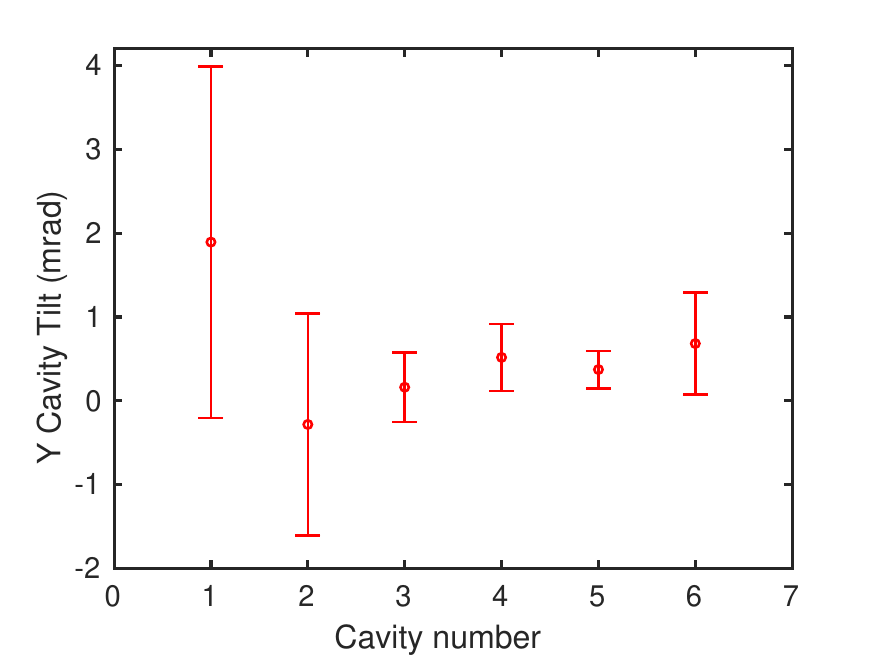}
        }\\%
        \subfigure[\hspace{0.2cm}Vertical cavity tilts.]{%
           \label{fig:cavtilts}
           \includegraphics[width=0.38\textwidth]{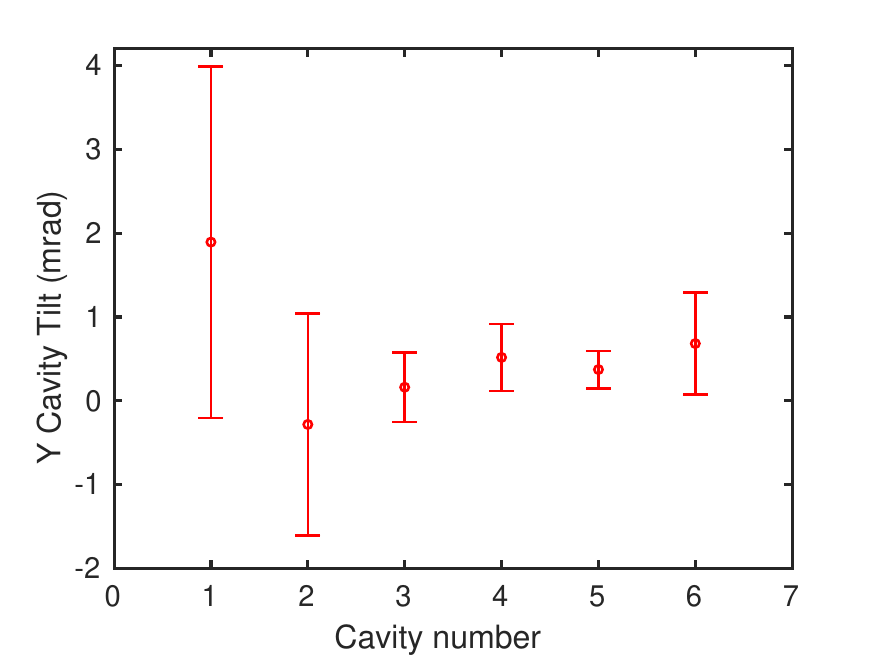}
        }
    \end{center}
    \caption{%
    \label{fig:cavofftilts}
    Vertical cavity offsets (a) and tilts (b).  The error estimates here include both systematic in fitting lines to the BPM data, as well as the random error in the BPM position readings.}%
\end{figure}

The resulting offsets and tilts are shown in Fig.~\ref{fig:cavofftilts}. All of the cavity offsets are in the positive vertical direction and have a weighted average of roughly 4.0 mm, very near the rough estimate provided by operators manually trying to center the beam through the linac cavities.  As a consequence, further surveying of the beamline, BPMs, and MLC will be performed, and the linac cryomodule will be lowered.  Additional beam based measurements will be performed to test for any remaining offsets of the cavities during the next commissioning period.

\subsection{S1 Splitter Line Commissioning\label{ssec:meas:s1}}


\subsubsection{Splitter Line BPM Nonlinearity Correction\label{sssec:bpm}}

In general, the particular method used to convert the BPM button signals into beam positions fundamentally limits the accuracy of BPM position data. While computationally simple, the well known ``difference over sum'' method for four button signals typically introduces large errors for beams significantly off axis.  As a consequence, many methods exist to correct this effect \cite{ref:tburger}. In the past, the CBETA/Cornell injector stripline BPMs made use of an analytic expression for the four BPM signals.  This approach approximates the beam as an infinite line charge located inside an infinite circular conducting pipe.  In this 2D approximation, the BPM signals are computed by integrating the induced surface charge density on the conductor over the angular width $\theta_b$ of the four striplines attached to the beam pipe \cite{ref:lowemitter}. The resulting expression depends only on the beam position $(x,y)$, pipe radius $R$, and $\theta_b$. Inversion of this expression for the beam position $(x,y)$ is achieved using a $\chi^2$ fit of the model to the real BPM signals.  A similar procedure allows for correcting the BPM intensity.  Note that as striplines are not actually 2D objects, we allow $\theta_b$ to vary in order to best correct the measured nonlinearity.  This procedure results in corrected BPM positions over nearly the entire enclosed beam pipe area.

Since the publication of \cite{ref:lowemitter}, modifications to the fitting process described have been made.  In particular, the original routine fits for all four individual BPM signals, along with an overall scale. While this fitting method has the benefit of easily extending to arbitrary numbers and locations of BPM buttons, it results in increased sensitivity to scale errors between the associated signal pairs (top-bottom, left-right) in the 4 BPM signal case.  Dividing out the overall amplitude of each associated signal pair removes this sensitivity:
\[
\begin{aligned}
S_x & = \frac{S_{\textmd{right}} - S_\textmd{left}}{S_\textmd{right} + S_\textmd{left}} \\
S_y & = \frac{S_{\textmd{top}} - S_\textmd{bottom}}{S_\textmd{top} + S_\textmd{bottom}}
\end{aligned}
\]
The new fitting method calculates these quantities and fits them to the measured values.
\begin{figure*}[ht!]
    \begin{center}
         \subfigure[\hspace{0.2cm}Beam positions using the difference/sum model.]{%
            \label{fig:diff_sum_pos}
            \includegraphics[width=0.3\textwidth]{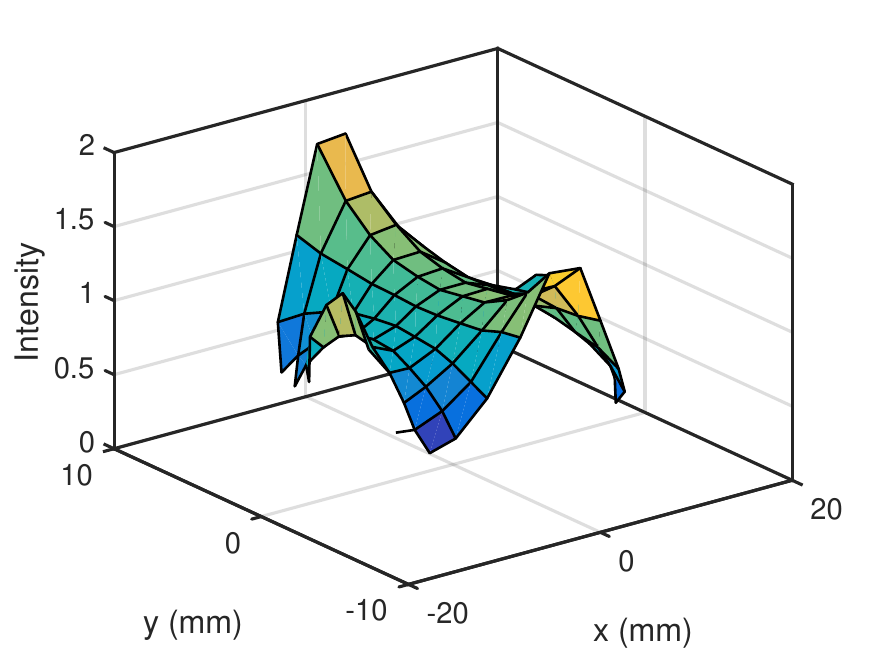}
        }%
        \subfigure[\hspace{0.2cm}Beam positions using the circular pipe approximation.]{%
           \label{fig:circle_pipe_pos}
           \includegraphics[width=0.3\textwidth]{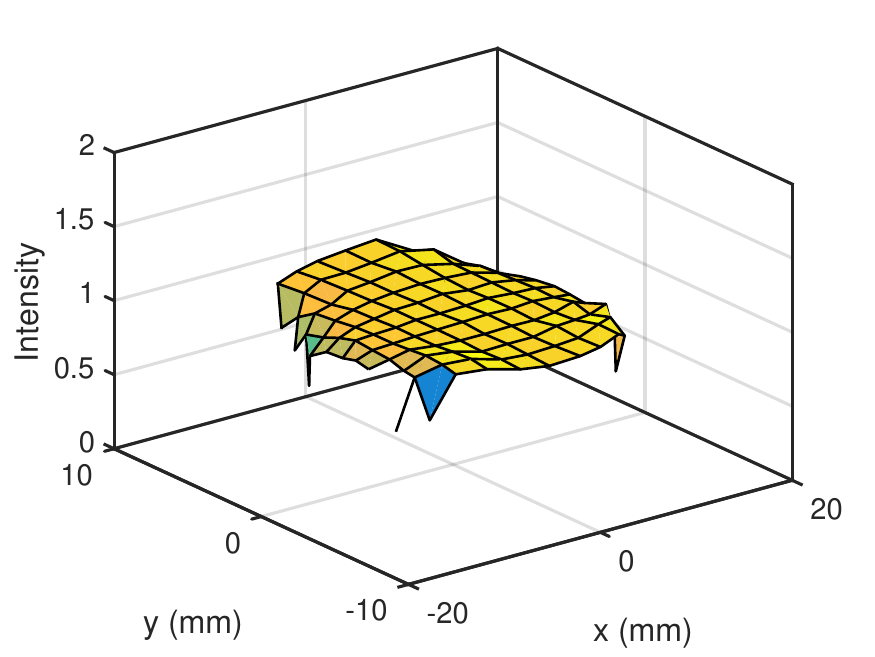}
        }
        \subfigure[\hspace{0.2cm}Beam positions using the Poisson model.]{%
           \label{fig:poisson_pos}
           \includegraphics[width=0.3\textwidth]{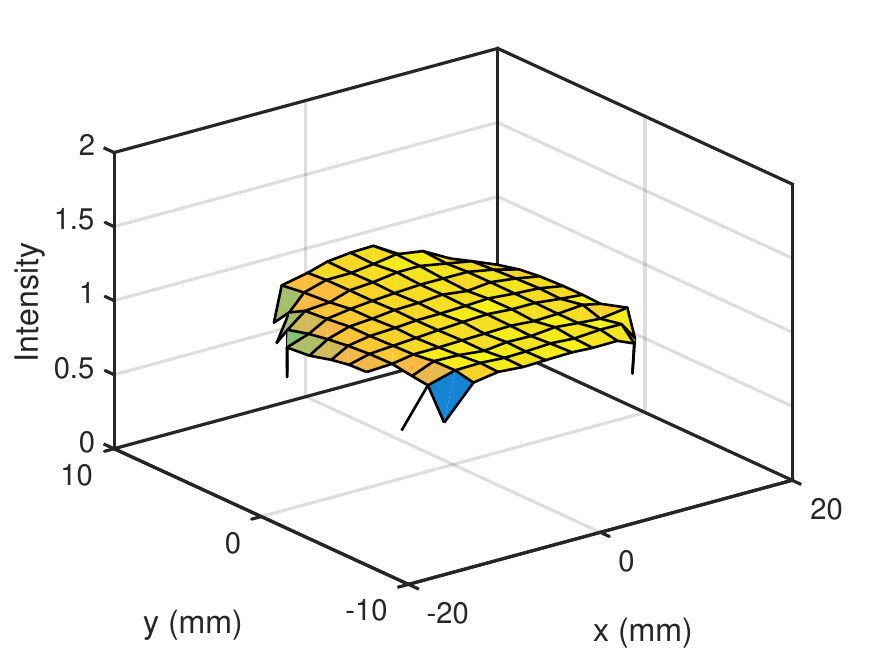}
        }\\ 
        \subfigure[\hspace{0.2cm}Beam intensity using the difference/sum model.]{%
            \label{fig:diff_sum_int}
            \includegraphics[width=0.3\textwidth]{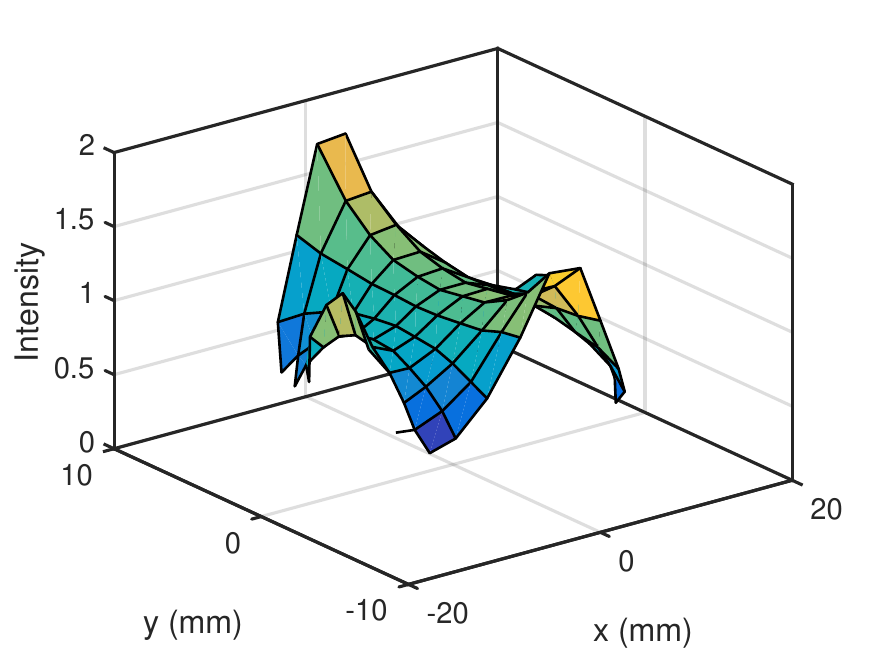}
        }%
        \subfigure[\hspace{0.2cm}Beam intensity using the circular pipe approximation.]{%
           \label{fig:circle_pipe_int}
           \includegraphics[width=0.3\textwidth]{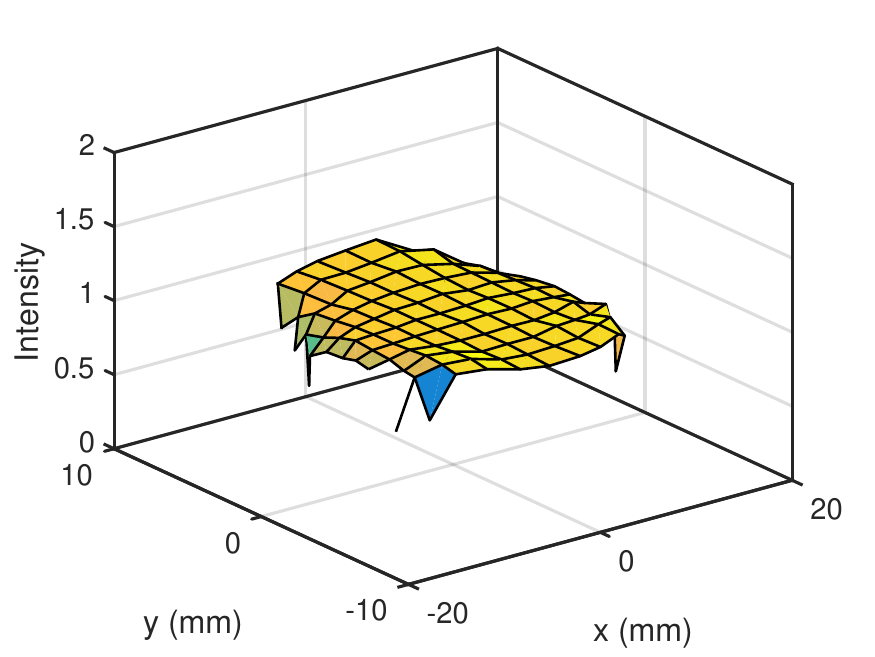}
        }
        \subfigure[\hspace{0.2cm}Beam intensity using the Poisson model.]{%
           \label{fig:poisson_int}
           \includegraphics[width=0.3\textwidth]{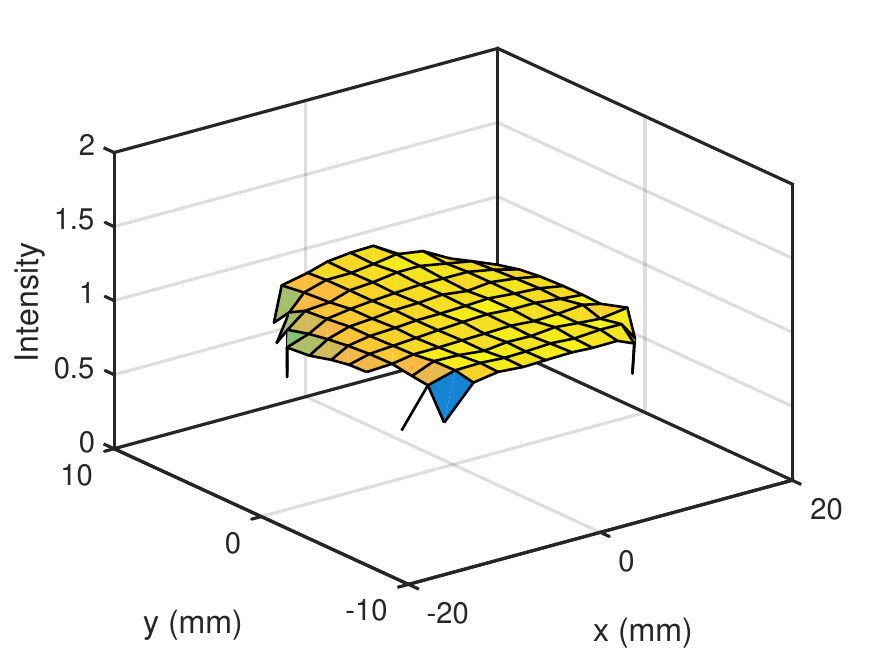}
        }
    \end{center}
    \caption{%
    \label{fig:position_intensity_bpm}
    Comparison of the beam positions and intensity using three different models to interpret the raw data from the BPM. First, a simple difference/sum model is used (a,d). Next, a correction is made using an approximation for a circular pipe (b,e). And finally, the model using a fieldmap from a Poisson calculation of the correct 2D pipe geometry is used (c,f).}%
\end{figure*}

The vacuum chamber in the S1 splitter section features a racetrack profile (see Fig.~\ref{fig:diff_sum_pos}-\ref{fig:poisson_pos}), with width of 36 mm and height of 24 mm. This violates the assumption of cylindrical symmetry made in the analytic model described above, resulting in a decreased range of validity for the resulting BPM positions. While the racetrack geometry precludes analytic treatment, programs such as Poisson \cite{ref:psfish} allow one to find a numerical solution to the 2D electrostatic BPM problem featuring the correct beam pipe cross-section using the method described by Helms and Hoffstaeter \cite{ref:bpmgeorg}. The generation of a look up table of beam positions from BPM signals based on this numerical approach provides a simple method to interpolate and invert (via $\chi^2$ fitting) the BPM signals for the BPM positions. Unfortunately the schedule of the FAT test prohibited the integration of this method into the control system, and thus its online evaluation.  Instead, it turns out that replacing the beam pipe radius $R$  with the average of the two inner radii of the pipe $R_\text{avg} = \frac{1}{2} (23.0/2 + 35.5/2) = 14.625$ mm significantly extends the range of validity of the analytic BPM method to a fraction of the beam pipe area suitable for initial commissioning efforts.  

We quantified the effects of these methods by performing a rectangular scan of the beam position on the first S1 BPM (IS1BPM01) using the first horizontal and vertical dipoles in the splitter line (MS1DPB01 and MS1CRV01) and recording the corresponding BPM button signals. The signal data was then inverted to find the predicted BPM positions for each of the three methods: the standard ``difference/sum'' model ($x = \frac{R}{2}\cdot S_x$, $y = \frac{R}{2}\cdot S_y$), the modified analytic approximation with average radius, and the model incorporating Poisson generated fieldmaps for the 2D racetrack pipe geometry. In the latter two cases, the effective size of the BPM buttons was allowed to change to best correct the non-linearity. The beam energy for all of these measurements was 6 MeV.

Fig.~\ref{fig:position_intensity_bpm} displays the resulting beam positions and intensities measured. The top row shows the predicted beam positions, while the bottom row shows a corrected intensity of the beam, which is proportional to the total bunch charge. The results make clear that the simplest model produces accurate beam positions only within a few millimeters of the center of the pipe, and results in a reported intensity that varies greatly over the scanned area. Interestingly, the modified analytic approximation more than doubles the useful region of the pipe, out to at least $\pm5$ mm in x and y. The corrected intensity is nearly flat, and a much better measure of the bunch charge. Finally, using the Poisson look-up table, the final amount of nonlinear curvature is nearly fully corrected, and the BPM intensity slightly more flat. The remaining slight curvature near the edge of the measured grid indicates beam clipping, as also seen in the decreased intensity.    


\subsubsection{Path Length Adjustment}

The establishment of single or multi-pass energy recovery requires precise control of the return phase(s) of the beam(s) at the MLC. To achieve this, the CBETA design makes use of two adjustable path length chicanes in for each beam energy.  Each chicane features a pair of remotely controlled translation stages for online control of the path length. In particular, the low energy S1 splitter line path length adjustment system provides up to 9.6 mm (or $15^{\text{o}}$ at the speed of light) total path length adjustment during beam operation.  Synchronous movement of the two stages ensures minimal stress is placed on the bellows connecting the chicane vacuum chambers.  In the S1 splitter line these two stages are located directly under S1 dipoles MS1DIP04 and MS1DIP05, as shown in Fig.~\ref{fig:fat_layout}. A change in translation stage position of $\Delta l$ results in a path length change $\Delta s$ experienced by the beam given by:
\begin{eqnarray}
\Delta s = 2\left(1-\cos\theta_{\text{bend}}\right)\Delta l,
\label{eqn:pathlength}
\end{eqnarray}
where $\theta_{\text{bend}} = 23.3^{\text{o}}$ is the bend angle of the inner splitter chicane.

\begin{figure}[hb!]
\includegraphics[width=0.38\textwidth]{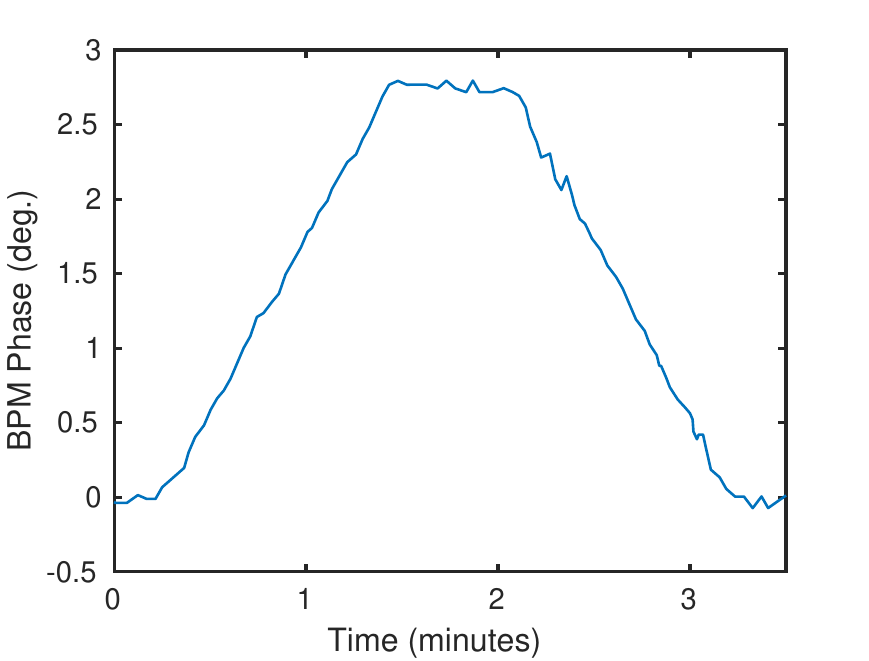}
\caption{\label{fig:path_length}The beam arrival phase on S1 BPM 5 during a splitter stage movement forward and back of 1 cm.}
\end{figure}

In order to test path length adjustment, we temporarily rewired the BPMs in the splitter line, making the fifth BPM in the S1 line capable of reporting both position and beam arrival phase. With all cavities in the MLC turned off, we then steered a 6 MeV electron beam through the splitter line and zeroed the reported BPM arrival phase. We then commanded the stage to move $\Delta l = +1$ cm and then back $-1$ cm, and recorded the measured BPM phase during the movement, as shown in Fig. \ref{fig:path_length}.
Eq.~(\ref{eqn:pathlength}) implies a $\Delta l = 1$ cm stage movement results in a 1.63 mm path length change seen by the beam. This corresponds to a predicted BPM phase change of 2.54$^{\text{o}}$. The measured BPM phase data in Fig.~\ref{fig:path_length} gives a measured BPM phase change of 2.8$^{\text{o}}$, roughly 10\% larger than expected.  Translated into angle, this corresponds to a bend of $\theta_b = 24.5^{\text{o}}$, or about 1 deg larger than the design value. 
While more work is needed to understand and resolve this discrepancy, the measured path length change with beam demonstrate the ability to tune path length during operation, and thus constitutes a successful preliminary test of the proposed path length control in CBETA.


\subsection{Measurements at 42 MeV\label{ssec:meas42}}

The nominal energy of the first pass is 42 MeV.  Consequently, reaching this energy represented a major milestone for the project, and allowed for quantification of the beam dynamics through both the splitter line in its design setting as well as the prototype permanent magnet girder. 


\subsubsection{\label{sssec:disp42}Horizontal Dispersion and $R_{56}$}

\begin{figure*}[ht!]
    \begin{center}
         \subfigure[\hspace{0.2cm}BPM position change with beam energy.]{%
            \label{fig:rawdisp42}
            \includegraphics[width=0.38\textwidth]{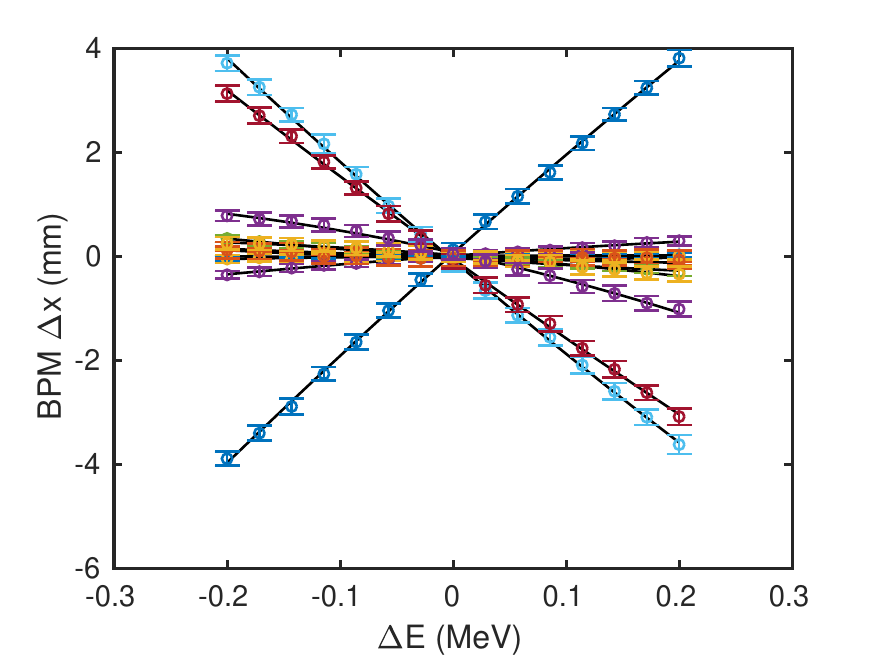}
        }%
        \subfigure[\hspace{0.2cm}Measured $R_{56}$ and model at 42 MeV.]{%
           \label{fig:disp42}
           \includegraphics[width=0.38\textwidth]{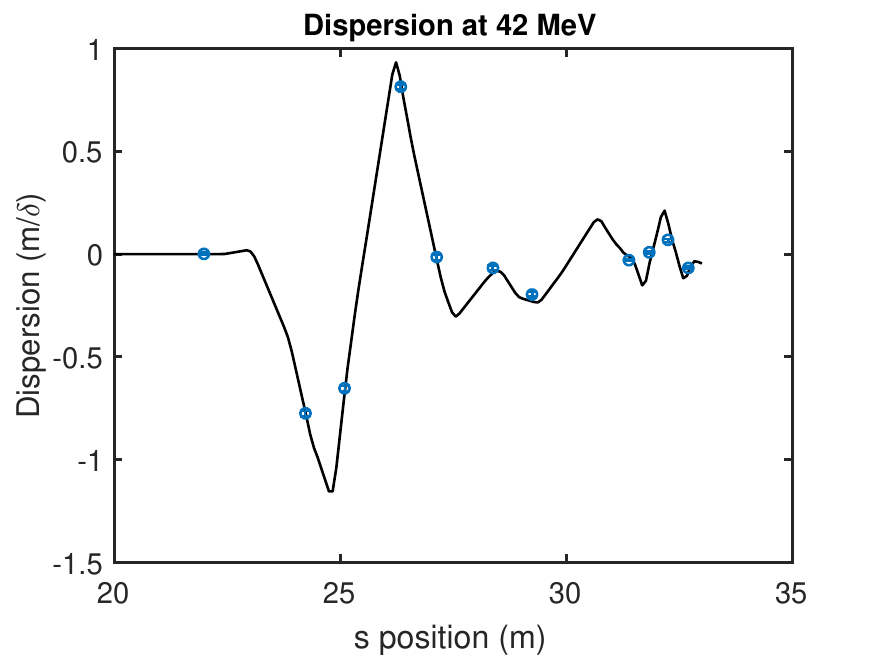}
        }\\ 
        \subfigure[\hspace{0.2cm}BPM phase change with beam energy.]{%
            \label{fig:rawR5642}
            \includegraphics[width=0.38\textwidth]{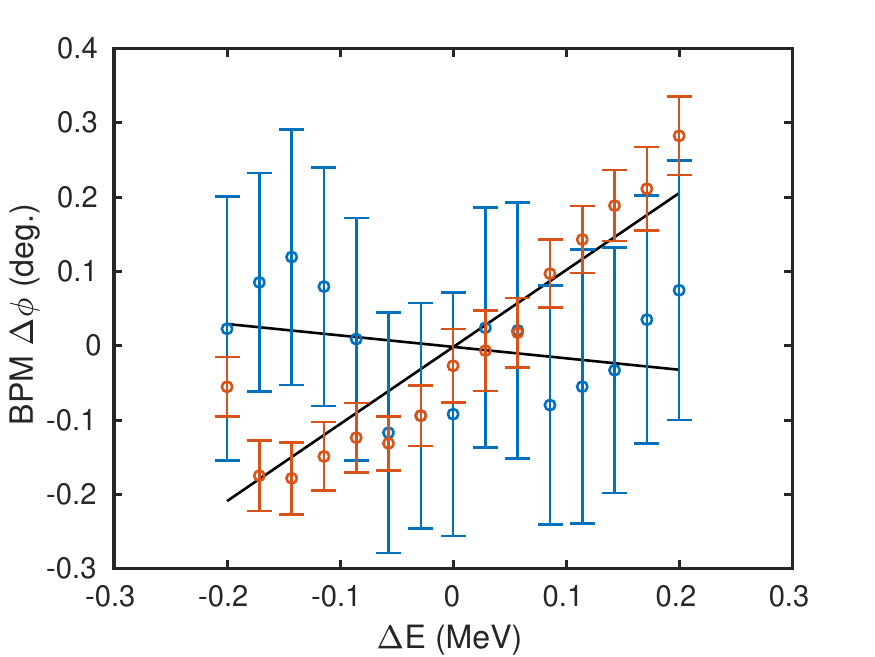}
        }%
        \subfigure[\hspace{0.2cm}Measured $R_{56}$ and model at 42 MeV.]{%
           \label{fig:r5642}
           \includegraphics[width=0.38\textwidth]{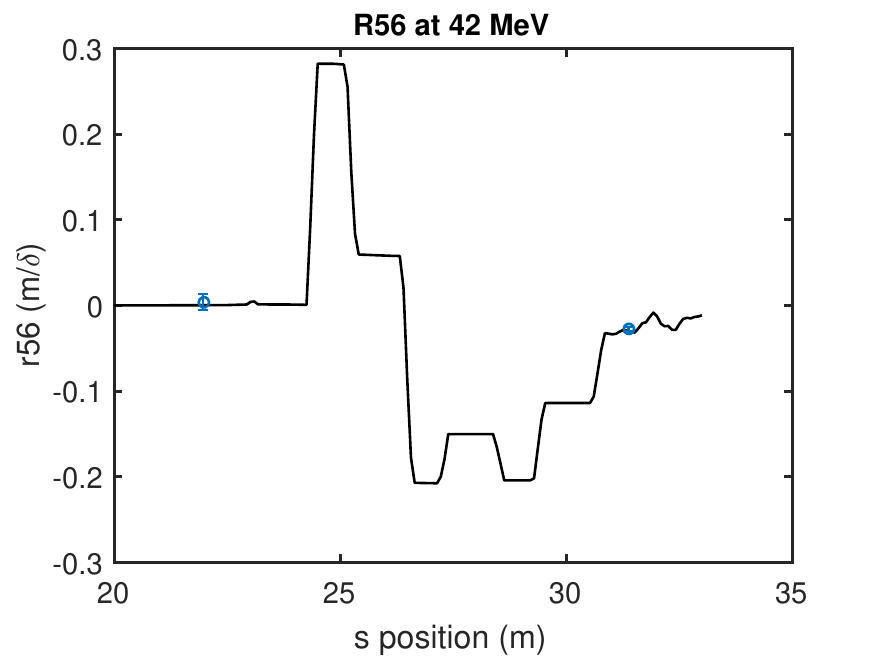}
        }
    \end{center}
    \caption{%
    \label{fig:disp_r56}
    Left: BPM position (a), and phase change (c) measured as a function of energy around the nominal 42 MeV energy. Each line corresponds to data on an individual BPM. Right: the resulting dispersion (b) in the S1 and FA line and $R_{56}$ in the S1 line (shown in blue) compared with the simulated values from the model (black line).}
\end{figure*}

The horizontal dispersion and $R_{56}$, defined here as $\eta_x = dx/d\delta$ and $R_{56} = (c/\omega)d\phi/d\delta$, where $\delta = (E \text{ [MeV]}- 42)/42$, $\omega$ is the angular cavity frequency, and $\phi$ is the BPM phase change in [rad], play important roles in controlling emittance dilution and establishing energy recovery in the full CBETA design, and thus necessitated experimental verification in the FAT. Scanning the voltage of the last MLC cavity allows for the simultaneous determination of both the dispersion $\eta_x$ and $R_{56}$ matrix element by measuring the orbit and arrival phase response on the downstream BPMs. Only the BPM directly after the linac (ID1BPC01) and first BPM in the FA arc (IFABPM01) were configured to read phase data for these measurements. Typical measurements scanned the voltage of the last MLC cavity in the range of $\pm$200 kV around the desired set-point in 7 steps. At each scan point, the beam position and phase was measured 10 times at 5 Hz and averaged.  The slope of the resulting orbit and phase response as a function of voltage, along with the beam momentum, determine $\eta_x$ and $R_{56}$.  Fig.~\ref{fig:disp_r56} shows an example data set, with the raw beam position and phase data, as well as the best linear fits to the data shown in Fig.~\ref{fig:rawdisp42} and \ref{fig:rawR5642} respectively. Fig.~\ref{fig:disp42} and \ref{fig:r5642} display the corresponding measured and simulated dispersion and $R_{56}$. The agreement seen here required tweaking the quad settings by a few percent using a scale single factor for all of these quads, indicating a small disagreement with the model.  
\begin{figure}[hb!]
\includegraphics[width=0.38\textwidth]{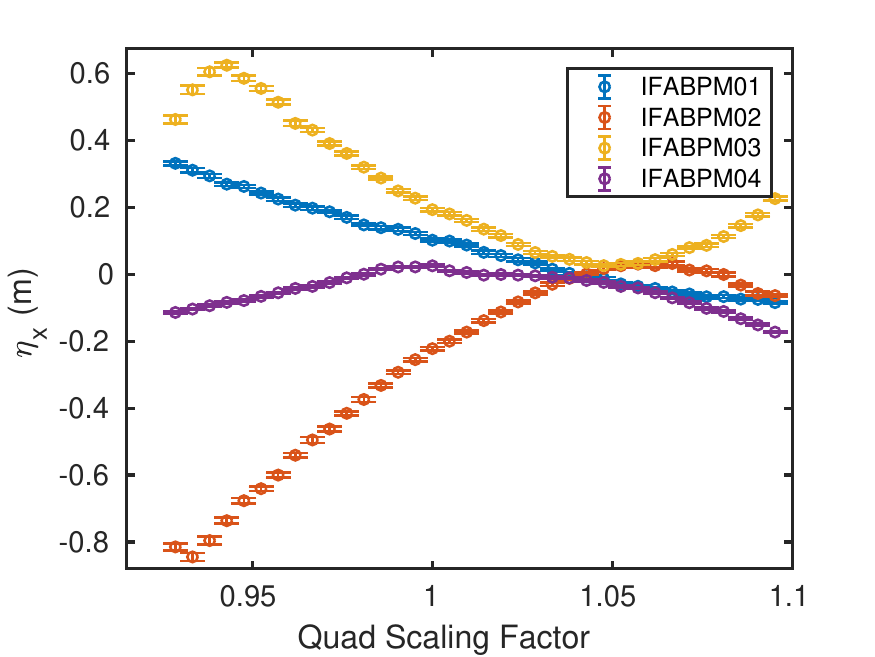}
\caption{\label{fig:qscan_disp}Dispersion measured in the FA section as a function of an overall scale of the S1 splitter quadrupoles. At around a factor of 1.04, the dispersion is minimized.}
\end{figure}

This discrepancy was further investigated by focusing on the dispersion values in the FA section.  Here, proper set up of the splitter magnets is designed to create a periodic dispersion solution in the FA section's repeating cells. At the FA BPMs, the model periodic value is $-11$~mm (see Fig.~\ref{fig:design_eta}). With the quadrupole currents at their design settings, we measured a non-periodic dispersion at the FA BPMs, with magnitudes as high as 0.4~m. To bring the dispersion values closer to periodic, we first scaled all of the quadrupoles by a single scaling factor. Fig.~\ref{fig:qscan_disp} shows the measured dispersion on the FA BPMs as a function of the single quadrupole scaling factor. The smallest magnitude for the dispersion occurred when the quadrupole scaling factor was about 1.04, with the resulting dispersion values having a maximum magnitude of about 0.1~m (still larger than anticipated).  We acknowledge the use of a single adjustment scale factor for the quads is likely too simplistic.  Attempts were made to use the dispersion response values from the model to further improve the dispersion. Unfortunately, these were unsuccessful, indicating a difference between the model and the experiment. To achieve the correct dispersion in the future, we will measure the response of the dispersion (and $R_{56}$) to quadrupole gradient changes and make corrections based on that.  Nonetheless, for the remainder of this work a scale factor of 1.04 was applied to the splitter quads automatically in order to keep the dispersion to a minimum. 


\subsubsection{Orbit Response \label{ssec:oresp}}

Measurement of the orbit response matrix provides a straightforward way of characterizing the single particle optics throughout the entire FAT layout, including verification of various magnet strength calibrations, and allows for comparison with the online simulation model.  Additionally, use of the response matrix features prominently in various feedback routines such orbit correction. 

The procedure for measuring the orbit response on each BPM begins by scanning the corrector and dipole currents over a broad range (the full range in the case of correctors), and recording the beam position and intensity on all downstream BPMs.  For each BPM, the data was truncated to include only those points within a small region around the BPM center, to best avoid BPM nonlinearity, and with intensity above a user defined BPM specific thresholds, to avoid cases of lost or partially scraped beam.  The slope of each line yields the corresponding orbit response in [mm/A].

\begin{figure*}[htb!]
    \begin{center}
         \subfigure[\hspace{0.2cm}Trim magnet on first injector merger dipole.]{%
            \label{fig:MB1CHE01}
            \includegraphics[width=0.3\textwidth]{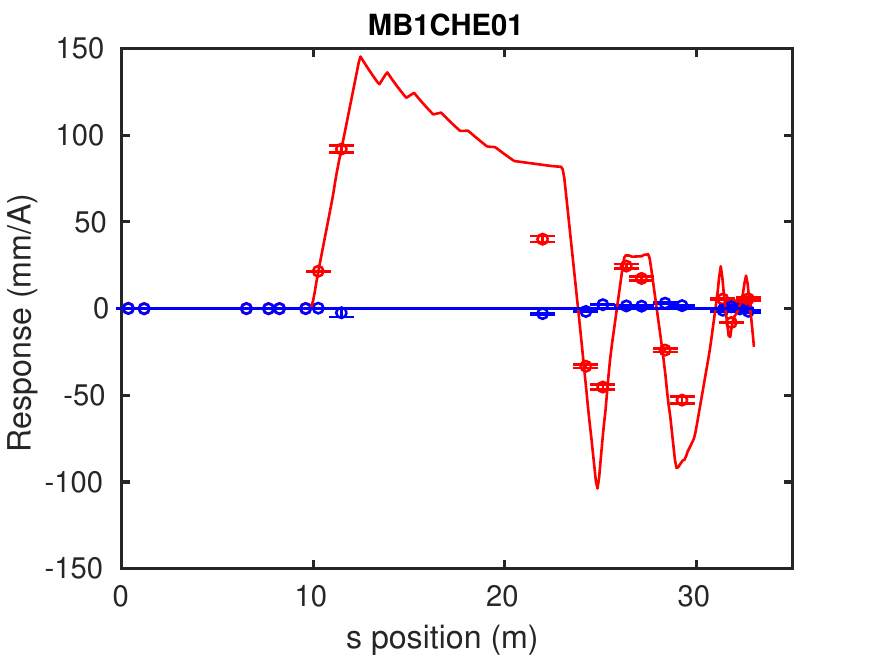}
        }%
        \subfigure[\hspace{0.2cm}Horizontal dipole windings on first injector merger quad.]{%
           \label{fig:MB1CHG01}
           \includegraphics[width=0.3\textwidth]{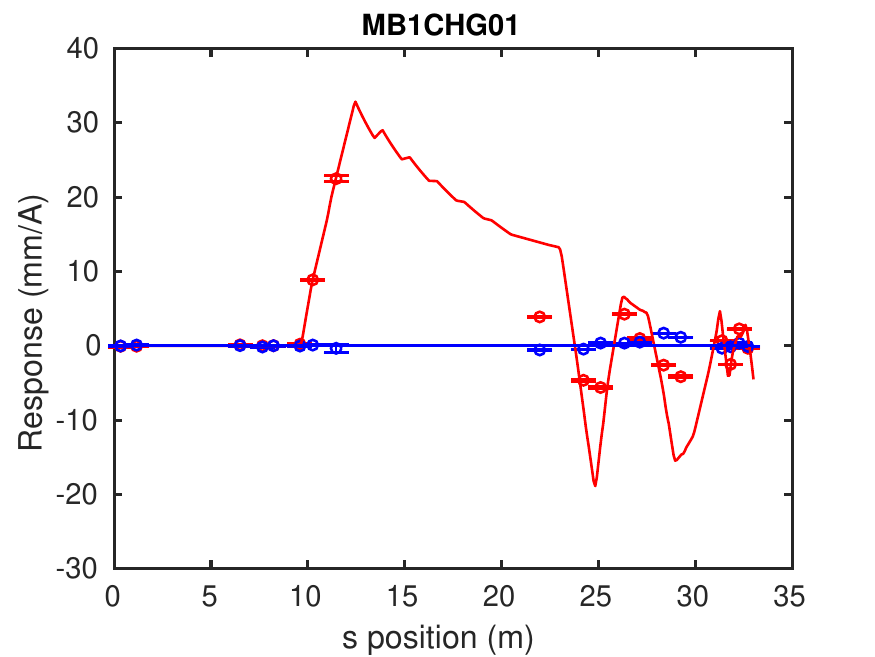}
        }
        \subfigure[\hspace{0.2cm}Vertical dipole windings on first injector merger quad.]{%
           \label{fig:MB1CVG01}
           \includegraphics[width=0.3\textwidth]{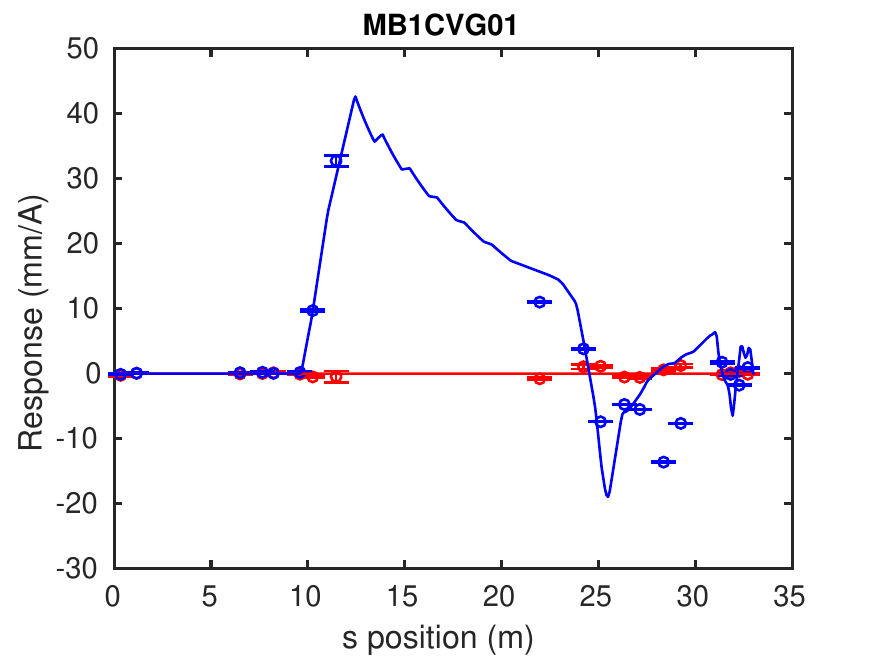}
        }\\ 
        \subfigure[\hspace{0.2cm}First S1 splitter dipole, the temporary replacement for the common magnet.]{%
            \label{fig:MS1DPB01}
            \includegraphics[width=0.3\textwidth]{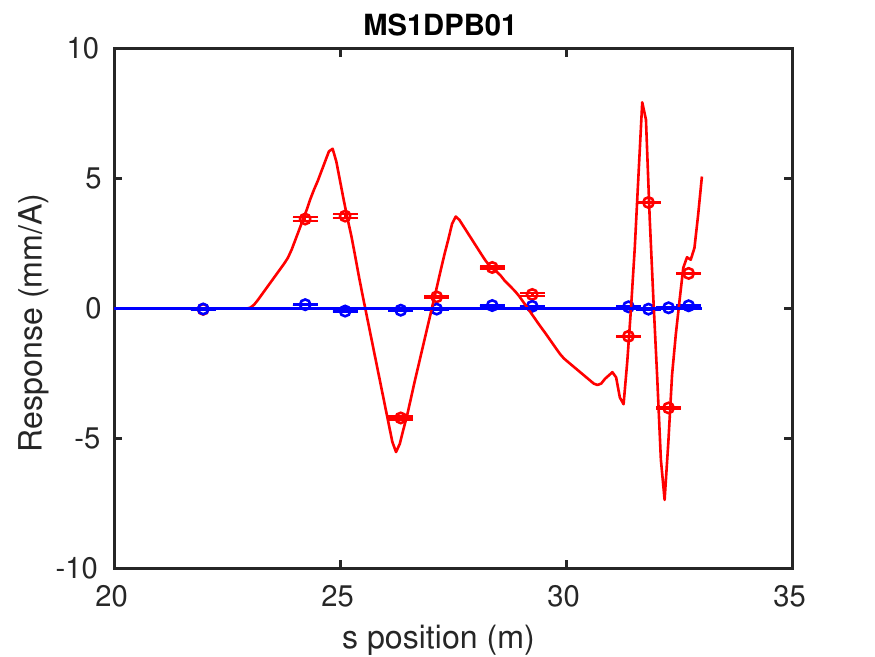}
        }%
        \subfigure[\hspace{0.2cm}Second S1 splitter dipole.]{%
           \label{fig:MS1DIP02}
           \includegraphics[width=0.3\textwidth]{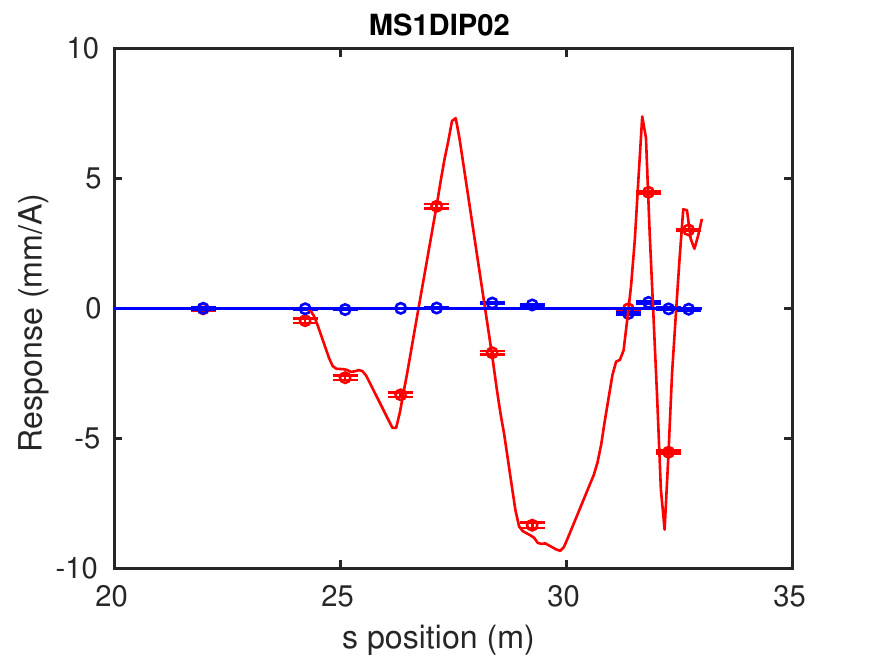}
        }
        \subfigure[\hspace{0.2cm}First S1 splitter vertical dipole corrector.]{%
           \label{fig:MS1CRV01}
           \includegraphics[width=0.3\textwidth]{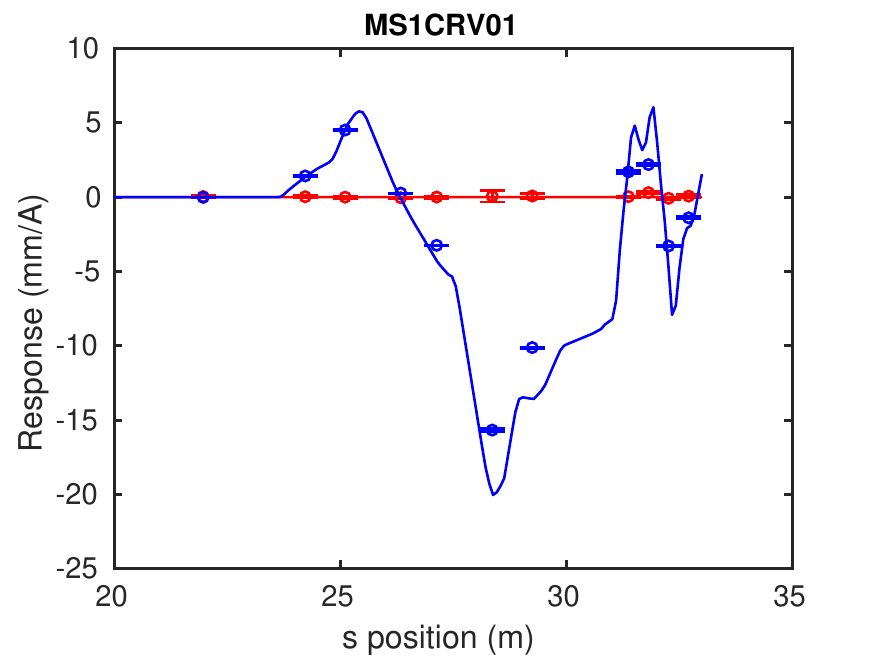}
        }
    \end{center}
    \caption{%
    \label{fig:orbit_response}
    Measured orbit response to dipole magnet kicks (points) compared to prediction from simulation (lines). Horizontal response is shown in red, and vertical in blue. An example magnet is shown for each type of dipole encountered after the injector merger.}%
\end{figure*}

Fig.~\ref{fig:orbit_response} displays several example orbit response data sets and their comparison to the response computed by the CBETA Virtual Machine. The top row, Fig.~\ref{fig:MB1CHE01}-\ref{fig:MB1CVG01}, shows the response from various correctors in the merger section before the main linac.  The measured and simulated response on the BPMs upstream of the MLC show good agreement, however the two disagree once the beam passes through the linac.  We point out that the Bmad lattice currently used in the online simulation makes use of a simple analytic cavity model instead of the realistic field maps for the cavities, and therefore may not accurately simulate the orbit through the MLC.  Figs.~\ref{fig:MS1DPB01}-\ref{fig:MS1CRV01} show the BPM response to several S1 magnets downstream of the MLC.  Here the measured data and simulated responses agree well, especially the horizontal orbit. For all of the simulations, the S1 splitter quad strengths were scaled in exactly the same manner as for the dispersion measurements discussed previously.

\subsubsection{Grid Scan in the Fractional FFA Arc\label{ssec:pcscan}}

Sending a grid of beam positions into the FFA arc and measuring the resulting positions on the FA BPMs provides a simple way of checking for BPM and magnet errors.    
\begin{figure*}[ht!]
    \begin{center}
         \subfigure[\hspace{0.2cm}Position scan measured on FA BPM 1.]{%
            \label{fig:gdBPM1}
            \includegraphics[width=0.42\textwidth]{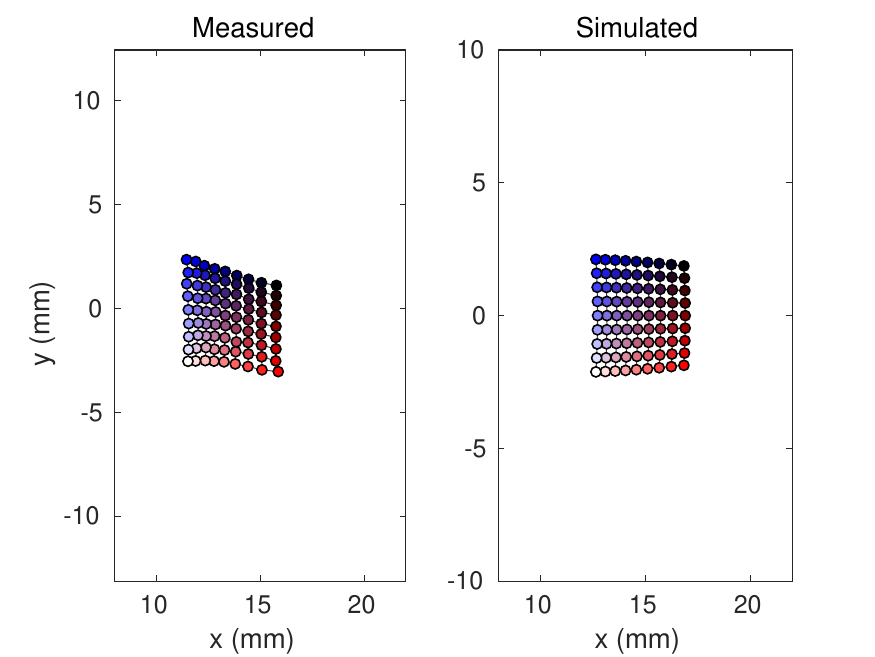}
        }%
        \subfigure[\hspace{0.2cm}Position scan measured on FA BPM 2.]{%
           \label{fig:gdBPM2}
           \includegraphics[width=0.42\textwidth]{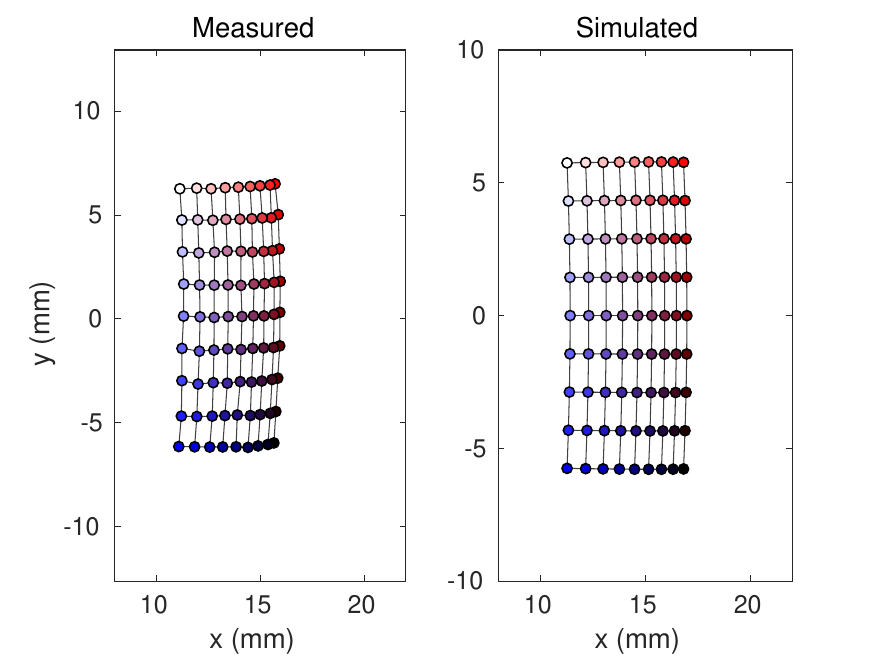}
        }\\ 
        \subfigure[\hspace{0.2cm}Position scan measured on FA BPM 3.]{%
            \label{fig:gdBPM3}
            \includegraphics[width=0.42\textwidth]{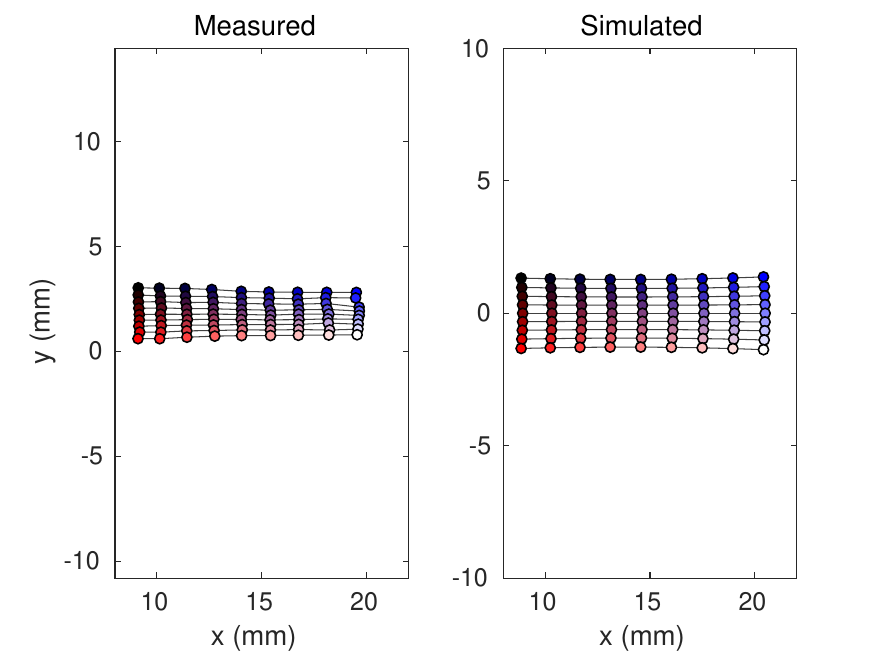}
        }%
        \subfigure[\hspace{0.2cm}Position scan measured on FA BPM 4.]{%
           \label{fig:gdBPM4}
           \includegraphics[width=0.42\textwidth]{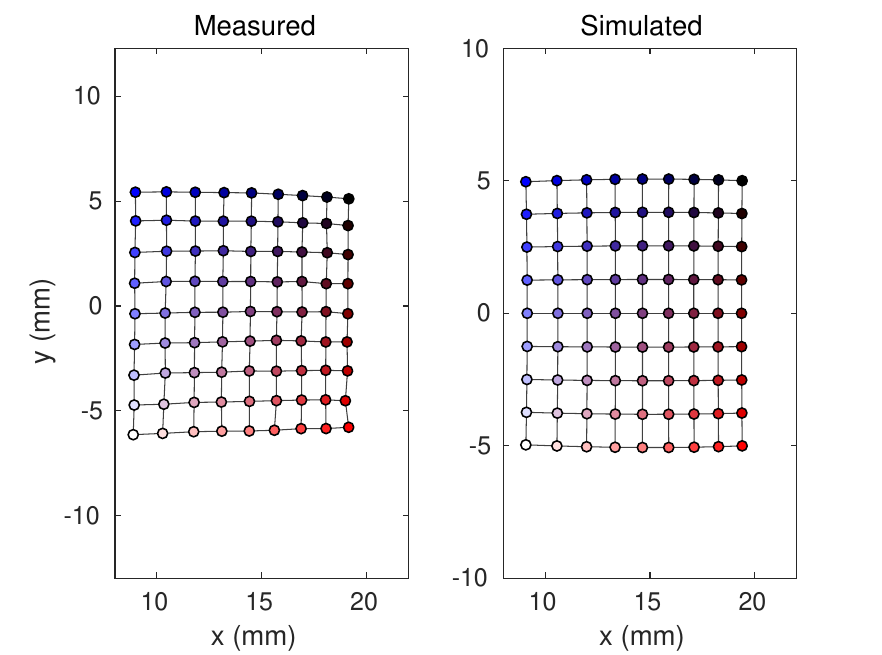}
        }
    \end{center}
    \caption{%
    \label{fig:gridscan}
    FA BPM positions resulting from a single $x,y$ grid scan going into the FA arc.  For each BPM, the measured data is shown on the left, and the simulated data shown on the right.}%
\end{figure*}
Using the Bmad model, we construct combinations of changes to the last two dipole currents that produce 1 mm changes in beam position and no change in angle at the first BPM in the fractional arc.  These combinations are shown in Table~\ref{tab:gridscales}. In the Bmad model of the machine, these pairs produce roughly 1 mm beam offsets with zero angle at the position of the first FA BPM.  

\begin{figure}
    \begin{center}
         \subfigure[\hspace{0.2cm}Example of fitting for the $m_{11}$ elements.]{%
            \label{fig:m11slopeFit}
            \includegraphics[width=0.38\textwidth]{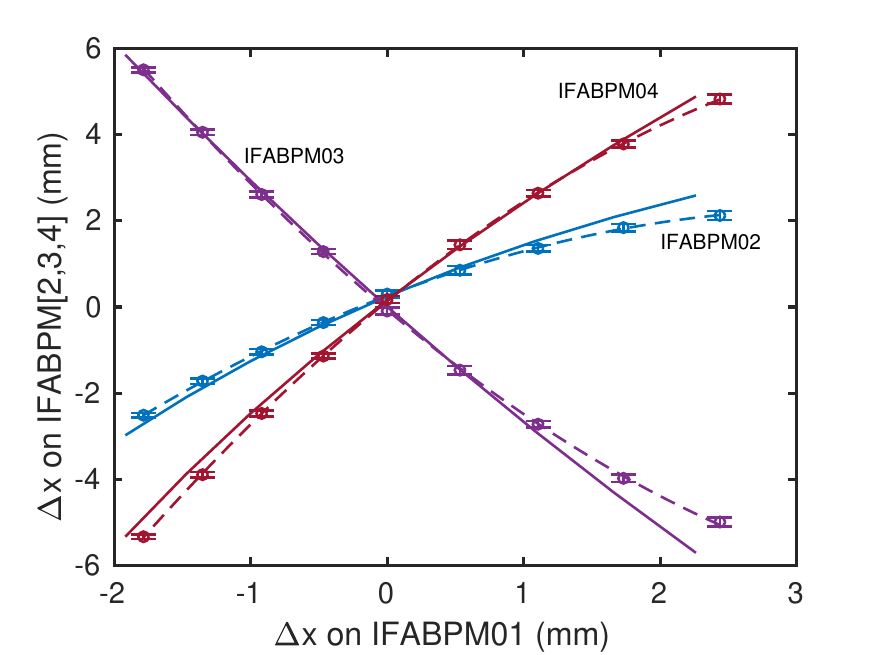}
        }\\
        \subfigure[\hspace{0.2cm}$m_{11}$ and $m_{33}$ elemnts as a funcion of $s$ though the fractional arc.]{%
           \label{fig:m11andm33}
           \includegraphics[width=0.38\textwidth]{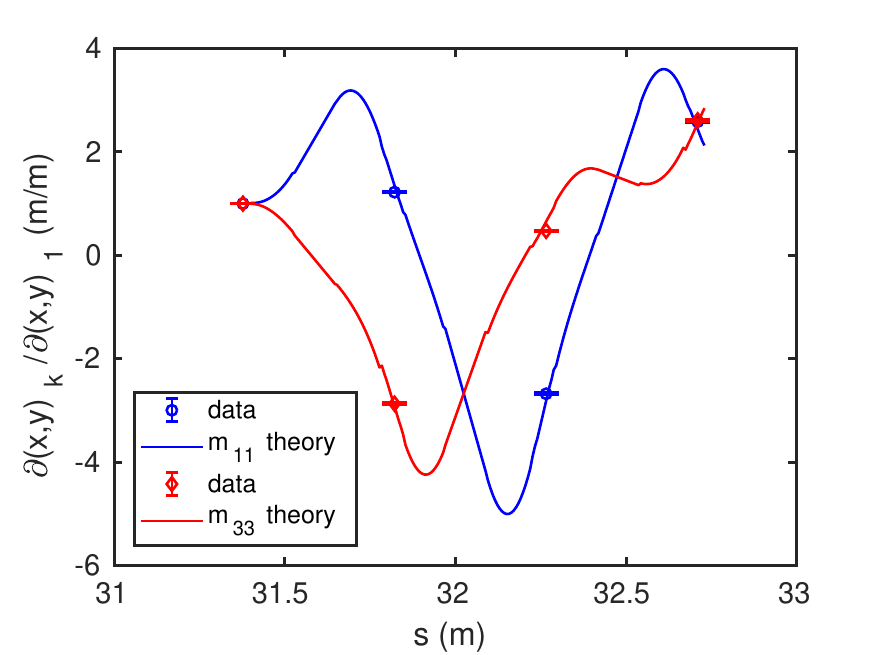}
        }
    \end{center}
    \caption{%
    \label{fig:mats} (a): Fitting the position of of the downstream FA BPMs to the position on first FA BPM. The dashed lines show the resulting fit to the data, the solid lines indicate the prediction from the model.  (b):
    The first derivative term for the x (blue) and y (red) data in (a).  The solid lines show the $m_{11}$ and $m_{33}$ transfer matrix elements as a function of $s$ through the FA line computed from the model.}%
\end{figure}

Fig.~\ref{fig:gridscan} shows the BPM positions when scanning the BPM offsets on the first FA BPM by $\pm$2 mm in nine steps.  For each BPM, both the measured and corresponding simulated BPM data are shown.  Reasonable qualitative agreement can be seen in all cases.
In order to provide a more quantitative comparison between the Bmad model of the FA magnets and measurement, the central ($x=0$ and $y=0$) data in each grid pattern on each of the last three BPMs were fit to a quadratic function of the beam position on the first FA BPM:
\begin{eqnarray}
x_k &=& x_{0,k} + b_x x_{1}  + a_x x_1^2 \label{eqn:matslope}\\
y_k &=& y_{0,k} + b_y y_{1}  + a_y y_1^2 \nonumber
\end{eqnarray}
The linear terms in the above expressions are given explicitly by $b_x = \partial x_k/\partial x_1(y=0)$ and $b_y = \partial y_k/\partial y_1(x=x_0)$ where $x_0$ and $y_0$ are near the design orbit.  Assuming that the beam angles at the first FA BPM remain fixed, these derivatives represent the $m_{11}$ and $m_{33}$ transfer matrix elements from the first BPM to the three downstream BPMs.  Fig.~\ref{fig:m11andm33} shows the comparison of the linear coefficients in Eq.~\ref{eqn:matslope} to the predicted $m_{11}$ and $m_{33}$ matrix elements from the Bmad model of the FA section. Defining the horizontal and vertical relative error in the measured quantities as $(\partial x_k/\partial x_1 - m_{11})/\text{max}(m_{11})$ and $(\partial y_k/\partial y_1 - m_{33})/\text{max}(m_{33})$ implies the measured slopes agree with the theoretical matrix elements to within 7\% for all three downstream BPMs. 

\begin{table}
\begin{ruledtabular}
\caption{\label{tab:gridscales}Magnet combinations used for grid scan.}
\begin{tabular}{c c | c c}
Horizontal Magnet & Scale &Vertical Magnet & Scale\\
\hline
MS1DIP07 & -0.9303 & MS1CRV03 & 0.1641 \\
MS1DPB08 & -4.0451 & MS1CRV04 & -1.2621 \\
\end{tabular}
\end{ruledtabular}
\end{table}


\subsection{Measurements over a Broad Energy Range\label{ssec:escan}} 

The CBETA design features four accelerating passes through the linac, each with 36 MeV gain, requiring a factor of about 3.6 in energy acceptance through the FFA return loop. To study the machine behavior over as much of this energy range as possible, we transported a beam through the FFA arc at a number of energies, ranging from the the minimum energy required to form a stable periodic orbit through the fractional arc (roughly 38 MeV), to the maximum energy deliverable from the main linac (59 MeV). This allowed us to measure intrinsic properties of the FFA arc (betatron phase advance per cell and periodic orbit location), study our model of the splitter line by measuring dispersion and $R_{56}$ near the end of the splitter line, estimate BPM offsets in the FFA arc, and determine a lower bound on the orbit distortions resulting from magnet errors in the FFA arc.

Performing the energy scan required computing a new periodic orbit and the set of matched optics settings in splitter line for each desired energy set point.  Appendix \ref{app:mstate} details the optimization algorithm for determining these new splitter line settings.  Table~\ref{tab:magsettings} shows the results of this procedure for the energy settings eventually used in the following measurements.  Additionally, Table~\ref{tab:mlcsets} shows the MLC cavity set points for the corresponding energies.  For each energy set-point, the appropriate settings from Table~\ref{tab:magsettings} were loaded into the machine.  The dipoles were then adjusted slightly to zero the BPM positions through the splitter line and to steer the beam onto the periodic orbit in the FFA magnets.

\subsubsection{\label{sssec:tunes}Betatron Oscillations}

Driving betatron oscillations at various amplitudes through the FFA arc and measuring the position response on the FFA bpms allows for determination of various lattice properties as a function of energy. In particular, these include intrinsic properties of the FFA arc lattice cell, namely the periodic orbit position at the BPMs and the betatron phase advance per cell (i.e., the tune per cell). In addition, because the phase advance varies as a function of energy, we will be able to estimate the offsets of the BPMs in that arc. These measurements provide a useful test of the validity of the our transport model through the downstream end of the splitter line and fractional arc.

At each of the energy setpoints specified in Table~\ref{tab:magsettings} and Table~\ref{tab:mlcsets}, the beam was kicked using two linear combinations of the last two S1 splitter dipoles (MS1DIP[06,07]) and the last two vertical correctors (MS1CRV[03,04]). These linear combinations were chosen to correspond to a betatron oscillation with a maximum amplitude at the FFA BPMs of 1~mm.  Table~\ref{tab:betavec} shows the linear combinations used as a function of energy. Two linear combinations were chosen for each plane, chosen to give betatron oscillations that are 90$^\circ$ apart in phase. Each linear combination was multiplied by a factor which was scanned from -2 to 2 in unit steps. Only one combination was scanned at a time. For each setting, the beam position on the four FFA BPMs (IFABPM[01-04]) were recorded.  This procedure was automated and tested with the online CBETA Virtual Machine before use. While taking the final measured data, the BPM readings were sampled 10 times at 5 Hz, and the average value and standard deviation saved for offline analysis.   A single second pause was used between magnet setpoints to allow the beam to stabilize.

Our measurements at the $m^{\textmd{th}}$ BPM and the $n^{\textmd{th}}$ energy were fit in the least squares sense to the following functions:
\begin{multline}
    x_{mn} = \left( s_x^{(1)}\cdot A_{x,n}^{(1)} + B_{x,n}^{(1)} \right)\cos\left(2\pi m\cdot\nu_{x,n} + \phi_{x,n}^{(1)} \right) \\
+\left( s_x^{(2)}\cdot A_{x,n}^{(2)} + B_{x,n}^{(2)} \right)\cos\left(2\pi m\cdot\nu_{x,n}  + \phi_{x,n}^{(2)} \right)\\ 
+ C_{x,m} + D_{n}
\label{eqn:tunefitx}
\end{multline}
\begin{multline}
    y_{mn} = \left( s_y^{(1)}\cdot A_{y,n}^{(1)} + B_{y,n}^{(1)}\right)\cos\left(2\pi m\cdot\nu_{y,n}\cdot  + \phi_{y,n}^{(1)} \right) \\
+\left( s_y^{(2)}\cdot A_{y,n}^{(2)} + B_{y,n}^{(2)} \right)\cos\left(2\pi m\cdot\nu_{y,n}  + \phi_{y,n}^{(2)} \right)\\ + C_{y,m}
\label{eqn:tunefity}
\end{multline}
$s_{x,y}^{(1,2)}$ are the scale factors of the kick, scanned from $-$2 to +2 in unit steps with only one of the four $s_{x,y}^{(1,2)}$ being nonzero. $A_{x,y,n}^{(1,2)}$ are the unit amplitudes of the two kicks used in each transverse plane at each energy; if our model were perfect they would be 1~mm. $B_{x,y,n}^{(1,2)}$ is the amplitude of the betatron oscillation with no additional kicks; if the orbit found by hand were the periodic orbit, these would be zero. $C_{x,y,m}$ are the (energy independent) BPM offsets. $D_{n}$ is the (energy dependent) horizontal periodic orbit position; the vertical periodic orbit is known to be zero, so this term only appears in the horizontal function. $\phi_{x,y,n}^{(1,2)}$ are phase offsets of each betatron oscillation;
$\phi_{x,y}^{(1)}$ and $\phi_{x,y}^{(2)}$ will differ by $\pi/2$ if out model were perfect. Finally $\nu_{x,y,n}$ are the tunes per cell of the periodic orbit at energy $E_n$. The initial guess for $A_{x,y,n}^{(1,2)}$ coefficients was set to 1~mm.   Similarly, the initial guesses for $B_{x,y,n}^{(1,2)}$, $\phi_{x,y,n}^{(1,2)}$, and, $C_{x,y,m}$ were set to zero, and the initial horizontal periodic orbit positions $D_m$ guesses set to roughly 15~mm (here the positive x-direction points towards arc center).

$\nu_{x,y}$ and $D_n$ are intrinsic to the arc design; they give a measure of how accurate the Bmad model of the FFA arc is. $C_{x,y,m}$ give an estimate of the BPM offsets. Note that in the horizontal plane, there is a redundancy between $C_{x,m}$ and $D_n$; one can add a given constant to all of the $C_{x,m}$ and subtract that same constant from the $D_n$. We adopt the convention that the average of the $C_{x,m}$ is zero. For full turn energy recovery operation, in particular for 150~MeV energy recovery where the beam passes through the arc with four different energies, having an estimate of the BPM offsets will be extremely helpful for orbit correction since they will help distinguish between orbit offsets caused by magnet errors and position reading errors arising from BPM offsets. Thus a similar energy scan and fit will be performed for the full ring to obtain an estimate of the BPM offsets via this method. $A_{x,y,n}^{(1,2)}$ and $\phi_{x,y,n}^{(1,2)}$ give an estimate of the error in the model from MS1DIP06 through IFABPM01.

\begin{figure*}[ht!]
    \begin{center}
         \subfigure[\hspace{0.2cm}Fit of model to horizontal 42 MeV data.]{%
            \label{fig:hfit}
            \includegraphics[width=0.38\textwidth]{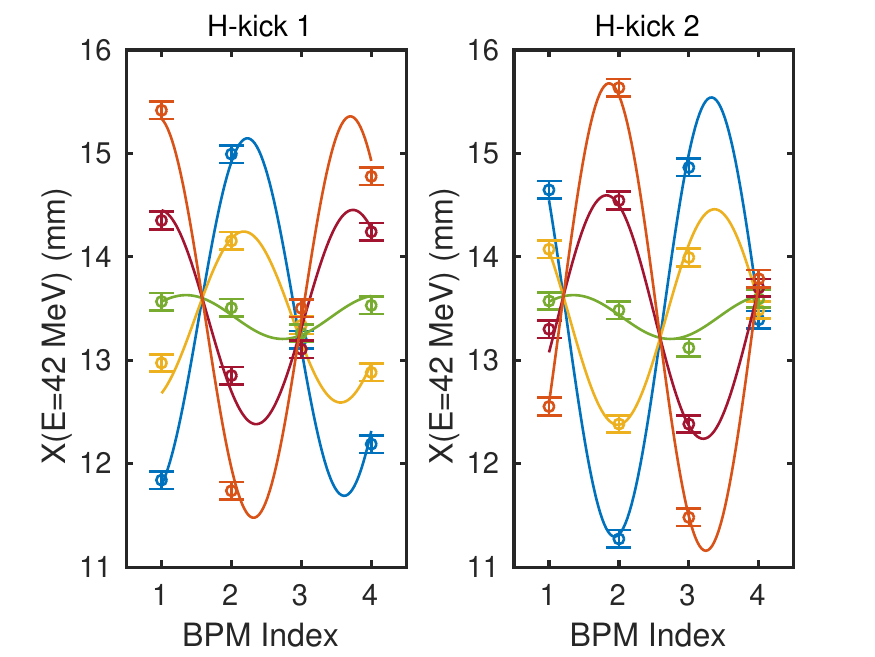}
        }%
        \subfigure[\hspace{0.2cm}Fit of model to vertical 42 MeV data.]{%
           \label{fig:vfit}
           \includegraphics[width=0.38\textwidth]{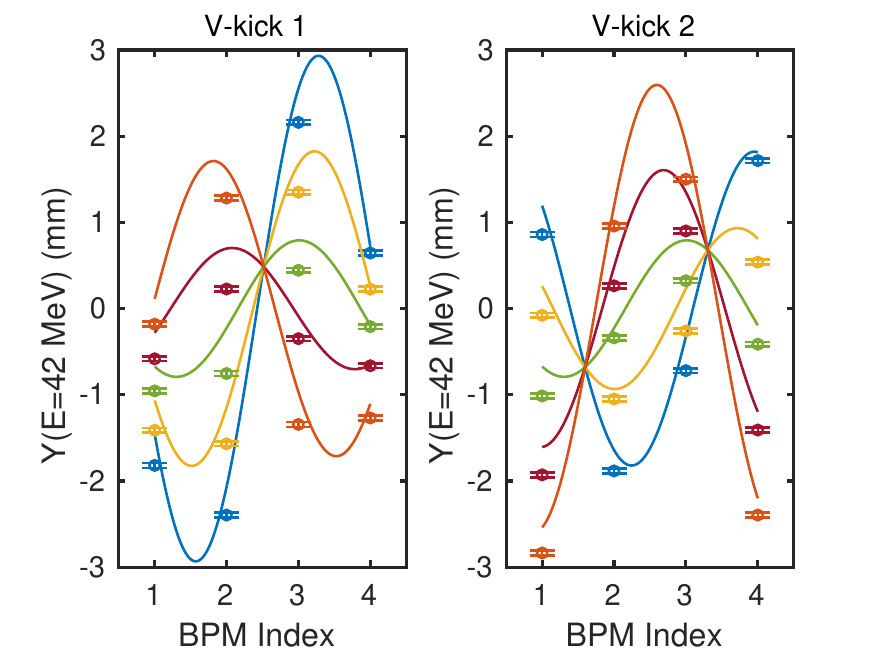}
        }\\ 
        \subfigure[\hspace{0.2cm}Horizontal residuals at 42 MeV.]{%
            \label{fig:hres}
            \includegraphics[width=0.38\textwidth]{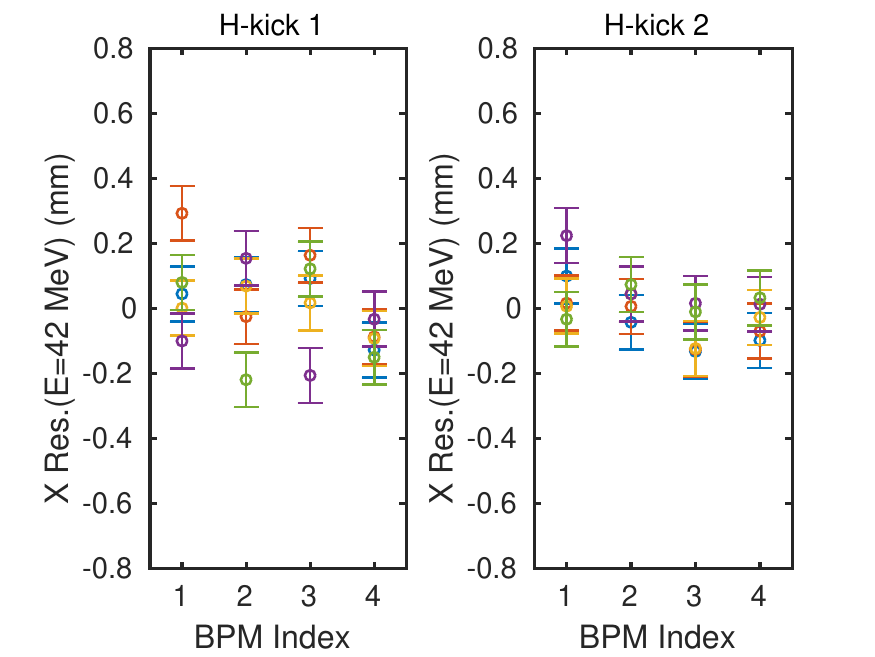}
        }%
        \subfigure[\hspace{0.2cm}Vertical residuals at 42 MeV.]{%
           \label{fig:vres}
           \includegraphics[width=0.38\textwidth]{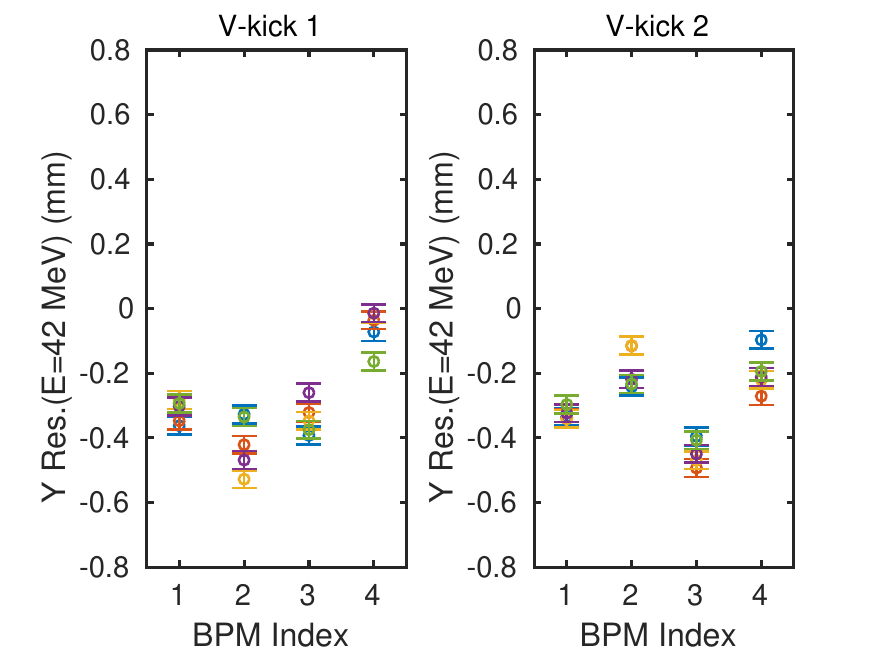}
        }\\         
    \end{center}
    \caption{%
    \label{fig:rawfit}
    Example data and model fits (top) and residuals (bottom) for the horizontal (left) and vertical (right) BPM positions on the FA BPMs for each value of the horizontal and vertical betatron oscillation kick(s) at 42 MeV.}%
\end{figure*}

Fig.~\ref{fig:rawfit} shows an example of the fit to the horizontal (Fig.~\ref{fig:hfit}) and vertical (Fig.~\ref{fig:vfit}) BPM data at 42 MeV for both betatron oscillations, as well as the corresponding resulting fit residuals (Fig.~\ref{fig:hres} and \ref{fig:vres}).  
For all horizontal/vertivcal positions we use an estimate of the error on the BPM positions of roughly 0.1 mm horizontally and 0.03 mm vertically, based on \emph{all} of the BPM measurements taken during the energy scan. The residuals in the fit that are beyond what one expect statistically are present due to magnet imperfections. Magnet imperfections can also change the parameters that we solve for, but having more cells and/or a wider range of energies will tend to make the parameters we solve for more accurate and transfer more of the imperfections into the residuals.

\begin{figure*}
    \begin{center}
        \subfigure[\hspace{0.2cm}Measured horizontal and vertical tunes, compared to prediction.]{%
        \label{fig:rtunes}
        \includegraphics[width=0.38\textwidth]{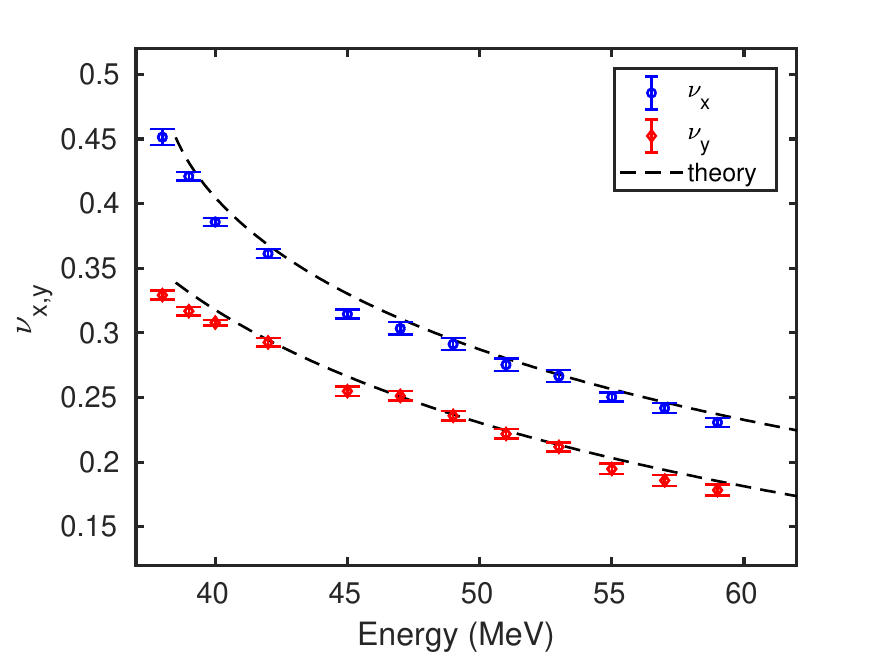}
        }%
        \subfigure[\hspace{0.2cm}Measured horizontal and vertical orbit offset.]{%
           \label{fig:bpmoffsets}
           \includegraphics[width=0.38\textwidth]{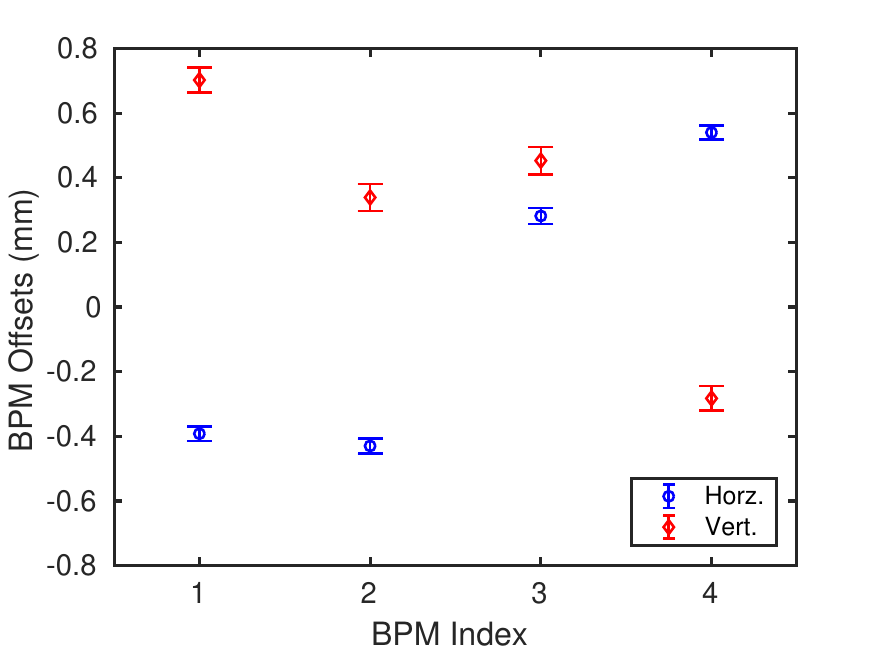}
        }\\ 
        \subfigure[\hspace{0.2cm}Amplitude of horizontal orbit kicks.]{%
        \label{fig:xamps}
        \includegraphics[width=0.38\textwidth]{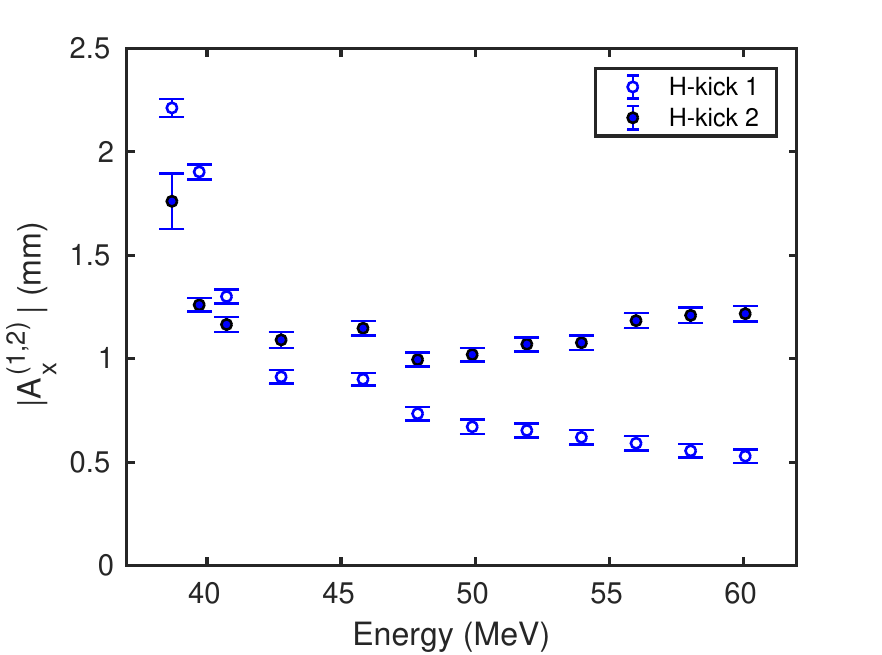}
        }%
        \subfigure[\hspace{0.2cm}Amplitude of vertical orbit kicks.]{%
        \label{fig:yamps}
        \includegraphics[width=0.38\textwidth]{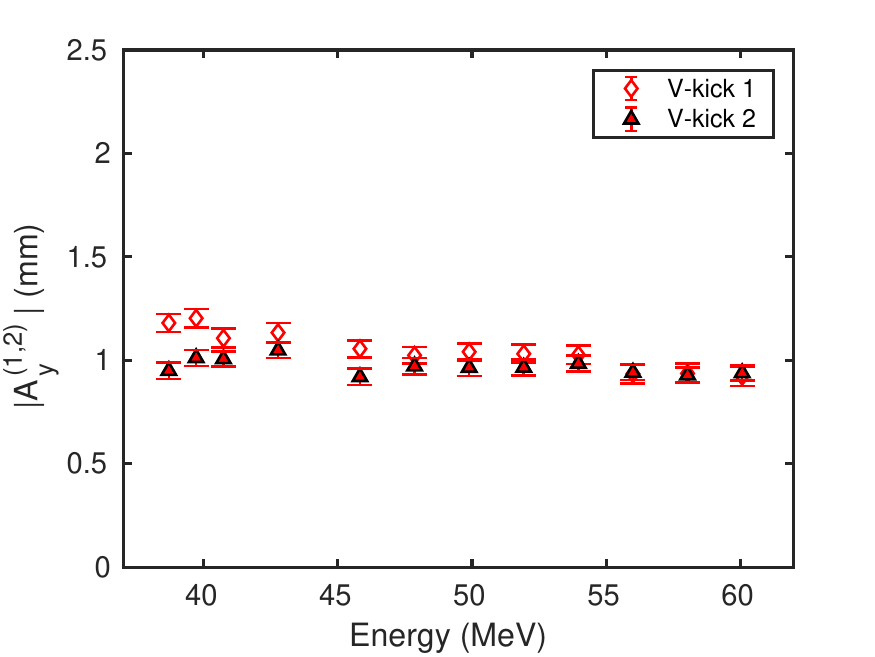}
        }\\ 
        \subfigure[\hspace{0.2cm}Phase difference between pairs of horizontal and vertical kicks.]{%
            \label{fig:phases}
            \includegraphics[width=0.38\textwidth]{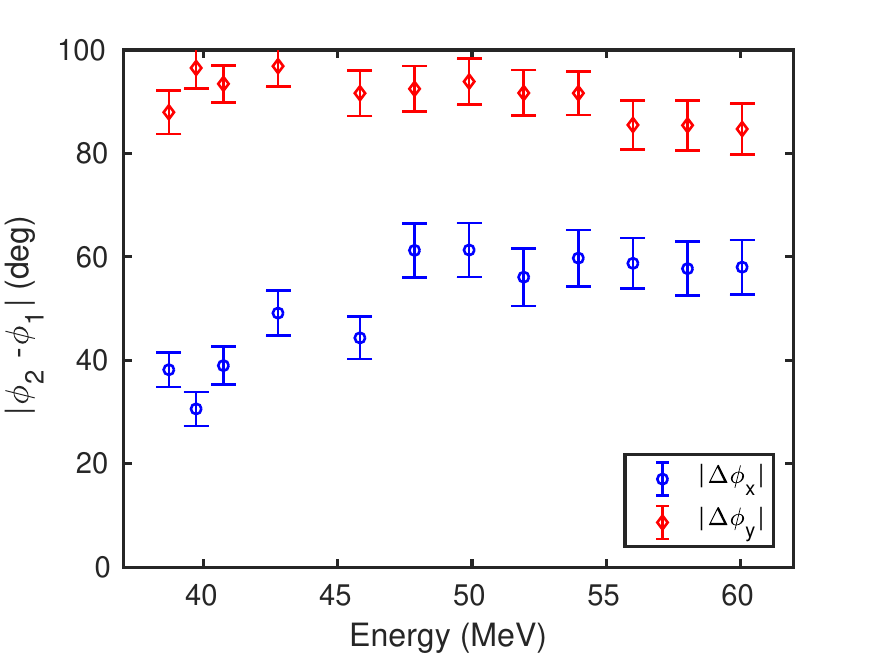}
        }%
        \subfigure[\hspace{0.2cm}Measured and simulated horizontal periodic orbit position.]{%
           \label{fig:xorb}
           \includegraphics[width=0.38\textwidth]{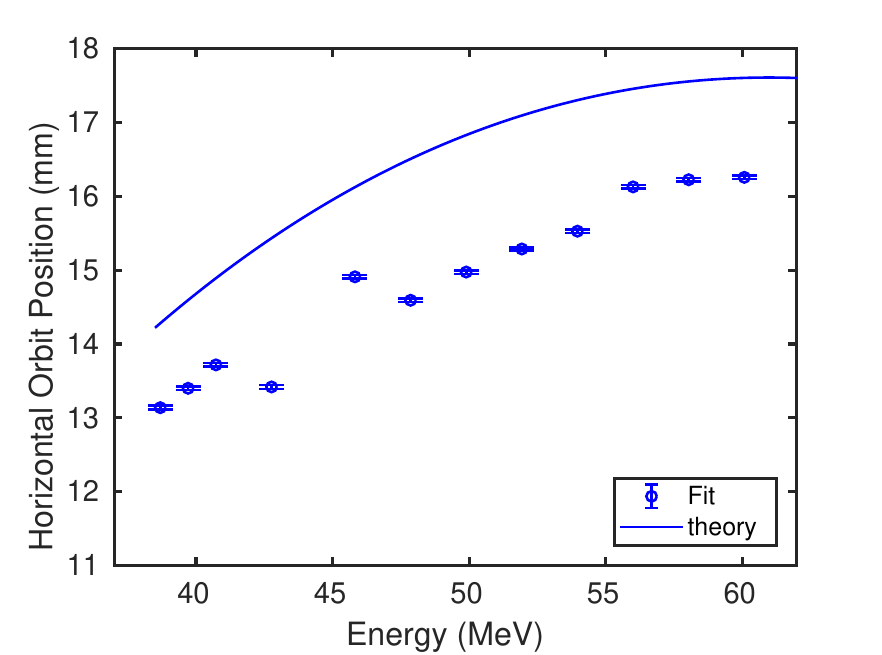}
        }\\ 
    \end{center}
    \caption{%
    \label{fig:rawtunes}
    Resulting fit parameters: tunes (a), BPM offsets (b), horizontal (c)  and vertical (d) unit amplitudes, oscillation phase difference (e), and horizontal periodic orbit position (f). The solid line in (f) shows the simulated prediction for the periodic orbit position.}%
\end{figure*}

The resulting fit parameters are displayed in Fig.~\ref{fig:rawtunes}.  These include the resulting tunes as a function of the machine set-point energy (Fig.~\ref{fig:rtunes}), the energy independent BPM offsets (Fig.~\ref{fig:bpmoffsets}, also in Table~\ref{tab:bpmoffsets}), the unit horizontal and vertical betatron amplitudes (Figs.~\ref{fig:xamps} and \ref{fig:yamps}), the phase difference between the two betatron oscillations in each direction (Fig.~\ref{fig:phases}), and the measured and simulated horizontal periodic orbit positions as a function of energy (Fig.~\ref{fig:xorb}). In Fig.~\ref{fig:rtunes} the dashed lines show the corresponding prediction for the tunes, determined by tracking particles through 3D fieldmaps of the FA magnets and solving for the closed orbit, and displays good quantitative agreement.
\begin{table}
\caption{\label{tab:bpmoffsets}BPM offsets (mm)}
\begin{tabular}{c | c | c}
\hline
\hline
BPM & X offset ($C_{x}$) (mm) & Y offset ($C_{y}$) (mm)  \\
\hline
IFABPM01 & $-0.40\pm0.02$ & $ 0.70\pm0.04$ \\
IFABPM02 & $-0.43\pm0.02$ & $ 0.34\pm0.04$ \\
IFABPM03 & $ 0.28\pm0.03$ & $ 0.45\pm0.04$ \\
IFABPM04 & $ 0.54\pm0.02$ & $-0.28\pm0.04$ \\
\hline
\hline
\end{tabular}
\end{table}

Both the horizontal unit betatron amplitudes and phases shown Fig.~\ref{fig:xamps} and Fig.~\ref{fig:phases} indicate an inaccuracy in the Bmad FAT model.  In the former case the unit amplitudes were designed to be 1~mm, in the later case the phase difference was intended to be 90 deg.  In the vertical plane, the measured amplitude and phase difference is in good agreement with the model as can be seen in the vertical unit amplitudes in Fig.~\ref{fig:yamps} and vertical phase shifts (red) in Fig.~\ref{fig:phases}. 

The horizontal periodic orbit position shown in Fig.~\ref{fig:xorb} has an average systematic error with respect to the theoretical prediction for the orbit position of roughly 1.5 mm. This could arise from a nonzero average in the BPM offsets (though it is unlikely to be this large), non-linearity in the BPM response, a systematic difference between the modeled and as-built magnets, and possibly other effects. 

All of these measurements relied on the semi-analytic model discussed in Sec.~\ref{sssec:bpm} for the BPM nonlinear correction in FA BPMs.  Unfortunately, direct measurement of the FA BPM non-linearity wasn't possible during the FAT as the beam offset here is purposely large (15 mm) which limits position range required to carefully quantify the effect.

\begin{figure}[ht!]
\includegraphics[width=0.38\textwidth]{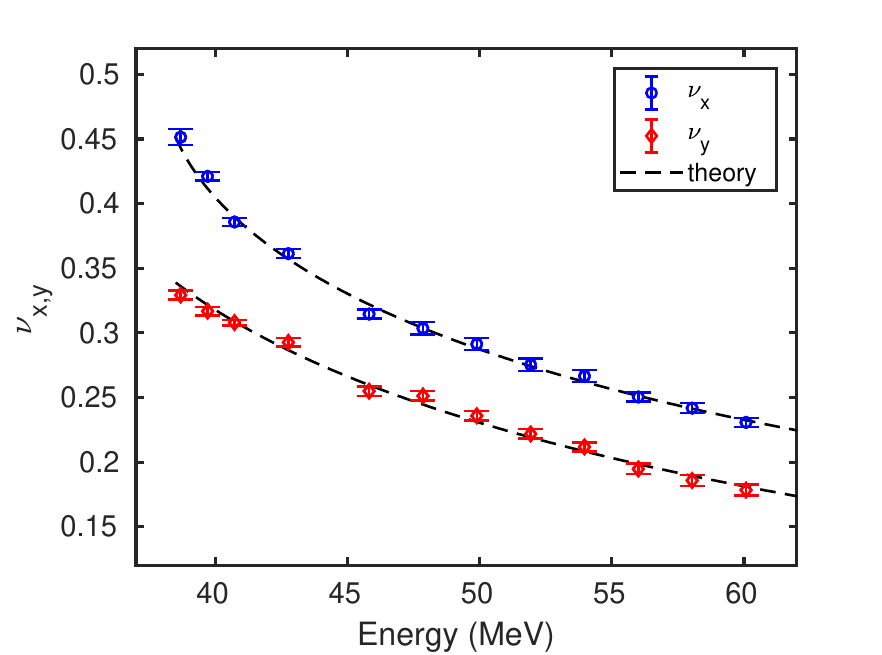}
\caption{\label{fig:tunes_fixed}Measured horizontal and vertical tunes, compared to simulation, after a systematic scaling of the energy. }
\end{figure}

Upon comparing the tune data with the model in Fig.~\ref{fig:rtunes}, we sought to determine whether some of the difference could arise from energies being off by a uniform scaling factor (as a result of an energy calibration error, for instance). This scale factor is determined by a least-squares fit in by exploiting the relationship between the tunes and the energy: $\cos(2\pi\nu) = \sum_n b_n E^{-n}$.  Inverting this relationship (and keeping the first six terms) gives:
\begin{eqnarray}
\frac{1}{E} = \sum_{n=1}^6 a_{x,y}^{(n)} \cos^n\left(2\pi\nu_{x,y}\right).
\end{eqnarray}
This relation allows for a closed form solution for the best energy scale factor $\alpha$ (in the least-squares sense) of
\begin{eqnarray}
\alpha &=& \frac{1}{2\langle E^2\rangle} \left[\left\langle\frac{E}{\sum_{n=1}^6a_{x}^{(n)}\cos^n(2\pi\nu_x)}\right\rangle\right.\\
&& \hspace{0.2cm} + \left.\left\langle\frac{E}{\sum_{n=1}^6a_y^{(n)}\cos^n(2\pi\nu_y)}\right\rangle\right].\nonumber
\end{eqnarray}
This results in an energy scale factor of $\alpha = 1.018$.  This puts the predicted tunes from 3D tracking nearly within all of the errorbars on the measured tune data, as seen in Fig. \ref{fig:tunes_fixed}.  Note that this scaling factor could arise not only from a systematic scaling of the energy, but also systematic error in the model of the quadruple magnets. 

\section{Conclusions\label{section:conclusion}}

The CBETA fractional arc test provided invaluable experience testing and commissioning many of the most critical CBETA subsystems.  Full beam commissioning of the MLC yielded beam energies up to roughly 60 MeV, nearly a factor of 1.6 times the 36 MeV necessary for the CBETA design.  Reaching these energies required development of many important measurement procedures including a method for cavity energy gain calibration using BPM time of arrival.   Similarly, commissioning of the low energy splitter line saw development of procedures for testing non-linearity correction of the splitter BPMs and the first successful test of the path length adjustment mechanism.  Additionally, we successfully transported the beam through the MLC and splitter, and ultimately injecting the beam onto the periodic orbit in the fractional arc. We succeeded in measuring the phase advance per cell and the periodic orbit location in that arc over a broad energy range. The results in Figs.~\ref{fig:rtunes} represent a significant milestone for the CBETA project. Many of the measurement techniques developed in this work will be used for commissioning the full return loop.  The FAT also proved extremely helpful in developing the CBETA Virtual Machine online modeling software. The simulation results here demonstrate its usefulness as a tool for not only displaying useful physics data to operators in real time, but also for debugging and testing measurement procedures before putting them to use in the real machine.  

In addition to these successes, the results of the FAT highlight several important issues that require addressing prior to future CBETA beam commissioning. Measurement of the MLC cavity offsets of roughly 4.0 mm (on average) represents a particularly important, if surprising, example.  Others include difficulty establishing a reproducible good match of the injector beam into the MLC and splitter line, lack of an automated MLC cavity phasing procedure, the inability currently to quantify the FA style BPM non-linearity, and lack of development for diagnostic procedures (other than beam profiles on viewscreens) to verify the match \emph{after} the linac. Additionally, while we did develop and test a measurement method for determining the corrector-to-BPM response matrix throughout the machine, we did not substantially test the orbit correction software under development for CBETA.  Fortunately, with the help of the CBETA Virtual Machine, this work has been able to continue past the end of the FAT experimental period, with promising initial results using Single Value Decomposition (SVD) \cite{ref:CBETAVM}.  We have not at this time fully accounted for the discrepancy between the quad settings giving the best fit to the dispersion in the permanent magnet arc in measurement and in the Bmad FAT model. Measurements of the tunes via induced betatron oscillations around the periodic orbit suggest an overall energy scale factor could explain the discrepancy between the measured and simulated tunes as a function of beam energy.  This could arise from an unknown systematic discrepancy between the 3D field maps for the permanent magnets and the fields in the magnets themselves or from an error in our energy calibration.  

Significant effort is currently underway to address many of these issues:  offline analysis of beam-based calibration data for the splitter magnets continues, as does orbit correction software development.  In addition to the direct measurement of the BPM non-linearity described in this work, 3D simulations of the splitter/FA BPM designs have been planned. Additionally, we plan to explore new injector machine, as well as investigating methods for verifying the match after the linac.  Perhaps most importantly, the entrance and exit beamlines to the MLC as well as the cryostat itself are currently being surveyed with the goal of centering the MLC for operation with the full return loop.  Given the demonstrated usefulness of the CBETA Virtual Machine online model, significant development of the software is ongoing, and we have begun to use the 42~MeV energy recovery lattice in the model.  With this work and more we hope to build on the results achieved during the CBETA FAT as we move forward to the final completion of CBETA construction and beam commissioning.

\acknowledgements
The authors would like to thank the hard work of the entire CBETA team as well as the growing group of international collaborators. This work was funded by NYSERDA, the New York State Energy Research and Development Agency. This manuscript has been authored by employees of Brookhaven Science Associates, LLC under Contract No. DE-SC0012704 with the U.S. Department of Energy.

\appendix

\section{Machine Settings used in the Energy Scan\label{app:mstate}}

To produce quadrupole and dipole settings which successfully match the beam from the splitter line into the fractional arc, we use an optimization process which starts from magnet settings at one energy and determines settings for a nearby energy. To do this requires a two-level nested optimization. The inner layer has a fixed set of quadrupole gradients and begins with the dipoles set to their 42~MeV settings multiplied by the ratio of the beam momentum at the desired energy to the momentum at 42~MeV/$c$. It then adjusts the fields of MS1DIP06 and MS1DIP07 to steer the beam onto the periodic orbit in the FFA. For each evaluation performed in the outer layer, it chooses a set of quadrupole gradients, calls the inner layer solution process, then computes the Betatron functions $\beta_{x,y}$ and $\alpha_{x,y}$, and the dispersion functions, $\eta_x$ and $\eta_{x}^{\prime}$, and subtracts the target values computed in the previous paragraph. 

To find the optimal quadrupole gradients, the outer layer applies the equivalent of a nonlinear SVD optimization. One step of the optimization involves a computation of the Betatron and dispersion functions, a computation of the derivative of those functions with respect to the quadrupole gradients, and finding a step in the quadrupole gradients that to linear order would match the Betatron and dispersion functions, while minimizing the sum of the squares of the quadrupole gradient steps. 

To determine usable settings for the energy scan performed during the FAT, we first determined the lowest linearly stable energy in the model. This resulted in a lowest stable energy of 38.5~MeV. Starting with our settings for 42~MeV, we scaled the quadrupole gradients from their 42~MeV design values by the momentum ratio of 38.5/42, and found the settings for 38.5~MeV. We then stepped in 0.1~MeV steps up to our highest energy, using the quadrupole settings from the previous energy. Note that the settings that this process finds for 42~MeV will be different from the initial 42~MeV design values. Because we were able to find stable orbits in the machine for 38~MeV, (below the minimum 38.5~MeV in the model), we scaled the quadrupole and dipole settings by a factor of 38/38.5 to find the settings to use for 38~MeV.
\begin{table*}[ht!]
\tiny
\begin{ruledtabular}
\caption{D1/S1 splitter magnet current setgtings (Amps) used during the energy scan}
\label{tab:magsettings}
\begin{tabular}{
>{\centering}m{0.8in}|
>{\centering}m{0.42in}
>{\centering}m{0.42in}
>{\centering}m{0.42in}
>{\centering}m{0.42in}
>{\centering}m{0.42in}
>{\centering}m{0.42in}
>{\centering}m{0.42in}
>{\centering}m{0.42in}
>{\centering}m{0.42in}
>{\centering}m{0.42in}
>{\centering}m{0.42in}
>{\centering\arraybackslash}m{0.42in}}
Energy (MeV) & 38 & 39 & 40 & 42 & 45 & 47 & 49 & 51 & 53 & 55 & 57 & 59 \\
D1DIP01 & 1.5502 & 1.5910 & 1.6318 & 1.7134 & 1.8358 & 1.9174 & 1.9990 & 2.0806 & 2.1622 & 2.2438 & 2.3254 & 2.4070 \\
\hline
S1DPB01 & 202.0905 & 207.4092 & 212.7283 & 223.3664 & 239.3234 & 249.9614 & 260.5992 & 271.2371 & 281.8749 & 292.5126 & 303.1504 & 313.7881 \\
S1DIP02 & 57.3341 & 58.8430 & 60.3521 & 63.3701 & 67.8972 & 70.9153 & 73.9333 & 76.9513 & 79.9693 & 82.9873 & 86.0052 & 89.0232 \\
S1DIP03 & 99.0435 & 101.6501 & 104.2569 & 109.4706 & 117.2911 & 122.5047 & 127.7182 & 132.9318 & 138.1453 & 143.3588 & 148.5723 & 153.7858 \\
S1DIP04 & 73.9750 & 75.9219 & 77.8689 & 81.7630 & 87.6041 & 91.4981 & 95.3921 & 99.2860 & 103.1800 & 107.0739 & 110.9678 & 114.8618 \\
S1DIP05 & 73.9750 & 75.9219 & 77.8689 & 81.7630 & 87.6041 & 91.4981 & 95.3921 & 99.2860 & 103.1800 & 107.0739 & 110.9678 & 114.8618 \\
S1DIP06 & 99.4919 & 102.1050 & 104.6062 & 109.4706 & 116.6253 & 121.3385 & 126.0165 & 130.6623 & 135.2771 & 139.8616 & 144.4155 & 148.9384 \\
S1DIP07 & 64.9532 & 66.6049 & 68.1571 & 71.1956 & 75.6898 & 78.6702 & 81.6471 & 84.6246 & 87.6055 & 90.5920 & 93.5859 & 96.5888 \\
S1DPB08 & 190.0498 & 195.0516 & 200.0537 & 210.0580 & 225.0643 & 235.0684 & 245.0725 & 255.0765 & 265.0805 & 275.0845 & 285.0884 & 295.0923 \\
\hline
\hline
S1QUA01 & 2.3022 & 2.4236 & 2.4628 & 2.5274 & 2.6445 & 2.7364 & 2.8378 & 2.9476 & 3.0645 & 3.1876 & 3.3161 & 3.4494 \\
S1QUA02 & 4.9717 & 5.1196 & 5.2379 & 5.4576 & 5.7773 & 5.9869 & 6.1942 & 6.3995 & 6.6029 & 6.8039 & 7.0024 & 7.1984 \\
S1QUA03 & 3.1103 & 3.1303 & 3.2062 & 3.3892 & 3.6685 & 3.8526 & 4.0344 & 4.2140 & 4.3918 & 4.5681 & 4.7433 & 4.9175 \\
S1QUA04 & 6.7818 & 6.8432 & 6.9538 & 7.2743 & 7.8138 & 8.1877 & 8.5669 & 8.9496 & 9.3347 & 9.7216 & 10.1098 & 10.4991 \\
S1QUA05 & 6.2332 & 6.5966 & 6.7918 & 7.0255 & 7.3366 & 7.5489 & 7.7687 & 7.9953 & 8.2278 & 8.4650 & 8.7059 & 8.9497 \\
S1QUA06 & 0.1935 & 0.7143 & 1.0475 & 1.3296 & 1.6045 & 1.7657 & 1.9211 & 2.0739 & 2.2255 & 2.3769 & 2.5283 & 2.6803 \\
S1QUA07 & 1.1752 & 1.9922 & 2.6517 & 3.2915 & 3.9222 & 4.2856 & 4.6341 & 4.9773 & 5.3202 & 5.6658 & 6.0159 & 6.3718 \\
S1QUA08 & 2.5738 & 2.2134 & 1.8072 & 1.4061 & 1.1062 & 0.9808 & 0.8821 & 0.7971 & 0.7180 & 0.6401 & 0.5604 & 0.4767 
\end{tabular}
\end{ruledtabular}
\end{table*}

A subset of these energies was used take to take beam data with.  Table~\ref{tab:mlcsets} collects the various energies data was taken at, along with the corresponding the cavity voltages.  All cavities were set to on-crest acceleration.
\begin{table*}
\begin{ruledtabular}
\caption{MLC cavity voltage settings (MV) used during the energy scan\label{tab:mlcsets}}
\begin{tabular}{
>{\centering}m{0.8in}|
>{\centering}m{0.42in}
>{\centering}m{0.42in}
>{\centering}m{0.42in}
>{\centering}m{0.42in}
>{\centering}m{0.42in}
>{\centering}m{0.42in}
>{\centering}m{0.42in}
>{\centering}m{0.42in}
>{\centering}m{0.42in}
>{\centering}m{0.42in}
>{\centering}m{0.42in}
>{\centering\arraybackslash}m{0.42in}}
Energy (MeV) & 38 & 39 & 40 & 42 & 45 & 47 & 49 & 51 & 53 & 55 & 57 & 59 \\
\hline
Cavity 6 & 8.00 & 8.00 & 8.00 & 6.00 & 9.00 & 11.00 & 11.00 & 11.00 & 11.00 & 11.00 & 11.00 & 11.35 \\
Cavity 5 & 6.00 & 6.00 & 6.00 & 6.00 & 6.00 & 6.00 & 6.00 & 6.00 & 8.00 & 8.00 & 8.00 & 8.50 \\
Cavity 4 & 5.00 & 5.00 & 5.00 & 6.00 & 6.00 & 6.00 & 8.00 & 8.00 & 8.00 & 8.00 & 8.00 & 8.40 \\
Cavity 3 & 4.00 & 4.00 & 5.00 & 6.00 & 6.00 & 6.00 & 6.00 & 6.00 & 6.00 & 7.50 & 7.50 & 7.50 \\
Cavity 2 & 4.00 & 5.00 & 5.00 & 6.00 & 6.00 & 6.00 & 6.00 & 8.00 & 8.00 & 8.00 & 8.00 & 8.00 \\
Cavity 1 & 5.00 & 5.00 & 5.00 & 6.00 & 6.00 & 6.00 & 6.00 & 6.00 & 6.00 & 6.50 & 8.50 & 9.25 \\
\end{tabular}
\end{ruledtabular}
\end{table*}
Finally, a simple response matrix calculation was used to generate linear combinations of the last two splitter dipoles and vertical correctors for use in inducing betatron oscillations around the peridic orbit established at the energies in Table~\ref{tab:mlcsets}. Two linear combinations were formed, corresponding to betatron oscillations 90 deg out of phase in the Bmad model of the fractional arc.
\begin{table*}
\tiny
\begin{ruledtabular}
\caption{S1 Orthogonal betatron excitation current settings (Amps) used during the energy scan}
\label{tab:betavec}
\begin{tabular}{
>{\centering}m{0.8in}|
>{\centering}m{0.42in}
>{\centering}m{0.42in}
>{\centering}m{0.42in}
>{\centering}m{0.42in}
>{\centering}m{0.42in}
>{\centering}m{0.42in}
>{\centering}m{0.42in}
>{\centering}m{0.42in}
>{\centering}m{0.42in}
>{\centering}m{0.42in}
>{\centering}m{0.42in}
>{\centering\arraybackslash}m{0.42in}}
Energy (MeV) & 38 & 39 & 40 & 42 & 45 & 47 & 49 & 51 & 53 & 55 & 57 & 59 \\
\hline
S1DIP06:H1 & -1.7061 & -1.7663 & -1.7594 & -1.7090 & -1.6360 & -1.5940 & -1.5572 & -1.5251 & -1.4971 & -1.4729 & -1.4519 & -1.4338 \\
S1DIP07:H1 & -0.6939 & -0.8341 & -0.9091 & -0.9219 & -0.8738 & -0.8325 & -0.7908 & -0.7503 & -0.7119 & -0.6757 & -0.6416 & -0.6095 \\
\hline
S1DIP06:H2 & 0.2240 & 0.3185 & 0.4418 & 0.6010 & 0.7607 & 0.8435 & 0.9159 & 0.9816 & 1.0427 & 1.1007 & 1.1568 & 1.2115 \\
S1DIP07:H2 & 0.1671 & 0.2300 & 0.3169 & 0.4264 & 0.5264 & 0.5722 & 0.6083 & 0.6377 & 0.6623 & 0.6833 & 0.7015 & 0.7174 \\
\hline
\hline
S1CRV03:V1 & -0.7991 & -0.7628 & -0.7289 & -0.7083 & -0.7024 & -0.7026 & -0.7036 & -0.7046 & -0.7054 & -0.7058 & -0.7059 & -0.7056 \\
S1CRV04:V1 & 1.3982 & 1.2964 & 1.1784 & 1.0744 & 1.0084 & 0.9832 & 0.9632 & 0.9449 & 0.9268 & 0.9078 & 0.8876 & 0.8659 \\
\hline
S1CRV03:V2 & 0.7814 & 0.7612 & 0.7498 & 0.7579 & 0.7785 & 0.7912 & 0.8024 & 0.8121 & 0.8202 & 0.8270 & 0.8325 & 0.8369 \\
S1CRV04:V2 & -1.8049 & -1.7516 & -1.6892 & -1.6532 & -1.6520 & -1.6605 & -1.6704 & -1.6798 & -1.6874 & -1.6926 & -1.6953 & -1.6952
\end{tabular}
\end{ruledtabular}
\end{table*}


\begin{thebibliography}{39}%
\makeatletter
\providecommand \@ifxundefined [1]{%
 \@ifx{#1\undefined}
}%
\providecommand \@ifnum [1]{%
 \ifnum #1\expandafter \@firstoftwo
 \else \expandafter \@secondoftwo
 \fi
}%
\providecommand \@ifx [1]{%
 \ifx #1\expandafter \@firstoftwo
 \else \expandafter \@secondoftwo
 \fi
}%
\providecommand \natexlab [1]{#1}%
\providecommand \enquote  [1]{``#1''}%
\providecommand \bibnamefont  [1]{#1}%
\providecommand \bibfnamefont [1]{#1}%
\providecommand \citenamefont [1]{#1}%
\providecommand \href@noop [0]{\@secondoftwo}%
\providecommand \href [0]{\begingroup \@sanitize@url \@href}%
\providecommand \@href[1]{\@@startlink{#1}\@@href}%
\providecommand \@@href[1]{\endgroup#1\@@endlink}%
\providecommand \@sanitize@url [0]{\catcode `\\12\catcode `\$12\catcode
  `\&12\catcode `\#12\catcode `\^12\catcode `\_12\catcode `\%12\relax}%
\providecommand \@@startlink[1]{}%
\providecommand \@@endlink[0]{}%
\providecommand \url  [0]{\begingroup\@sanitize@url \@url }%
\providecommand \@url [1]{\endgroup\@href {#1}{\urlprefix }}%
\providecommand \urlprefix  [0]{URL }%
\providecommand \Eprint [0]{\href }%
\providecommand \doibase [0]{http://dx.doi.org/}%
\providecommand \selectlanguage [0]{\@gobble}%
\providecommand \bibinfo  [0]{\@secondoftwo}%
\providecommand \bibfield  [0]{\@secondoftwo}%
\providecommand \translation [1]{[#1]}%
\providecommand \BibitemOpen [0]{}%
\providecommand \bibitemStop [0]{}%
\providecommand \bibitemNoStop [0]{.\EOS\space}%
\providecommand \EOS [0]{\spacefactor3000\relax}%
\providecommand \BibitemShut  [1]{\csname bibitem#1\endcsname}%
\let\auto@bib@innerbib\@empty
\bibitem [{\citenamefont {{The 2015 Nuclear Science Advisory
  Committee}}()}]{ref:LongRange1}%
  \BibitemOpen
  \bibfield  {author} {\bibinfo {author} {\bibnamefont {{The 2015 Nuclear
  Science Advisory Committee}}},\ }\href@noop {} {\ }\bibinfo {note}
  {\url{https://science.energy.gov/np/nsac/}}\BibitemShut {NoStop}%
\bibitem [{\citenamefont {{National Academies of Sciences, Engineering, and
  Medicine}}(2018)}]{ref:NAS}%
  \BibitemOpen
  \bibfield  {author} {\bibinfo {author} {\bibnamefont {{National Academies of
  Sciences, Engineering, and Medicine}}},\ }\href {\doibase 10.17226/25171}
  {\emph {\bibinfo {title} {An Assessment of U.S.-Based Electron-Ion Collider
  Science}}}\ (\bibinfo  {publisher} {The National Academies Press},\ \bibinfo
  {address} {Washington, DC},\ \bibinfo {year} {2018})\ \bibinfo {note}
  {doi:10.17226/25171}\BibitemShut {NoStop}%
\bibitem [{\citenamefont {{Hoffstaetter}}\ \emph {et~al.}(2017)\citenamefont
  {{Hoffstaetter}}, \citenamefont {{Trbojevic}}, \citenamefont {{Mayes}},
  \citenamefont {{Banerjee}}, \citenamefont {{Barley}}, \citenamefont
  {{Bazarov}}, \citenamefont {{Bartnik}}, \citenamefont {{Berg}}, \citenamefont
  {{Brooks}}, \citenamefont {{Burke}}, \citenamefont {{Crittenden}},
  \citenamefont {{Cultrera}}, \citenamefont {{Dobbins}}, \citenamefont
  {{Douglas}}, \citenamefont {{Dunham}}, \citenamefont {{Eichhorn}},
  \citenamefont {{Full}}, \citenamefont {{Furuta}}, \citenamefont {{Franck}},
  \citenamefont {{Gallagher}}, \citenamefont {{Ge}}, \citenamefont
  {{Gulliford}}, \citenamefont {{Heltsley}}, \citenamefont {{Jusic}},
  \citenamefont {{Kaplan}}, \citenamefont {{Kostroun}}, \citenamefont {{Li}},
  \citenamefont {{Liepe}}, \citenamefont {{Liu}}, \citenamefont {{Lou}},
  \citenamefont {{Mahler}}, \citenamefont {{Meot}}, \citenamefont {{Michnoff}},
  \citenamefont {{Minty}}, \citenamefont {{Patterson}}, \citenamefont
  {{Peggs}}, \citenamefont {{Ptitsyn}}, \citenamefont {{Quigley}},
  \citenamefont {{Roser}}, \citenamefont {{Sabol}}, \citenamefont {{Sagan}},
  \citenamefont {{Sears}}, \citenamefont {{Shore}}, \citenamefont {{Smith}},
  \citenamefont {{Smolenski}}, \citenamefont {{Thieberger}}, \citenamefont
  {{Trabocchi}}, \citenamefont {{Tuozzolo}}, \citenamefont {{Tsoupas}},
  \citenamefont {{Veshcherevich}}, \citenamefont {{Widger}}, \citenamefont
  {{Wang}}, \citenamefont {{Willeke}},\ and\ \citenamefont
  {{Xu}}}]{ref:CBETAdr}%
  \BibitemOpen
  \bibfield  {author} {\bibinfo {author} {\bibfnamefont {G.~H.}\ \bibnamefont
  {{Hoffstaetter}}}, \bibinfo {author} {\bibfnamefont {D.}~\bibnamefont
  {{Trbojevic}}}, \bibinfo {author} {\bibfnamefont {C.}~\bibnamefont
  {{Mayes}}}, \bibinfo {author} {\bibfnamefont {N.}~\bibnamefont {{Banerjee}}},
  \bibinfo {author} {\bibfnamefont {J.}~\bibnamefont {{Barley}}}, \bibinfo
  {author} {\bibfnamefont {I.}~\bibnamefont {{Bazarov}}}, \bibinfo {author}
  {\bibfnamefont {A.}~\bibnamefont {{Bartnik}}}, \bibinfo {author}
  {\bibfnamefont {J.~S.}\ \bibnamefont {{Berg}}}, \bibinfo {author}
  {\bibfnamefont {S.}~\bibnamefont {{Brooks}}}, \bibinfo {author}
  {\bibfnamefont {D.}~\bibnamefont {{Burke}}}, \bibinfo {author} {\bibfnamefont
  {J.}~\bibnamefont {{Crittenden}}}, \bibinfo {author} {\bibfnamefont
  {L.}~\bibnamefont {{Cultrera}}}, \bibinfo {author} {\bibfnamefont
  {J.}~\bibnamefont {{Dobbins}}}, \bibinfo {author} {\bibfnamefont
  {D.}~\bibnamefont {{Douglas}}}, \bibinfo {author} {\bibfnamefont
  {B.}~\bibnamefont {{Dunham}}}, \bibinfo {author} {\bibfnamefont
  {R.}~\bibnamefont {{Eichhorn}}}, \bibinfo {author} {\bibfnamefont
  {S.}~\bibnamefont {{Full}}}, \bibinfo {author} {\bibfnamefont
  {F.}~\bibnamefont {{Furuta}}}, \bibinfo {author} {\bibfnamefont
  {C.}~\bibnamefont {{Franck}}}, \bibinfo {author} {\bibfnamefont
  {R.}~\bibnamefont {{Gallagher}}}, \bibinfo {author} {\bibfnamefont
  {M.}~\bibnamefont {{Ge}}}, \bibinfo {author} {\bibfnamefont {C.}~\bibnamefont
  {{Gulliford}}}, \bibinfo {author} {\bibfnamefont {B.}~\bibnamefont
  {{Heltsley}}}, \bibinfo {author} {\bibfnamefont {D.}~\bibnamefont {{Jusic}}},
  \bibinfo {author} {\bibfnamefont {R.}~\bibnamefont {{Kaplan}}}, \bibinfo
  {author} {\bibfnamefont {V.}~\bibnamefont {{Kostroun}}}, \bibinfo {author}
  {\bibfnamefont {Y.}~\bibnamefont {{Li}}}, \bibinfo {author} {\bibfnamefont
  {M.}~\bibnamefont {{Liepe}}}, \bibinfo {author} {\bibfnamefont
  {C.}~\bibnamefont {{Liu}}}, \bibinfo {author} {\bibfnamefont
  {W.}~\bibnamefont {{Lou}}}, \bibinfo {author} {\bibfnamefont
  {G.}~\bibnamefont {{Mahler}}}, \bibinfo {author} {\bibfnamefont
  {F.}~\bibnamefont {{Meot}}}, \bibinfo {author} {\bibfnamefont
  {R.}~\bibnamefont {{Michnoff}}}, \bibinfo {author} {\bibfnamefont
  {M.}~\bibnamefont {{Minty}}}, \bibinfo {author} {\bibfnamefont
  {R.}~\bibnamefont {{Patterson}}}, \bibinfo {author} {\bibfnamefont
  {S.}~\bibnamefont {{Peggs}}}, \bibinfo {author} {\bibfnamefont
  {V.}~\bibnamefont {{Ptitsyn}}}, \bibinfo {author} {\bibfnamefont
  {P.}~\bibnamefont {{Quigley}}}, \bibinfo {author} {\bibfnamefont
  {T.}~\bibnamefont {{Roser}}}, \bibinfo {author} {\bibfnamefont
  {D.}~\bibnamefont {{Sabol}}}, \bibinfo {author} {\bibfnamefont
  {D.}~\bibnamefont {{Sagan}}}, \bibinfo {author} {\bibfnamefont
  {J.}~\bibnamefont {{Sears}}}, \bibinfo {author} {\bibfnamefont
  {C.}~\bibnamefont {{Shore}}}, \bibinfo {author} {\bibfnamefont
  {E.}~\bibnamefont {{Smith}}}, \bibinfo {author} {\bibfnamefont
  {K.}~\bibnamefont {{Smolenski}}}, \bibinfo {author} {\bibfnamefont
  {P.}~\bibnamefont {{Thieberger}}}, \bibinfo {author} {\bibfnamefont
  {S.}~\bibnamefont {{Trabocchi}}}, \bibinfo {author} {\bibfnamefont
  {J.}~\bibnamefont {{Tuozzolo}}}, \bibinfo {author} {\bibfnamefont
  {N.}~\bibnamefont {{Tsoupas}}}, \bibinfo {author} {\bibfnamefont
  {V.}~\bibnamefont {{Veshcherevich}}}, \bibinfo {author} {\bibfnamefont
  {D.}~\bibnamefont {{Widger}}}, \bibinfo {author} {\bibfnamefont
  {G.}~\bibnamefont {{Wang}}}, \bibinfo {author} {\bibfnamefont
  {F.}~\bibnamefont {{Willeke}}}, \ and\ \bibinfo {author} {\bibfnamefont
  {W.}~\bibnamefont {{Xu}}},\ }\href@noop {} {\bibfield  {journal} {\bibinfo
  {journal} {ArXiv e-prints}\ } (\bibinfo {year} {2017})},\ \Eprint
  {http://arxiv.org/abs/1706.04245} {arXiv:1706.04245 [physics.acc-ph]}
  \BibitemShut {NoStop}%
\bibitem [{\citenamefont {Symon}\ \emph {et~al.}(1956)\citenamefont {Symon},
  \citenamefont {Kerst}, \citenamefont {Jones}, \citenamefont {Laslett},\ and\
  \citenamefont {Terwilliger}}]{ref:ffag1}%
  \BibitemOpen
  \bibfield  {author} {\bibinfo {author} {\bibfnamefont {K.~R.}\ \bibnamefont
  {Symon}}, \bibinfo {author} {\bibfnamefont {D.~W.}\ \bibnamefont {Kerst}},
  \bibinfo {author} {\bibfnamefont {L.~W.}\ \bibnamefont {Jones}}, \bibinfo
  {author} {\bibfnamefont {L.~J.}\ \bibnamefont {Laslett}}, \ and\ \bibinfo
  {author} {\bibfnamefont {K.~M.}\ \bibnamefont {Terwilliger}},\ }\href
  {\doibase 10.1103/PhysRev.103.1837} {\bibfield  {journal} {\bibinfo
  {journal} {Phys. Rev.}\ }\textbf {\bibinfo {volume} {103}},\ \bibinfo {pages}
  {1837} (\bibinfo {year} {1956})}\BibitemShut {NoStop}%
\bibitem [{\citenamefont {Ohkawa}(1958)}]{ref:ffag2}%
  \BibitemOpen
  \bibfield  {author} {\bibinfo {author} {\bibfnamefont {T.}~\bibnamefont
  {Ohkawa}},\ }\href {\doibase 10.1063/1.1716114} {\bibfield  {journal}
  {\bibinfo  {journal} {Review of Scientific Instruments}\ }\textbf {\bibinfo
  {volume} {29}},\ \bibinfo {pages} {108} (\bibinfo {year} {1958})},\ \Eprint
  {http://arxiv.org/abs/https://doi.org/10.1063/1.1716114}
  {https://doi.org/10.1063/1.1716114} \BibitemShut {NoStop}%
\bibitem [{\citenamefont {Dunham}\ \emph {et~al.}(2013)\citenamefont {Dunham},
  \citenamefont {Barley}, \citenamefont {Bartnik}, \citenamefont {Bazarov},
  \citenamefont {Cultrera}, \citenamefont {Dobbins}, \citenamefont
  {Hoffstaetter}, \citenamefont {Johnson}, \citenamefont {Kaplan},
  \citenamefont {Karkare}, \citenamefont {Kostroun}, \citenamefont {Li},
  \citenamefont {Liepe}, \citenamefont {Liu}, \citenamefont {Loehl},
  \citenamefont {Maxson}, \citenamefont {Quigley}, \citenamefont {Reilly},
  \citenamefont {Rice}, \citenamefont {Sabol}, \citenamefont {Smith},
  \citenamefont {Smolenski}, \citenamefont {Tigner}, \citenamefont
  {Vesherevich}, \citenamefont {Widger},\ and\ \citenamefont
  {Zhao}}]{ref:hcrecord}%
  \BibitemOpen
  \bibfield  {author} {\bibinfo {author} {\bibfnamefont {B.}~\bibnamefont
  {Dunham}}, \bibinfo {author} {\bibfnamefont {J.}~\bibnamefont {Barley}},
  \bibinfo {author} {\bibfnamefont {A.}~\bibnamefont {Bartnik}}, \bibinfo
  {author} {\bibfnamefont {I.}~\bibnamefont {Bazarov}}, \bibinfo {author}
  {\bibfnamefont {L.}~\bibnamefont {Cultrera}}, \bibinfo {author}
  {\bibfnamefont {J.}~\bibnamefont {Dobbins}}, \bibinfo {author} {\bibfnamefont
  {G.}~\bibnamefont {Hoffstaetter}}, \bibinfo {author} {\bibfnamefont
  {B.}~\bibnamefont {Johnson}}, \bibinfo {author} {\bibfnamefont
  {R.}~\bibnamefont {Kaplan}}, \bibinfo {author} {\bibfnamefont
  {S.}~\bibnamefont {Karkare}}, \bibinfo {author} {\bibfnamefont
  {V.}~\bibnamefont {Kostroun}}, \bibinfo {author} {\bibfnamefont
  {Y.}~\bibnamefont {Li}}, \bibinfo {author} {\bibfnamefont {M.}~\bibnamefont
  {Liepe}}, \bibinfo {author} {\bibfnamefont {X.}~\bibnamefont {Liu}}, \bibinfo
  {author} {\bibfnamefont {F.}~\bibnamefont {Loehl}}, \bibinfo {author}
  {\bibfnamefont {J.}~\bibnamefont {Maxson}}, \bibinfo {author} {\bibfnamefont
  {P.}~\bibnamefont {Quigley}}, \bibinfo {author} {\bibfnamefont
  {J.}~\bibnamefont {Reilly}}, \bibinfo {author} {\bibfnamefont
  {D.}~\bibnamefont {Rice}}, \bibinfo {author} {\bibfnamefont {D.}~\bibnamefont
  {Sabol}}, \bibinfo {author} {\bibfnamefont {E.}~\bibnamefont {Smith}},
  \bibinfo {author} {\bibfnamefont {K.}~\bibnamefont {Smolenski}}, \bibinfo
  {author} {\bibfnamefont {M.}~\bibnamefont {Tigner}}, \bibinfo {author}
  {\bibfnamefont {V.}~\bibnamefont {Vesherevich}}, \bibinfo {author}
  {\bibfnamefont {D.}~\bibnamefont {Widger}}, \ and\ \bibinfo {author}
  {\bibfnamefont {Z.}~\bibnamefont {Zhao}},\ }\href {\doibase
  10.1063/1.4789395} {\bibfield  {journal} {\bibinfo  {journal} {Applied
  Physics Letters}\ }\textbf {\bibinfo {volume} {102}},\ \bibinfo {eid}
  {034105} (\bibinfo {year} {2013})}\BibitemShut {NoStop}%
\bibitem [{\citenamefont {Gulliford}\ \emph {et~al.}(2013)\citenamefont
  {Gulliford}, \citenamefont {Bartnik}, \citenamefont {Bazarov}, \citenamefont
  {Cultrera}, \citenamefont {Dobbins}, \citenamefont {Dunham}, \citenamefont
  {Gonzalez}, \citenamefont {Karkare}, \citenamefont {Lee}, \citenamefont {Li},
  \citenamefont {Li}, \citenamefont {Liu}, \citenamefont {Maxson},
  \citenamefont {Nguyen}, \citenamefont {Smolenski},\ and\ \citenamefont
  {Zhao}}]{ref:lowemitter}%
  \BibitemOpen
  \bibfield  {author} {\bibinfo {author} {\bibfnamefont {C.}~\bibnamefont
  {Gulliford}}, \bibinfo {author} {\bibfnamefont {A.}~\bibnamefont {Bartnik}},
  \bibinfo {author} {\bibfnamefont {I.}~\bibnamefont {Bazarov}}, \bibinfo
  {author} {\bibfnamefont {L.}~\bibnamefont {Cultrera}}, \bibinfo {author}
  {\bibfnamefont {J.}~\bibnamefont {Dobbins}}, \bibinfo {author} {\bibfnamefont
  {B.}~\bibnamefont {Dunham}}, \bibinfo {author} {\bibfnamefont
  {F.}~\bibnamefont {Gonzalez}}, \bibinfo {author} {\bibfnamefont
  {S.}~\bibnamefont {Karkare}}, \bibinfo {author} {\bibfnamefont
  {H.}~\bibnamefont {Lee}}, \bibinfo {author} {\bibfnamefont {H.}~\bibnamefont
  {Li}}, \bibinfo {author} {\bibfnamefont {Y.}~\bibnamefont {Li}}, \bibinfo
  {author} {\bibfnamefont {X.}~\bibnamefont {Liu}}, \bibinfo {author}
  {\bibfnamefont {J.}~\bibnamefont {Maxson}}, \bibinfo {author} {\bibfnamefont
  {C.}~\bibnamefont {Nguyen}}, \bibinfo {author} {\bibfnamefont
  {K.}~\bibnamefont {Smolenski}}, \ and\ \bibinfo {author} {\bibfnamefont
  {Z.}~\bibnamefont {Zhao}},\ }\href {\doibase 10.1103/PhysRevSTAB.16.073401}
  {\bibfield  {journal} {\bibinfo  {journal} {Phys. Rev. ST Accel. Beams}\
  }\textbf {\bibinfo {volume} {16}},\ \bibinfo {pages} {073401} (\bibinfo
  {year} {2013})}\BibitemShut {NoStop}%
\bibitem [{\citenamefont {Gulliford}\ \emph
  {et~al.}(2015{\natexlab{a}})\citenamefont {Gulliford}, \citenamefont
  {Bartnik}, \citenamefont {Bazarov}, \citenamefont {Dunham},\ and\
  \citenamefont {Cultrera}}]{ref:lowemitter2}%
  \BibitemOpen
  \bibfield  {author} {\bibinfo {author} {\bibfnamefont {C.}~\bibnamefont
  {Gulliford}}, \bibinfo {author} {\bibfnamefont {A.}~\bibnamefont {Bartnik}},
  \bibinfo {author} {\bibfnamefont {I.}~\bibnamefont {Bazarov}}, \bibinfo
  {author} {\bibfnamefont {B.}~\bibnamefont {Dunham}}, \ and\ \bibinfo {author}
  {\bibfnamefont {L.}~\bibnamefont {Cultrera}},\ }\href {\doibase
  http://dx.doi.org/10.1063/1.4913678} {\bibfield  {journal} {\bibinfo
  {journal} {Applied Physics Letters}\ }\textbf {\bibinfo {volume} {106}},\
  \bibinfo {eid} {094101} (\bibinfo {year} {2015}{\natexlab{a}})}\BibitemShut
  {NoStop}%
\bibitem [{\citenamefont {Bartnik}\ \emph {et~al.}(2015)\citenamefont
  {Bartnik}, \citenamefont {Gulliford}, \citenamefont {Bazarov}, \citenamefont
  {Cultera},\ and\ \citenamefont {Dunham}}]{ref:lowemitter3}%
  \BibitemOpen
  \bibfield  {author} {\bibinfo {author} {\bibfnamefont {A.}~\bibnamefont
  {Bartnik}}, \bibinfo {author} {\bibfnamefont {C.}~\bibnamefont {Gulliford}},
  \bibinfo {author} {\bibfnamefont {I.}~\bibnamefont {Bazarov}}, \bibinfo
  {author} {\bibfnamefont {L.}~\bibnamefont {Cultera}}, \ and\ \bibinfo
  {author} {\bibfnamefont {B.}~\bibnamefont {Dunham}},\ }\href {\doibase
  10.1103/PhysRevSTAB.18.083401} {\bibfield  {journal} {\bibinfo  {journal}
  {Phys. Rev. ST Accel. Beams}\ }\textbf {\bibinfo {volume} {18}},\ \bibinfo
  {pages} {083401} (\bibinfo {year} {2015})}\BibitemShut {NoStop}%
\bibitem [{\citenamefont {Eichhorn}\ \emph {et~al.}(2015)\citenamefont
  {Eichhorn}, \citenamefont {Bullock}, \citenamefont {Elmore}, \citenamefont
  {Clasby}, \citenamefont {Furuta}, \citenamefont {He}, \citenamefont
  {Hoffstaetter}, \citenamefont {Liepe}, \citenamefont {O?Connell},
  \citenamefont {Conway}, \citenamefont {Quigley}, \citenamefont {Sabol},
  \citenamefont {Sears}, \citenamefont {Smith},\ and\ \citenamefont
  {Veshcherevich}}]{ref:mlcfab}%
  \BibitemOpen
  \bibfield  {author} {\bibinfo {author} {\bibfnamefont {R.}~\bibnamefont
  {Eichhorn}}, \bibinfo {author} {\bibfnamefont {B.}~\bibnamefont {Bullock}},
  \bibinfo {author} {\bibfnamefont {B.}~\bibnamefont {Elmore}}, \bibinfo
  {author} {\bibfnamefont {B.}~\bibnamefont {Clasby}}, \bibinfo {author}
  {\bibfnamefont {F.}~\bibnamefont {Furuta}}, \bibinfo {author} {\bibfnamefont
  {Y.}~\bibnamefont {He}}, \bibinfo {author} {\bibfnamefont {G.}~\bibnamefont
  {Hoffstaetter}}, \bibinfo {author} {\bibfnamefont {M.}~\bibnamefont {Liepe}},
  \bibinfo {author} {\bibfnamefont {T.}~\bibnamefont {O?Connell}}, \bibinfo
  {author} {\bibfnamefont {J.}~\bibnamefont {Conway}}, \bibinfo {author}
  {\bibfnamefont {P.}~\bibnamefont {Quigley}}, \bibinfo {author} {\bibfnamefont
  {D.}~\bibnamefont {Sabol}}, \bibinfo {author} {\bibfnamefont
  {J.}~\bibnamefont {Sears}}, \bibinfo {author} {\bibfnamefont
  {E.}~\bibnamefont {Smith}}, \ and\ \bibinfo {author} {\bibfnamefont
  {V.}~\bibnamefont {Veshcherevich}},\ }\href {\doibase
  https://doi.org/10.1016/j.phpro.2015.06.133} {\bibfield  {journal} {\bibinfo
  {journal} {Physics Procedia}\ }\textbf {\bibinfo {volume} {67}},\ \bibinfo
  {pages} {785 } (\bibinfo {year} {2015})},\ \bibinfo {note} {proceedings of
  the 25th International Cryogenic Engineering Conference and International
  Cryogenic Materials Conference 2014}\BibitemShut {NoStop}%
\bibitem [{\citenamefont {Furuta}\ \emph {et~al.}(2016)\citenamefont {Furuta},
  \citenamefont {Banerjee}, \citenamefont {Dobbins}, \citenamefont {Eichhorn},
  \citenamefont {Hoffstaetter}, \citenamefont {Liepe}, \citenamefont
  {O?Connell}, \citenamefont {Quigley}, \citenamefont {Sabol}, \citenamefont
  {Sears}, \citenamefont {Smith},\ and\ \citenamefont
  {Veshcherevich}}]{ref:mlcperf}%
  \BibitemOpen
  \bibfield  {author} {\bibinfo {author} {\bibfnamefont {F.}~\bibnamefont
  {Furuta}}, \bibinfo {author} {\bibfnamefont {N.}~\bibnamefont {Banerjee}},
  \bibinfo {author} {\bibfnamefont {J.}~\bibnamefont {Dobbins}}, \bibinfo
  {author} {\bibfnamefont {R.}~\bibnamefont {Eichhorn}}, \bibinfo {author}
  {\bibfnamefont {G.}~\bibnamefont {Hoffstaetter}}, \bibinfo {author}
  {\bibfnamefont {M.}~\bibnamefont {Liepe}}, \bibinfo {author} {\bibfnamefont
  {T.}~\bibnamefont {O?Connell}}, \bibinfo {author} {\bibfnamefont
  {P.}~\bibnamefont {Quigley}}, \bibinfo {author} {\bibfnamefont
  {D.}~\bibnamefont {Sabol}}, \bibinfo {author} {\bibfnamefont
  {J.}~\bibnamefont {Sears}}, \bibinfo {author} {\bibfnamefont
  {E.}~\bibnamefont {Smith}}, \ and\ \bibinfo {author} {\bibfnamefont
  {V.}~\bibnamefont {Veshcherevich}},\ }\href@noop {} {\ ,\ \bibinfo {pages}
  {204 } (\bibinfo {year} {2016})},\ \bibinfo {note} {proceedings of the North
  American Particle Accelerator Conference}\BibitemShut {NoStop}%
\bibitem [{\citenamefont {Brooks}(2016)}]{ref:magdes}%
  \BibitemOpen
  \bibfield  {author} {\bibinfo {author} {\bibfnamefont {S.}~\bibnamefont
  {Brooks}},\ }\href@noop {} {\emph {\bibinfo {title} {Magnet and Lattice
  Specifications for the CBETA First Girder}}},\ \bibinfo {type} {Tech. Rep.}\
  \bibinfo {number} {CBETA-001}\ (\bibinfo  {institution} {Brookhaven National
  Lab},\ \bibinfo {year} {2016})\ \bibinfo {note}
  {\url{https://www.classe.cornell.edu/CBETA_PM/notes/CBETA001.pdf}}\BibitemShut
  {NoStop}%
\bibitem [{\citenamefont {Brooks}(2017)}]{ref:1stprod}%
  \BibitemOpen
  \bibfield  {author} {\bibinfo {author} {\bibfnamefont {S.}~\bibnamefont
  {Brooks}},\ }\href
  {https://public.bnl.gov/docs/cad/Documents/Measurement%20of%20first%20magnet%20made%20with%20production%20material.pdf}
  {\emph {\bibinfo {title} {Measurement of First Magnet made with Production
  Material}}},\ \bibinfo {type} {Tech. Rep.}\ \bibinfo {number} {CBETA-023}\
  (\bibinfo  {institution} {Brookhaven National Lab},\ \bibinfo {year} {2017})\
  \bibinfo {note}
  {\url{https://www.classe.cornell.edu/CBETA_PM/notes/CBETA023.pdf}}\BibitemShut
  {NoStop}%
\bibitem [{\citenamefont {Witte}\ \emph
  {et~al.}(2017{\natexlab{a}})\citenamefont {Witte}, \citenamefont {Berg},
  \citenamefont {Cintorino}, \citenamefont {Mahler}, \citenamefont {Tsoupas},\
  and\ \citenamefont {Wanderer}}]{ref:halbach2}%
  \BibitemOpen
  \bibfield  {author} {\bibinfo {author} {\bibfnamefont {H.}~\bibnamefont
  {Witte}}, \bibinfo {author} {\bibfnamefont {J.}~\bibnamefont {Berg}},
  \bibinfo {author} {\bibfnamefont {J.}~\bibnamefont {Cintorino}}, \bibinfo
  {author} {\bibfnamefont {G.}~\bibnamefont {Mahler}}, \bibinfo {author}
  {\bibfnamefont {N.}~\bibnamefont {Tsoupas}}, \ and\ \bibinfo {author}
  {\bibfnamefont {P.}~\bibnamefont {Wanderer}},\ }in\ \href {\doibase
  https://doi.org/10.18429/JACoW-IPAC2017-TUPIK130} {{\selectlanguage
  {english}\emph {\bibinfo {booktitle} {Proc. of International Particle
  Accelerator Conference (IPAC'17), Copenhagen, Denmark, 14–19 May,
  2017}}}},\ \bibinfo {series and number} {\bibinfo {series} {International
  Particle Accelerator Conference}\ No.~\bibinfo {number} {8}}\ (\bibinfo
  {publisher} {JACoW},\ \bibinfo {address} {Geneva, Switzerland},\ \bibinfo
  {year} {2017})\ pp.\ \bibinfo {pages} {2016--2018},\ \bibinfo {note}
  {https://doi.org/10.18429/JACoW-IPAC2017-TUPIK130}\BibitemShut {NoStop}%
\bibitem [{\citenamefont {Witte}\ \emph
  {et~al.}(2017{\natexlab{b}})\citenamefont {Witte}, \citenamefont {Berg},\
  and\ \citenamefont {Parker}}]{ref:halbach3}%
  \BibitemOpen
  \bibfield  {author} {\bibinfo {author} {\bibfnamefont {H.}~\bibnamefont
  {Witte}}, \bibinfo {author} {\bibfnamefont {J.}~\bibnamefont {Berg}}, \ and\
  \bibinfo {author} {\bibfnamefont {B.}~\bibnamefont {Parker}},\ }in\ \href
  {\doibase https://doi.org/10.18429/JACoW-IPAC2017-THPVA151} {{\selectlanguage
  {english}\emph {\bibinfo {booktitle} {Proc. of International Particle
  Accelerator Conference (IPAC'17), Copenhagen, Denmark, 14–19 May,
  2017}}}},\ \bibinfo {series and number} {\bibinfo {series} {International
  Particle Accelerator Conference}\ No.~\bibinfo {number} {8}}\ (\bibinfo
  {publisher} {JACoW},\ \bibinfo {address} {Geneva, Switzerland},\ \bibinfo
  {year} {2017})\ pp.\ \bibinfo {pages} {4814--4816},\ \bibinfo {note}
  {https://doi.org/10.18429/JACoW-IPAC2017-THPVA151}\BibitemShut {NoStop}%
\bibitem [{\citenamefont {Brooks}\ \emph {et~al.}(2017)\citenamefont {Brooks},
  \citenamefont {Cintorino}, \citenamefont {Jain},\ and\ \citenamefont
  {Mahler}}]{ref:halbach4}%
  \BibitemOpen
  \bibfield  {author} {\bibinfo {author} {\bibfnamefont {S.}~\bibnamefont
  {Brooks}}, \bibinfo {author} {\bibfnamefont {J.}~\bibnamefont {Cintorino}},
  \bibinfo {author} {\bibfnamefont {A.}~\bibnamefont {Jain}}, \ and\ \bibinfo
  {author} {\bibfnamefont {G.}~\bibnamefont {Mahler}},\ }in\ \href {\doibase
  https://doi.org/10.18429/JACoW-IPAC2017-THPIK007} {{\selectlanguage
  {english}\emph {\bibinfo {booktitle} {Proc. of International Particle
  Accelerator Conference (IPAC'17), Copenhagen, Denmark, 14–19 May,
  2017}}}},\ \bibinfo {series and number} {\bibinfo {series} {International
  Particle Accelerator Conference}\ No.~\bibinfo {number} {8}}\ (\bibinfo
  {publisher} {JACoW},\ \bibinfo {address} {Geneva, Switzerland},\ \bibinfo
  {year} {2017})\ pp.\ \bibinfo {pages} {4118--4120},\ \bibinfo {note}
  {https://doi.org/10.18429/JACoW-IPAC2017-THPIK007}\BibitemShut {NoStop}%
\bibitem [{\citenamefont {Lou}\ \emph {et~al.}(2018)\citenamefont {Lou},
  \citenamefont {Bartnik}, \citenamefont {Berg}, \citenamefont {Brooks},
  \citenamefont {Crittenden}, \citenamefont {Gulliford}, \citenamefont
  {Hoffstaetter}, \citenamefont {Meot}, \citenamefont {Sagan}, \citenamefont
  {Trbojevic},\ and\ \citenamefont {Tsoupas}}]{ref:ipac18}%
  \BibitemOpen
  \bibfield  {author} {\bibinfo {author} {\bibfnamefont {W.}~\bibnamefont
  {Lou}}, \bibinfo {author} {\bibfnamefont {A.}~\bibnamefont {Bartnik}},
  \bibinfo {author} {\bibfnamefont {J.~S.}\ \bibnamefont {Berg}}, \bibinfo
  {author} {\bibfnamefont {S.}~\bibnamefont {Brooks}}, \bibinfo {author}
  {\bibfnamefont {J.~A.}\ \bibnamefont {Crittenden}}, \bibinfo {author}
  {\bibfnamefont {C.}~\bibnamefont {Gulliford}}, \bibinfo {author}
  {\bibfnamefont {G.~H.}\ \bibnamefont {Hoffstaetter}}, \bibinfo {author}
  {\bibfnamefont {F.}~\bibnamefont {Meot}}, \bibinfo {author} {\bibfnamefont
  {D.}~\bibnamefont {Sagan}}, \bibinfo {author} {\bibfnamefont
  {D.}~\bibnamefont {Trbojevic}}, \ and\ \bibinfo {author} {\bibfnamefont
  {N.}~\bibnamefont {Tsoupas}},\ }in\ \href {\doibase
  doi:10.18429/JACoW-IPAC2018-THPAF023} {{\selectlanguage {english}\emph
  {\bibinfo {booktitle} {Proc. 9th International Particle Accelerator
  Conference (IPAC'18), Vancouver, BC, Canada, April 29-May 4, 2018}}}},\
  \bibinfo {series and number} {\bibinfo {series} {International Particle
  Accelerator Conference}\ No.~\bibinfo {number} {9}}\ (\bibinfo  {publisher}
  {JACoW Publishing},\ \bibinfo {address} {Geneva, Switzerland},\ \bibinfo
  {year} {2018})\ pp.\ \bibinfo {pages} {3000--3003},\ \bibinfo {note}
  {https://doi.org/10.18429/JACoW-IPAC2018-THPAF023}\BibitemShut {NoStop}%
\bibitem [{\citenamefont {Méot}\ \emph {et~al.}(2018)\citenamefont {Méot},
  \citenamefont {Brooks}, \citenamefont {Trbojevic},\ and\ \citenamefont
  {Tsoupas}}]{ref:ffacheck}%
  \BibitemOpen
  \bibfield  {author} {\bibinfo {author} {\bibfnamefont {F.}~\bibnamefont
  {Méot}}, \bibinfo {author} {\bibfnamefont {S.}~\bibnamefont {Brooks}},
  \bibinfo {author} {\bibfnamefont {D.}~\bibnamefont {Trbojevic}}, \ and\
  \bibinfo {author} {\bibfnamefont {N.}~\bibnamefont {Tsoupas}},\ }in\ \href
  {\doibase doi:10.18429/JACoW-IPAC2018-TUPMF024} {{\selectlanguage
  {english}\emph {\bibinfo {booktitle} {Proc. 9th International Particle
  Accelerator Conference (IPAC'18), Vancouver, BC, Canada, April 29-May 4,
  2018}}}},\ \bibinfo {series and number} {\bibinfo {series} {International
  Particle Accelerator Conference}\ No.~\bibinfo {number} {9}}\ (\bibinfo
  {publisher} {JACoW Publishing},\ \bibinfo {address} {Geneva, Switzerland},\
  \bibinfo {year} {2018})\ pp.\ \bibinfo {pages} {1304--1305},\ \bibinfo {note}
  {https://doi.org/10.18429/JACoW-IPAC2018-TUPMF024}\BibitemShut {NoStop}%
\bibitem [{\citenamefont {Full}\ \emph {et~al.}(2016)\citenamefont {Full},
  \citenamefont {Bartnik}, \citenamefont {Bazarov}, \citenamefont {Dobbins},
  \citenamefont {Dunham},\ and\ \citenamefont {Hoffstaetter}}]{ref:FullIons}%
  \BibitemOpen
  \bibfield  {author} {\bibinfo {author} {\bibfnamefont {S.}~\bibnamefont
  {Full}}, \bibinfo {author} {\bibfnamefont {A.}~\bibnamefont {Bartnik}},
  \bibinfo {author} {\bibfnamefont {I.~V.}\ \bibnamefont {Bazarov}}, \bibinfo
  {author} {\bibfnamefont {J.}~\bibnamefont {Dobbins}}, \bibinfo {author}
  {\bibfnamefont {B.}~\bibnamefont {Dunham}}, \ and\ \bibinfo {author}
  {\bibfnamefont {G.~H.}\ \bibnamefont {Hoffstaetter}},\ }\href {\doibase
  10.1103/PhysRevAccelBeams.19.034201} {\bibfield  {journal} {\bibinfo
  {journal} {Phys. Rev. Accel. Beams}\ }\textbf {\bibinfo {volume} {19}},\
  \bibinfo {pages} {034201} (\bibinfo {year} {2016})}\BibitemShut {NoStop}%
\bibitem [{\citenamefont {Maxson}\ \emph
  {et~al.}(2014{\natexlab{a}})\citenamefont {Maxson}, \citenamefont {Bazarov},
  \citenamefont {Dunham}, \citenamefont {Dobbins}, \citenamefont {Liu},\ and\
  \citenamefont {Smolenski}}]{ref:jaredRSI}%
  \BibitemOpen
  \bibfield  {author} {\bibinfo {author} {\bibfnamefont {J.}~\bibnamefont
  {Maxson}}, \bibinfo {author} {\bibfnamefont {I.}~\bibnamefont {Bazarov}},
  \bibinfo {author} {\bibfnamefont {B.}~\bibnamefont {Dunham}}, \bibinfo
  {author} {\bibfnamefont {J.}~\bibnamefont {Dobbins}}, \bibinfo {author}
  {\bibfnamefont {X.}~\bibnamefont {Liu}}, \ and\ \bibinfo {author}
  {\bibfnamefont {K.}~\bibnamefont {Smolenski}},\ }\href {\doibase
  http://dx.doi.org/10.1063/1.4895641} {\bibfield  {journal} {\bibinfo
  {journal} {Review of Scientific Instruments}\ }\textbf {\bibinfo {volume}
  {85}},\ \bibinfo {eid} {093306} (\bibinfo {year}
  {2014}{\natexlab{a}})}\BibitemShut {NoStop}%
\bibitem [{\citenamefont {Maxson}\ \emph {et~al.}(2015)\citenamefont {Maxson},
  \citenamefont {Cultrera}, \citenamefont {Gulliford},\ and\ \citenamefont
  {Bazarov}}]{ref:redmte}%
  \BibitemOpen
  \bibfield  {author} {\bibinfo {author} {\bibfnamefont {J.}~\bibnamefont
  {Maxson}}, \bibinfo {author} {\bibfnamefont {L.}~\bibnamefont {Cultrera}},
  \bibinfo {author} {\bibfnamefont {C.}~\bibnamefont {Gulliford}}, \ and\
  \bibinfo {author} {\bibfnamefont {I.}~\bibnamefont {Bazarov}},\ }\href
  {\doibase http://dx.doi.org/10.1063/1.4922146} {\bibfield  {journal}
  {\bibinfo  {journal} {Applied Physics Letters}\ }\textbf {\bibinfo {volume}
  {106}},\ \bibinfo {eid} {234102} (\bibinfo {year} {2015})}\BibitemShut
  {NoStop}%
\bibitem [{\citenamefont {Maxson}\ \emph
  {et~al.}(2014{\natexlab{b}})\citenamefont {Maxson}, \citenamefont {Lee},
  \citenamefont {Bartnik}, \citenamefont {Kiefer},\ and\ \citenamefont
  {Bazarov}}]{ref:lasershaping2}%
  \BibitemOpen
  \bibfield  {author} {\bibinfo {author} {\bibfnamefont {J.}~\bibnamefont
  {Maxson}}, \bibinfo {author} {\bibfnamefont {H.}~\bibnamefont {Lee}},
  \bibinfo {author} {\bibfnamefont {A.}~\bibnamefont {Bartnik}}, \bibinfo
  {author} {\bibfnamefont {J.}~\bibnamefont {Kiefer}}, \ and\ \bibinfo {author}
  {\bibfnamefont {I.}~\bibnamefont {Bazarov}},\ }\href@noop {} {\bibfield
  {journal} {\bibinfo  {journal} {submitted to Phys. Rev. ST Accel. Beams}\ }
  (\bibinfo {year} {2014}{\natexlab{b}})}\BibitemShut {NoStop}%
\bibitem [{\citenamefont {Zhao}\ \emph {et~al.}(2014)\citenamefont {Zhao},
  \citenamefont {Bartnik}, \citenamefont {Wise}, \citenamefont {Bazarov},\ and\
  \citenamefont {Dunham}}]{ref:GHzMHzLasers}%
  \BibitemOpen
  \bibfield  {author} {\bibinfo {author} {\bibfnamefont {Z.}~\bibnamefont
  {Zhao}}, \bibinfo {author} {\bibfnamefont {A.}~\bibnamefont {Bartnik}},
  \bibinfo {author} {\bibfnamefont {F.~W.}\ \bibnamefont {Wise}}, \bibinfo
  {author} {\bibfnamefont {I.~V.}\ \bibnamefont {Bazarov}}, \ and\ \bibinfo
  {author} {\bibfnamefont {B.~M.}\ \bibnamefont {Dunham}},\ }\href {\doibase
  10.1103/PhysRevSTAB.17.053501} {\bibfield  {journal} {\bibinfo  {journal}
  {Phys. Rev. ST Accel. Beams}\ }\textbf {\bibinfo {volume} {17}},\ \bibinfo
  {pages} {053501} (\bibinfo {year} {2014})}\BibitemShut {NoStop}%
\bibitem [{\citenamefont {Bazarov}\ \emph {et~al.}(2008)\citenamefont
  {Bazarov}, \citenamefont {Dunham}, \citenamefont {Gulliford}, \citenamefont
  {Li}, \citenamefont {Liu}, \citenamefont {Sinclair}, \citenamefont {Soong},\
  and\ \citenamefont {Hannon}}]{ref:bmsc}%
  \BibitemOpen
  \bibfield  {author} {\bibinfo {author} {\bibfnamefont {I.~V.}\ \bibnamefont
  {Bazarov}}, \bibinfo {author} {\bibfnamefont {B.~M.}\ \bibnamefont {Dunham}},
  \bibinfo {author} {\bibfnamefont {C.}~\bibnamefont {Gulliford}}, \bibinfo
  {author} {\bibfnamefont {Y.}~\bibnamefont {Li}}, \bibinfo {author}
  {\bibfnamefont {X.}~\bibnamefont {Liu}}, \bibinfo {author} {\bibfnamefont
  {C.~K.}\ \bibnamefont {Sinclair}}, \bibinfo {author} {\bibfnamefont
  {K.}~\bibnamefont {Soong}}, \ and\ \bibinfo {author} {\bibfnamefont
  {F.}~\bibnamefont {Hannon}},\ }\href {\doibase 10.1103/PhysRevSTAB.11.100703}
  {\bibfield  {journal} {\bibinfo  {journal} {Phys. Rev. ST Accel. Beams}\
  }\textbf {\bibinfo {volume} {11}},\ \bibinfo {pages} {100703} (\bibinfo
  {year} {2008})}\BibitemShut {NoStop}%
\bibitem [{\citenamefont {Li}(2012)}]{ref:heng}%
  \BibitemOpen
  \bibfield  {author} {\bibinfo {author} {\bibfnamefont {H.}~\bibnamefont
  {Li}},\ }\emph {\bibinfo {title} {Mutli-dimensional Characterization of the
  Laser and Electron Beams of the Cornell Energy Recovery Linac Photoinjector
  Prototype}},\ \href@noop {} {Ph.D. thesis},\ \bibinfo  {school} {Cornell
  University} (\bibinfo {year} {2012})\BibitemShut {NoStop}%
\bibitem [{\citenamefont {Belomestnykh}\ \emph {et~al.}(2010)\citenamefont
  {Belomestnykh}, \citenamefont {Bazarov}, \citenamefont {Shemelin},
  \citenamefont {Sikora}, \citenamefont {Smolenski},\ and\ \citenamefont
  {Veshcherevich}}]{ref:defcav}%
  \BibitemOpen
  \bibfield  {author} {\bibinfo {author} {\bibfnamefont {S.}~\bibnamefont
  {Belomestnykh}}, \bibinfo {author} {\bibfnamefont {I.}~\bibnamefont
  {Bazarov}}, \bibinfo {author} {\bibfnamefont {V.}~\bibnamefont {Shemelin}},
  \bibinfo {author} {\bibfnamefont {J.}~\bibnamefont {Sikora}}, \bibinfo
  {author} {\bibfnamefont {K.}~\bibnamefont {Smolenski}}, \ and\ \bibinfo
  {author} {\bibfnamefont {V.}~\bibnamefont {Veshcherevich}},\ }\href {\doibase
  10.1016/j.nima.2009.12.063} {\bibfield  {journal} {\bibinfo  {journal}
  {Nuclear Instruments and Methods in Physics Research Section A: Accelerators,
  Spectrometers, Detectors and Associated Equipment}\ }\textbf {\bibinfo
  {volume} {614}},\ \bibinfo {pages} {179 } (\bibinfo {year}
  {2010})}\BibitemShut {NoStop}%
\bibitem [{ref(2019)}]{ref:Elytt}%
  \BibitemOpen
  \href {http://www.elytt.com/} {}\bibinfo {howpublished} {Elytt Energy, Orense
  11, 28020 Madrid, Spain} (\bibinfo {year} {2019})\BibitemShut {NoStop}%
\bibitem [{\citenamefont {Halbach}(1980)}]{ref:halbach}%
  \BibitemOpen
  \bibfield  {author} {\bibinfo {author} {\bibfnamefont {K.}~\bibnamefont
  {Halbach}},\ }\href {\doibase https://doi.org/10.1016/0029-554X(80)90094-4}
  {\bibfield  {journal} {\bibinfo  {journal} {Nuclear Instruments and Methods}\
  }\textbf {\bibinfo {volume} {169}},\ \bibinfo {pages} {1 } (\bibinfo {year}
  {1980})}\BibitemShut {NoStop}%
\bibitem [{\citenamefont {Urban}\ \emph {et~al.}(2005)\citenamefont {Urban},
  \citenamefont {Fields},\ and\ \citenamefont {Sagan}}]{ref:BMAD1}%
  \BibitemOpen
  \bibfield  {author} {\bibinfo {author} {\bibfnamefont {J.}~\bibnamefont
  {Urban}}, \bibinfo {author} {\bibfnamefont {L.}~\bibnamefont {Fields}}, \
  and\ \bibinfo {author} {\bibfnamefont {D.}~\bibnamefont {Sagan}},\ }in\ \href
  {\doibase 10.1109/PAC.2005.1590964} {\emph {\bibinfo {booktitle} {Proceedings
  of the 2005 Particle Accelerator Conference}}}\ (\bibinfo {year} {2005})\
  pp.\ \bibinfo {pages} {1937--1939}\BibitemShut {NoStop}%
\bibitem [{\citenamefont {Sagan}\ and\ \citenamefont {Smith}(2005)}]{ref:TAO1}%
  \BibitemOpen
  \bibfield  {author} {\bibinfo {author} {\bibfnamefont {D.}~\bibnamefont
  {Sagan}}\ and\ \bibinfo {author} {\bibfnamefont {J.~C.}\ \bibnamefont
  {Smith}},\ }in\ \href {\doibase 10.1109/PAC.2005.1591750} {\emph {\bibinfo
  {booktitle} {Proceedings of the 2005 Particle Accelerator Conference}}}\
  (\bibinfo {year} {2005})\ pp.\ \bibinfo {pages} {4159--4161}\BibitemShut
  {NoStop}%
\bibitem [{ref(2011)}]{ref:gpt1}%
  \BibitemOpen
  \href {http://www.pulsar.nl/gpt/} {\enquote {\bibinfo {title} {Pulsar website
  for gpt},}\ }\bibinfo {howpublished} {\url{http://www.pulsar.nl/gpt/}}
  (\bibinfo {year} {2011})\BibitemShut {NoStop}%
\bibitem [{\citenamefont {van~der Geer}\ \emph {et~al.}(2005)\citenamefont
  {van~der Geer}, \citenamefont {O.J.~Luiten}, \citenamefont {P\"{o}plau},\
  and\ \citenamefont {van Rienen}}]{ref:gpt2}%
  \BibitemOpen
  \bibfield  {author} {\bibinfo {author} {\bibfnamefont {S.}~\bibnamefont
  {van~der Geer}}, \bibinfo {author} {\bibfnamefont {M.~d.~L.}\ \bibnamefont
  {O.J.~Luiten}}, \bibinfo {author} {\bibfnamefont {G.}~\bibnamefont
  {P\"{o}plau}}, \ and\ \bibinfo {author} {\bibfnamefont {U.}~\bibnamefont {van
  Rienen}}\ }(\bibinfo {year} {2005})\ p.\ \bibinfo {pages} {101}\BibitemShut
  {NoStop}%
\bibitem [{\citenamefont {Gulliford}\ \emph {et~al.}(2018)\citenamefont
  {Gulliford}, \citenamefont {Bartnik}, \citenamefont {Dobbins}, \citenamefont
  {Sagan}, \citenamefont {Berg},\ and\ \citenamefont
  {Nunez-delPrado}}]{ref:CBETAVM}%
  \BibitemOpen
  \bibfield  {author} {\bibinfo {author} {\bibfnamefont {C.}~\bibnamefont
  {Gulliford}}, \bibinfo {author} {\bibfnamefont {A.}~\bibnamefont {Bartnik}},
  \bibinfo {author} {\bibfnamefont {J.}~\bibnamefont {Dobbins}}, \bibinfo
  {author} {\bibfnamefont {D.}~\bibnamefont {Sagan}}, \bibinfo {author}
  {\bibfnamefont {J.}~\bibnamefont {Berg}}, \ and\ \bibinfo {author}
  {\bibfnamefont {A.}~\bibnamefont {Nunez-delPrado}},\ }in\ \href@noop {}
  {\emph {\bibinfo {booktitle} {Proceedings of the 13th International
  Computational Accelerator Physics Conference}}}\ (\bibinfo {year}
  {2018})\BibitemShut {NoStop}%
\bibitem [{ref()}]{ref:EPICS}%
  \BibitemOpen
  \href@noop {} {}\bibinfo {note} {\url{https://epics.anl.gov/}}\BibitemShut
  {NoStop}%
\bibitem [{\citenamefont {Gulliford}\ \emph
  {et~al.}(2015{\natexlab{b}})\citenamefont {Gulliford}, \citenamefont
  {Bartnik},\ and\ \citenamefont {Bazarov}}]{ref:coldgun}%
  \BibitemOpen
  \bibfield  {author} {\bibinfo {author} {\bibfnamefont {C.}~\bibnamefont
  {Gulliford}}, \bibinfo {author} {\bibfnamefont {A.}~\bibnamefont {Bartnik}},
  \ and\ \bibinfo {author} {\bibfnamefont {I.}~\bibnamefont {Bazarov}},\ }\href
  {http://arxiv.org/abs/1510.07738} {\enquote {\bibinfo {title} {Cold electron
  beams from cryo-cooled, alkali antimonide photocathodes},}\ }\bibinfo
  {howpublished} {\url{http://arxiv.org/abs/1510.07738}} (\bibinfo {year}
  {2015}{\natexlab{b}})\BibitemShut {NoStop}%
\bibitem [{\citenamefont {Gulliford}\ \emph {et~al.}(2017)\citenamefont
  {Gulliford}, \citenamefont {Bartnik}, \citenamefont {Bazarov},\ and\
  \citenamefont {Maxson}}]{ref:RFgunUED}%
  \BibitemOpen
  \bibfield  {author} {\bibinfo {author} {\bibfnamefont {C.}~\bibnamefont
  {Gulliford}}, \bibinfo {author} {\bibfnamefont {A.}~\bibnamefont {Bartnik}},
  \bibinfo {author} {\bibfnamefont {I.}~\bibnamefont {Bazarov}}, \ and\
  \bibinfo {author} {\bibfnamefont {J.}~\bibnamefont {Maxson}},\ }\href
  {\doibase 10.1103/PhysRevAccelBeams.20.033401} {\bibfield  {journal}
  {\bibinfo  {journal} {Phys. Rev. Accel. Beams}\ }\textbf {\bibinfo {volume}
  {20}},\ \bibinfo {pages} {033401} (\bibinfo {year} {2017})}\BibitemShut
  {NoStop}%
\bibitem [{\citenamefont {Thieberger}\ \emph {et~al.}(2018)\citenamefont
  {Thieberger}, \citenamefont {Gassner}, \citenamefont {Hulsart}, \citenamefont
  {Michnoff}, \citenamefont {Miller}, \citenamefont {Minty}, \citenamefont
  {Sorrell},\ and\ \citenamefont {Bartnik}}]{ref:tburger}%
  \BibitemOpen
  \bibfield  {author} {\bibinfo {author} {\bibfnamefont {P.}~\bibnamefont
  {Thieberger}}, \bibinfo {author} {\bibfnamefont {D.}~\bibnamefont {Gassner}},
  \bibinfo {author} {\bibfnamefont {R.}~\bibnamefont {Hulsart}}, \bibinfo
  {author} {\bibfnamefont {R.}~\bibnamefont {Michnoff}}, \bibinfo {author}
  {\bibfnamefont {T.}~\bibnamefont {Miller}}, \bibinfo {author} {\bibfnamefont
  {M.}~\bibnamefont {Minty}}, \bibinfo {author} {\bibfnamefont
  {Z.}~\bibnamefont {Sorrell}}, \ and\ \bibinfo {author} {\bibfnamefont
  {A.}~\bibnamefont {Bartnik}},\ }\href {\doibase 10.1063/1.5021607} {\bibfield
   {journal} {\bibinfo  {journal} {Review of Scientific Instruments}\ }\textbf
  {\bibinfo {volume} {89}},\ \bibinfo {pages} {043303} (\bibinfo {year}
  {2018})},\ \Eprint {http://arxiv.org/abs/https://doi.org/10.1063/1.5021607}
  {https://doi.org/10.1063/1.5021607} \BibitemShut {NoStop}%
\bibitem [{ref(2012)}]{ref:psfish}%
  \BibitemOpen
  \href {http://laacg1.lanl.gov/laacg/services/download_sf.phtml} {}\bibinfo
  {howpublished}
  {\url{http://laacg1.lanl.gov/laacg/services/download_sf.phtml}} (\bibinfo
  {year} {2012})\BibitemShut {NoStop}%
\bibitem [{\citenamefont {{Helms}}\ and\ \citenamefont
  {{Hoffstaetter}}(2005)}]{ref:bpmgeorg}%
  \BibitemOpen
  \bibfield  {author} {\bibinfo {author} {\bibfnamefont {R.~W.}\ \bibnamefont
  {{Helms}}}\ and\ \bibinfo {author} {\bibfnamefont {G.~H.}\ \bibnamefont
  {{Hoffstaetter}}},\ }\href {\doibase 10.1103/PhysRevSTAB.8.062802} {\bibfield
   {journal} {\bibinfo  {journal} {Physical Review Accelerators and Beams}\
  }\textbf {\bibinfo {volume} {8}},\ \bibinfo {eid} {062802} (\bibinfo {year}
  {2005})},\ \Eprint {http://arxiv.org/abs/physics/0406145} {physics/0406145}
  \BibitemShut {NoStop}%
\end{thebibliography}
\end{document}